\documentclass[12pt]{article}
\usepackage{amsmath,amssymb}
\usepackage{graphicx,psfrag,epsf}
\usepackage{enumerate}
\usepackage{color}
\usepackage{amsfonts}
\usepackage{amsthm}
\usepackage{multirow}
\usepackage{xurl}
\usepackage{comment}

\usepackage{xr}

\makeatletter
\newcommand*{\addFileDependency}[1]{
  \typeout{(#1)}
  \@addtofilelist{#1}
  \IfFileExists{#1}{}{\typeout{No file #1.}}
}
\makeatother

\newcommand*{\myexternaldocument}[1]{%
    \externaldocument{#1}%
    \addFileDependency{#1.tex}%
    \addFileDependency{#1.aux}%
}
\myexternaldocument{supplement}

\DeclareMathOperator*{\argmax}{argmax}
\DeclareMathOperator*{\argmin}{argmin}

\usepackage{algorithm} 
\usepackage{algpseudocode} 

\newlength\myindent
\newcommand{\blind}{0}

\newtheorem{definition}{Definition}

\newtheorem{theorem}{Theorem}

\usepackage[backend=bibtex,
style=authoryear,
natbib=true,
maxcitenames=2,
firstinits=true,
maxbibnames=99]
{biblatex}
\newcommand\restr[2]{{
		\left.\kern-\nulldelimiterspace 
		#1 
		\right|_{#2} 
}}

\begin{document}

\def\spacingset#1{\renewcommand{\baselinestretch}%
{#1}\small\normalsize} \spacingset{1}


\if0\blind
{
  \title{\bf \emph{funBIalign}: a hierachical algorithm for functional motif discovery based on mean squared residue scores}
  \author{Jacopo Di Iorio 
        \thanks{
            J.~Di Iorio acknowledges the support of the Penn State Eberly College of Science.}\hspace{.2cm}\\
    Dept. of Statistics, Penn State University\\
    \\
    Marzia A. Cremona  
        \thanks{
            M.A.~Cremona is also affiliated to CHU de Qu\'ebec – Universit\'e Laval Research Center.
            M.A.~Cremona acknowledges the support of the Natural Sciences and Engineering Research Council of Canada (NSERC), of the Fonds de recherche du Québec Health (FRQS), and of FSA, Université Laval.}\hspace{.2cm}\\
    Dept. of Operations and Decision Systems, Université Laval \\
    \\
    Francesca Chiaromonte 
        \thanks{F.~Chiaromonte is also affiliated to the Inst.~of Economics and L'EMbeDS, Sant'Anna School of Advanced Studies.
        F.~Chiaromonte acknowledges support from the Huck Institutes of the Life Sciences, Penn State.}\hspace{.2cm}\\
    Dept. of Statistics, Penn State University}
  \maketitle
} \fi

\if1\blind
{
  \bigskip
  \bigskip
  \bigskip
  \begin{center}
    {\LARGE\bf Title}
\end{center}
  \medskip
} \fi

\bigskip
\begin{abstract}

Motif discovery is gaining increasing attention in the domain of functional data analysis. 
Functional motifs are typical ``shapes'' or ``patterns'' that recur multiple times in different portions of a single curve and/or in misaligned portions of multiple curves. 
In this paper, we define functional motifs using an additive model and we propose \emph{funBIalign} for their discovery and evaluation. 
Inspired by clustering and biclustering techniques, \emph{funBIalign} is a multi-step procedure which uses agglomerative hierarchical clustering with complete linkage and a functional distance based on mean squared residue scores to discover functional motifs, both in a single curve (e.g., time series) and in a set of curves. 
We assess its performance and compare it to other 
recent methods through extensive simulations. Moreover, we use {\em funBIalign} for discovering motifs in two real-data case studies; one on food price inflation and one on temperature changes. 
\end{abstract}

\noindent%

\spacingset{1.5}
\section{Background and motivation}
\label{sec:intro}


The last decades have seen an increasing interest in the analysis of functional data, i.e.,~data that can be represented as smooth curves. 
Functional Data Analysis (FDA) methods \citep[see, e.g.,][]{ramsay2010fda,ferraty2006nonparametric,kokoszka2017introduction} have been applied in a variety of scientific fields. These include biology, medicine, and genetics, where FDA has been employed to analyze, e.g.,~the genomic landscape of ``jumping genes'' and COVID-19 epidemics \citep{chen2020,boschi2021};
neurosciences and psychometrics, where it has been used, e.g.,~to map cognitive processes analyzing response times and brain imaging data \citep{buckner2004unified, lila2017functional}; 
economics and environmental sciences, where it has been used to explore patterns and generate predictions over space or time concerning air pollution, climate indicators, stock market prices, etc (e.g.,~ \cite{das2019effect,ghumman2020functional,}).

In this paper, we 
focus on \emph{functional motif discovery}; that is, the identification of typical “shapes” or “patterns” that recur multiple times in different portions of a single curve and/or in misaligned portions of multiple curves. 
While the ability to identify such patterns offers great promise in multiple scientific fields \citep{cremona2022}, to the best of our knowledge, the notion of functional motif still lacks a rigorous statistical formalization. 
We define a functional motif $Q$ of length $l$ as a collection of $n_Q$ curve portions $p_k(t)$ $k = 1, \dots, n_Q$  with $t \in [0,l]$ (the occurrences of $Q$) obeying the additive model
$$
    p_k(t) = \mu^{Q} + \alpha_{k}^{Q} + \beta^{Q}(t) + \varepsilon_{k}(t) \hspace{10pt} \forall t \in [0,l]
$$ 
where $\mu^{Q}$ is the mean of the motif, $\alpha_{k}^{Q}$ its portion-specific adjustment, $\beta^{Q}(t)$ its t-varying adjustment and $\varepsilon_k(t)$ an error term (see Fig.~\ref{fig:fourtypes}). 

In order to discover functional motifs, both in a single curve or in a set of curves, we develop \emph{funBIalign}, a multi-step algorithm that requires as input, in addition to the curve(s) themselves, only the length $\ell$ and the minimum cardinality $n_{min}$ of the motifs to be discovered. 
\emph{funBIalign} performs a comprehensive scan of all portions of length $\ell$ of all curves in the data, and arranges them in a dendrogram employing agglomerative hierarchical clustering with complete linkage and a functional generalization of the Mean Squared Residue Score (MRS) -- a measure typically employed by biclustering techniques \citep{pontes2015biclustering}. 
The dendrogram is dynamically cut to identify a set of candidate functional motifs, which are then post-processed to select the most interesting.
While hierarchical agglomerations are commonly used for functional clustering (see
\cite{ferreira2009comparison} for a comparison, and \cite{jacques2014functional} for a survey on functional clustering techniques), the use of means squared residues still represents a novelty in the functional framework. 
The MSR was originally introduced by \cite{cheng2000}, for discovery and validation of biclusters, i.e.~of subsets of rows and columns of a data matrix, and has been widely used since in the multivariate setting (e.g.,~\cite{liu2007computing, angiulli2008random, yang2005improved}), but its functional generalization, fMRS, has only very recently been employed in functional clustering problems by \citet{galvani2021funcc} and \cite{diiorio2023funloci}. 

Functional motif discovery, while gaining increasing attention, is itself an under-explored area. 
To the best of our knowledge, the only other approach that deals specifically with it in the FDA domain is probabilistic $K$-means with local alignment (\emph{probKMA}, \cite{cremona2022}).
Following a stream of literature devoted to the simultaneous alignment and clustering of curves at the global level \citep[see, e.g.,][]{liu2009simultaneous, sangalli2010k}, and to the simultaneous domain selection and clustering of curves \citep{fraiman2016feature, floriello2017sparse, vitelli2019novel}, \emph{probKMA} identifies candidate functional motifs combining a probabilistic $K$-means algorithm and local alignment (or domain selection) techniques. 
A strong point of {\em probKMA} is the fact that, while requiring the specification of a minimum motif length, it can extend such length in a motif-specific and data-driven fashion. 
However, it is designed to operate on a set of curves and it requires extensive post-processing. 
As a counterpart to its efficacy, the need for multiple initializations of the $K$-means algorithm and the complexity of post-processing make {\em probKMA} computationally expensive.
Outside the FDA domain, a problem similar to functional motif discovery has been tackled by the data mining community -- seeking patterns embedded multiple times within a single time series \citep[see, e.g.][]{lonardi2002finding,mueen2009exact,yeh2016matrix}.
This work are based on a $k$ Nearest Neighbors algorithm and requires users to specify several parameters, some of which may not be intuitive and significantly affect outcomes (newer versions of these procedures have been published since 2016, but they primarily enhance computational performance on large data sets, rather than modifying required inputs or algorithmic approach).

As mentioned above, and in contrast to existing methods, \emph{funBIalign} can naturally handle both single curves and sets of curves. 
In the latter case, thanks to the hierarchy it creates, it can also highlight relationships among curves harboring the same motif.
On a different front, although the comprehensive scan employed by {\em funBIalign} can be computationally demanding, it must be performed only once -- without the need for multiple runs or initializations as in {\em probKMA}. 
More generally, {\em funBIalign} can effectively tackle applications where functional alignment fails or may be inadequate, such as identifying motifs embedded consecutively, or appearing only in one curve. 
We also note that the two input parameters required by our procedure, the motif length $\ell$ and the minimum number of portions $n_{min}$, while impacting outcomes, are user-friendly and intuitive. 

The remainder of the paper is organized as follows. 
Section \ref{sec:meth} presents the theoretical setting of \emph{funBIalign}, including the rigorous definition of functional motifs and of the dissimilarity measure employed. 
The algorithmic implementation is described in Section \ref{sec:funBIalign_algo}. 
Finally, the performance of our proposal is assessed through simulations, comparisons with other available methods, and real-data case studies in Sections \ref{sec:simulation}, \ref{sec:comparison_main} and \ref{sec:case}.

\section{Theoretical setting}
\label{sec:meth}

\subsection{Model-based definition of functional motifs}
\label{sec:defin_additive}

Consider a set of real-valued curves $f_{i}(t)$, $i = 1, \dots, N$, each defined on a compact interval which we assume to be $[0,T_i]$ without loss of generality. 
Intuitively, a functional motif 
is a ``shape'' or ``pattern'', defined on a domain interval of given length $l$, which occurs multiple times within the set of curves -- possibly with noise. 
A motif can occur at different positions within a curve and/or in misaligned portions of different curves in the set ($N=1$ corresponds to the 
case in which motifs are sought within a single curve, e.g.,~a time series). 
For each curve $f_{i}(t)$, we consider all possible overlapping portions of length $l$, and we align them on the interval $[0,l]$. 
In symbols, a generic portion is $f_{i,L}(t) = \restr{f_i}{L} \circ h(t)$ with $t \in [0,l]$, where $L$ is a sub-interval of length $l$ of $[0,T_i]$ and $h$ is a shift transformation from $[0,l]$ to $L$.
Let $I$ be the collection of all portions of length $l$ of all curves in the set, i.e.~$I = \{p(t) =  f_{i,L}(t)  \mid i = 1,\dots,N \text{ and } L 
\subseteq [0,T_i] \text{ with } |L| = l \}$. 
We provide a rigorous definition of functional motif using an additive model as follows.

\begin{figure}[!b]
    \centering
    \includegraphics[width=0.5\linewidth]{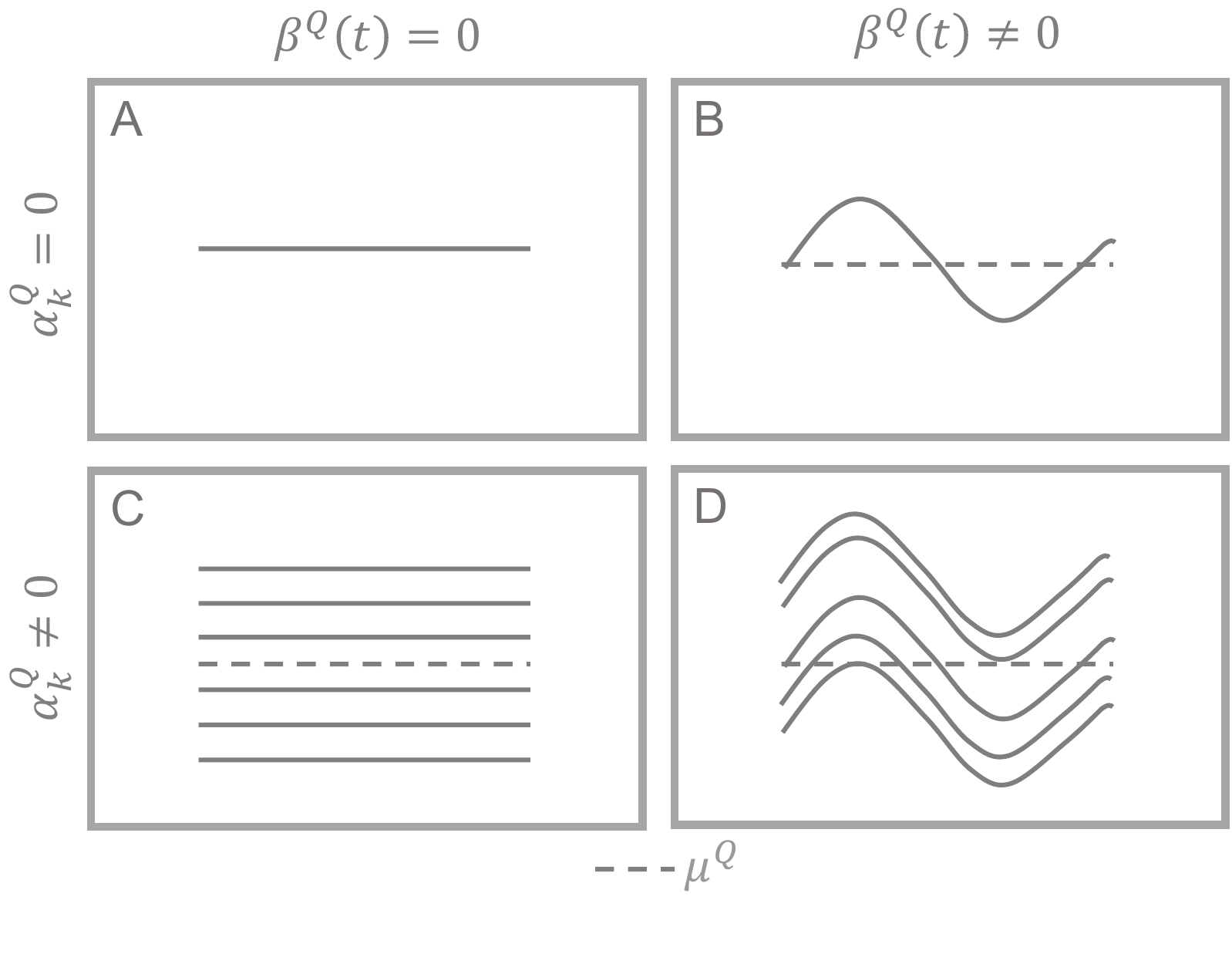}
    \caption{
        Examples of ideal functional motifs, obtained by (\ref{eq:additive_model}) with $\varepsilon_{k}(t) = 0$. 
        In panel A 
        all portions are identical and constant; $p_{k}(t) = \mu^Q$ ($\alpha_{k}^Q=0$ and $\beta^{Q}(t)=0$).
        In panel B all portions are identical; $p_{k}(t) = \mu^Q +\beta^{Q}(t)$ ($\alpha_{k}^Q=0$).
        In panel C portions are constant and parallel; $p_{k}(t) = \mu^Q + \alpha_{k}^Q$ ($\beta^{Q}(t)=0$).
        Panel D illustrates the general case, with parallel portions sharing the same shape; $p_{k}(t) = \mu^Q + \alpha_{k}^Q +\beta^{Q}(t)$.
        In all panels, the dashed line represents the motif mean $\mu^Q$. }
	\label{fig:fourtypes}
\end{figure}

\begin{definition}
    A functional motif $Q$ is a collection $I_Q \subset I$ of $n_Q > 1$ curve portions $p_{k}(t)$ $k = 1, \dots, n_Q$ such that 
    \begin{equation}
        \label{eq:additive_model}
        p_{k}(t) = \mu^{Q} + \alpha_{k}^{Q} + \beta^{Q}(t) + \varepsilon_{k}(t) \hspace{10pt} \forall t \in [0,l]
    \end{equation}
    where $\mu^{Q}$ is 
    a mean level, $\alpha_{k}^{Q}$ a portion-specific adjustment, $\beta^{Q}(t)$ a t-varying adjustment, and $\varepsilon_{k}(t)$ an error term.
\end{definition}

\noindent To have unique and identifiable parameters, we impose $\sum_{k=1}^{n_Q}\alpha_{k}^{Q} = 0$, \mbox{$\int_{0}^{l}\beta^{Q}(t)dt = 0$}, and $\mathbb{E}[\varepsilon_{k}(t)]= 0$ for $\forall k = 1,\ldots, n_Q$ and $\forall t \in [0,l]$.
It is important to remark that, even if the definition above 
does not consider explicitly the curve to which each portion belongs to, this information is crucial in the algorithm 
we introduce in Section~\ref{sec:funBIalign_algo}.

Definition~\ref{eq:additive_model} 
is inspired by the biclustering literature, and in particular by the definition of multivariate coherent evolution biclusters \citep{madeira2004biclustering} utilized in the seminal paper by \citet{cheng2000}.
In the functional framework, \citet{galvani2021funcc} used a model similar to the one in (\ref{eq:additive_model}) to discover functional biclusters in a data matrix whose cells correspond to curves, while \cite{diiorio2023funloci} employed the model in (\ref{eq:additive_model}) to identify local clusters in subsets of aligned curves. 
Following the biclustering literature, we call a functional motif ``ideal'' when $\varepsilon_{k}(t) = 0$, i.e.~in the absence of noise.
Such a motif is composed by perfectly parallel portions sharing the same shape (Fig. \ref{fig:fourtypes}D). 
Among ideal motifs, special cases can be obtained setting $\alpha_{k}^Q$ and/or $\beta^{Q}(t)$ to $0$. 
If both $\alpha_{k}^Q = 0$ and $\beta^{Q}(t) = 0$ we have a constant motif, whose portions are all $p_{k}(t) = \mu^Q$ (Fig. \ref{fig:fourtypes}A).
Setting $\beta^{Q}(t) = 0$ but allowing $\alpha_{k}^Q \ne 0$ we have parallel constant portions $p_k(t) = \mu^Q +\alpha_{k}^{Q}$ (Fig. \ref{fig:fourtypes}C), 
while setting $\alpha_{k}^Q = 0$ but allowing $\beta^{Q}(t) \neq 0$ all portions will be identical to $p_{k}(t) = \mu^Q + \beta^{Q}(t)$ (Fig. \ref{fig:fourtypes}B).

\subsection{Evaluating the coherence of functional motifs}

To evaluate the coherence of a candidate motif, i.e.~of a collection of portions, to an 
(unknown) additive model, we gauge the error term $\varepsilon_k(t)$ using a functional version of the Mean Squared Residue score (MSR), or H-score, first introduced by \citet{cheng2000} to seek large biclusters among the rows (genes) and columns (experimental conditions) of a gene expression data matrix. 
\citet{galvani2021funcc} first proposed a generalization of the MSR to the functional framework to seek large biclusters in matrices of curves.
Here, we define the functional Mean Squared Residue score (fMRS) as follows.

\begin{definition}
    The functional Mean Squared Residue score (fMSR) 
    of a functional motif $Q$ is 
    \begin{equation}
        \label{hscore}
        H(Q) = \frac{1}{n_Q}\frac{1}{l}
            \sum_{k = 1}^{n_Q} \int_{0}^{l}
    	\left(
    		p_{k}(t) - \left(\hat{\mu}^Q + \hat{\alpha}^{Q}_{k} + \hat{\beta}^{Q}(t)\right)
    	\right)^{2}dt,
    \end{equation}
    where the estimates $\hat{\mu}^Q$, $\hat{\alpha}^{Q}_{k}$ and $\hat{\beta}^{Q}(t)$ of the terms 
    in (\ref{eq:additive_model}) are 
    \begin{equation*}
        \hat{\mu}^Q = \overline{p} = \frac{1}{n_Q}\frac{1}{l}\sum_{k = 1}^{n_Q}\int_{0}^{l}p_{k}(t)dt, 
    \end{equation*}
    \begin{equation*}
        \hat{\alpha}^{Q}_{k} = 
        \overline{p}_{k} - \hat{\mu}^Q = \frac{1}{l}\int_{0}^{l}p_{k}(t)dt \hspace{5pt} - \hat{\mu}^{Q}, 
    \end{equation*}
    \begin{equation*}
        \hat{\beta}^{Q}(t) = \overline{p}(t) - \hat{\mu}^{Q} = \frac{1}{n_Q}\sum_{k=1}^{n_Q}p_{k}(t) \hspace{5pt} - \hat{\mu}^{Q}.
    \end{equation*}
\end{definition}

\noindent
In forming the estimates, $\overline{p}_{k}$, $\overline{p}(t)$ and $\overline{p}$ represent, respectively, the mean value of each portion $p_{k}(t)$, the functional mean of all portions $p_{k}(t)$, $k= 1, \ldots, n_Q$, and the mean value of $\overline{p}(t)$ (or equivalently, the mean of the $\overline{p}_{k}$'s).
Using them,
we can rewrite (\ref{hscore}) as 
\begin{equation}
    \label{hscore_practioners}
    H(Q) = \frac{1}{n_Q}\frac{1}{l}
    	\sum_{i=1}^{n_Q} \int_{0}^{l}
    	\left(
    		p_{k}(t) - \overline{p}_{k} - \overline{p}(t) + \overline{p}
    	\right)^{2}dt
\end{equation}
which 
allows us to easily implement the score calculation.
We observe that the fMSR of an ideal, i.e.~noiseless, functional motif $Q$ 
is $H(Q)=0$. 
Thus, similar to biclustering methods which look for biclusters with low MSR, we look for functional motifs with low fMSR. 

\citet{di2020bias} recently proved that the 
MSR is biased towards small biclusters (i.e.,~biclusters with a small number of rows and/or columns). 
Here, we prove that the fMSR suffers from a similar bias towards small functional motifs; that is, motifs comprising a small number of portions.
This can hinder motif discovery, as it distorts the comparisons of motifs comprising different numbers of portions.
The following theorem fully characterizes this bias
and suggests a way of correcting it (a proof is provided in Section \ref{sec:s1_proof}). 

\begin{theorem}
\label{theorem:bias}
    Let $Q$ be a functional motif.
    For $n=2,3,\dots, n_Q$ let $\overline{H}_{n}$ be the average 
    fMRS of all sub-motifs of $Q$ obtained selecting exactly $n$ of the $n_Q$ portions $p_{k}(t)$ belonging to $Q$.
    Then
    \begin{equation}
        \overline{H}_{n+1} = \overline{H}_{n} \frac{n^2}{n^2-1}.
    \label{eq:ratio}
    \end{equation}
\end{theorem}

\noindent
An implication of Theorem \ref{theorem:bias} is that the ratio $r_{n,n+1} = \frac{\overline{H}_{n+1}}{\overline{H}_{n}}$ depends on the number of portions $n$ included in the sub-motif. 
It is also straightforward to verify that
\begin{equation}
    \overline{H}_{n+m} = \overline{H}_{n}\prod_{x=n}^{n+m-1}\frac{x^2}{x^2-1}
\label{eq:evo}
\end{equation}
for every $m>0$, allowing one to compute $\overline{H}_{n+m}$ using $\overline{H}_{n}$ alone. 
Hence, we have that
\begin{equation}
    r_{n,n+m}=\frac{\overline{H}_{n+m}}{\overline{H}_{n}}
    \xrightarrow[m \to \infty]{} \prod_{x=n}^{\infty} \frac{x^2}{x^2-1}.
\label{eq:infprod}
\end{equation}
The infinite product in (\ref{eq:infprod}) converges and we have $r_{2,2+m} \to 2$. 
As a consequence, the bias can be at most $1$, and it decreases when considering motifs comprising a large number of portions \citep[see][for simulations and additional details in the multivariate case]{di2020bias}. 
However, its effects can be troublesome in comparisons involving seldom motifs, which will be non-negligibly favored.
For this reason, we propose to correct the bias defining an {\em adjusted} version of the fMSR as follows.

\begin{definition}
Let $Q$ be a functional motif comprising $n_Q \geq 2$ portions. Its adjusted functional mean squared residue score
is defined as:
\begin{equation}
    H_{adj}(Q) = 
    \begin{cases}
    	H(Q) & \text{if } n_Q=2, \\
    	\frac{H(Q)}{\prod_{r=2}^{n_Q-1} \frac{r^2}{r^2-1}} & \text{if } n_Q>2.
    \end{cases}
\end{equation}
\end{definition}

\noindent
The adjusted fMRS is the measure we employ in the remainder of the paper.


\subsection{An fMSR-based dissimilarity measure}

The adjusted fMSR can be used to construct a dissimilarity measure between two curve portions $p_{1}(t)$ and $p_{2}(t)$ in $[0,l]$ as 
\begin{equation}
    d_{{\bf fMSR}}(p_{1}, p_{2}) = H_{adj}(W),
\end{equation}
where $W = \{p_1(t), p_2(t)\}$ is the functional motif composed only by the portions $p_1(t)$ and $p_2(t)$. 
According to this definition, $d_{{\bf fMSR}}(p_1, p_2) = 0$ if and only if $W$ is an ideal motif. 

\section{The \emph{funBIalign} algorithm}
\label{sec:funBIalign_algo}

Given a set of $N \geq 1$ real-valued curves $f_i(t)$ defined on $[0,T_i]$, $i = 1, \dots, N$, 
\emph{funBIalign} discovers recurrent and coherent 
motifs as defined by (\ref{eq:additive_model}). 
The algorithm considers the evaluation of the curves over a grid of equally spaced points $t_0=0, t_1=\delta t, t_2=2 \delta t, \dots$
and requires as input the {\em discretized} motif length $\ell$ (the number of grid points corresponding to the length $l$ in (\ref{eq:additive_model})) and the minimum number of portions $n_{min}$.
It comprises four steps, as described below 
(a schematic of the first three is provided in Fig.~\ref{sup:algo_scheme}).
\vspace{-0.1in}
\paragraph{Step 1 - Portion creation and alignment.}
For every curve $f_i(t), i = 1, \dots, N$, we create all $n_i$ portions
of length $\ell$ (starting at $t_0, t_1, t_2$, etc.), and align them so that their domains all start at $t=0$. 
We indicate the resulting overall set of aligned portions with $p_j(t)$, $j=1,\dots,n$, where $n = \sum_{i=1}^N n_i$.
For each portion, we keep track of the originating curve $i(j)$ and of the grid points occupied $t_j,\dots,t_{j+\ell}$. 
\vspace{-0.2in}
\paragraph{Step 2 - Hierarchical clustering based on the fMRS dissimilarity.}
For every portion pair $(p_{j_1}, p_{j_2}), 1 \leq j_1, j_2 \leq n$, we compute the fMRS-based dissimilarity $d_{\bf fMSR}(p_{j_1}, p_{j_2})$.
If two portions originate from the same curve, $i(j_1)=i(j_2)$, and share at least $50\%$ of their grid points, we name them ``acolytes'' and artificially increase their dissimilarity by $M = \max_{j_1,j_2}\left( d_{\bf fMSR}(p_{j_1}, p_{j_2}) \right)$ to prevent obvious similarities among curve portions with large overlaps from dominating the agglomeration.
Our dissimilarities are thus
\begin{equation}
	d_{j_1,j_2} = 
	\begin{cases}
		d_{\bf fMSR}(p_{j_1}, p_{j_2})+M & \text{if } p_{j_1} \text{ and } p_{j_2} \text{ are acolytes,} \\
		d_{\bf fMSR}(p_{j_1}, p_{j_2}) & \text{otherwise}
	\end{cases}
\end{equation}
and we use their matrix $D_{\bf fMRS} = [d_{j_1, j_2}]_{1 \leq j_1 \leq n, 1 \leq j_2 \leq n}$ to perform agglomerative hierarchical clustering with complete linkage \citep[see, e.g.,][]{murtagh2012algorithms}.
Let $Tree$ indicate the resulting dendrogram, 
whose nodes represent 
clusters of portions. 
Due to complete linkage and the increased acolyte dissimilarity, the longest branches occur when merging nodes comprising acolytes.
We use such longest branches to cut the dendrogram, generating  $N_{tree}$ sub-trees $tree_{s}$, $s = 1, \dots, N_{tree}$, which do not contain acolytes.
\vspace{-0.2in}
\paragraph{Step 3 - Collection of candidate functional motifs.}

For every sub-tree $tree_{s}$, $s=1,\ldots, N_{tree}$, we consider the sets of all nodes $nodes_s$ and all leaf nodes $leaves_s$ ($leaves_s \subseteq nodes_s$), and for every $x \in nodes_s$ we consider the sets of all descendants $des(x)$ and ascendants $asc(x)$.
We define the seeds of $tree_{s}$ as the set of nodes with at least $n_{min}$ leaf descendants, and with no descendant meeting the same criterion; in symbols
\begin{equation}
    seed_s = \{x \in nodes_s : |des(x) \in leaves_s| \geq n_{min} \wedge \forall y \in des(x), y \notin seed_s \}.
\end{equation}
Note that a sub-tree can have zero, one or multiple seeds. 
Finally, we define the family of $v \in seed_s$ as the union of $v$ and all its ascendants that are not shared with other seeds: 
\begin{equation}
    family(v) = \{v\} \cup \{x \in asc(v) :  \nexists y \in seed_s, y\neq v \hspace{5pt} s.t. \hspace{5pt} x\in asc(y)\}.
\end{equation}
Next, we select a ``recommended representative'' for the family. 
If $|family(v)| = 1$, this is trivially the one family member.
If $|family(v)| > 1$, we sort the nodes $x \in family(v)$ in order of increasing cardinality $|x|$ (i.e.~number of portions) and consider $H_{adj}(x)$ (adjusted $fMSR$ value computed on the portions).
If $H_{adj}(x)$ increases with $rank(|x|)$, we use an ``elbow'' approach and select the node just before the maximum increase.
Otherwise, we select the node with minimum $H_{adj}(x)$. 
Our collection of candidate motifs ${\cal C}$ is composed of recommended representatives of all families from all sub-trees. 
%
\vspace{-0.2in}
\paragraph{Step 4 - Post-processing.}
We sort candidate motifs $x \in {\cal C}$ 
based either on adjusted fMSR, i.e.~$rank(H_{adj}(x))$, or on a combination of adjusted fMRS and inverse cardinality, i.e.~$rank(H_{adj}(x)) + rank(-|x|)$ (alternative ranking criteria can be used, see Section \ref{sec:case}). 
Starting from the top, we compare each motif to those with higher rank. 
If all portions of $x$ are acolytes to portions of an higher ranking motif, we filter it out; otherwise we retain it. This produces a final collection of discovered motifs ${\cal D} \subseteq {\cal C}$. 

\vspace{0.1in}
\noindent
We note that the dynamic cut in Step 3 selects candidate motifs controlling a trade off between cohesiveness (small adjusted fMRS) and prevalence (large number of occurrences).
In contrast, the post-processing in Step 4 eliminates overlapping candidates which might have been selected in different sub-trees. In addition, it provides an ``importance ranking'' which facilitates further exploration and selection among discovered motifs.

\section{Simulations}
\label{sec:simulation}

We assess the performance of \emph{funBIalign} through an extensive simulation study.
To simulate a smooth curve embedding multiple occurrences of functional motifs we use the flexible B-spline-based model proposed by \citet{cremona2022}.
Briefly, this model generates a smooth curve as $f(t) = \sum_{j=1}^{J}c_j\phi_{j}(t)$ where $\{\phi_{j}\}^{J}_{j=1}$ is a B-spline basis of order $n$ with equally spaced knots $t_1, . . . , t_{J - n + 2}$ and $c_j$, $j = 1, \dots, J$ are real coefficients.
In every simulation conducted in our study, we generate a single smooth curve of length $7000$ using order $n = 3$ and knots at distance $T_{knots} = 10$. The curve is then evaluated across a grid of 7001 equally spaced points ($t_0 = 0, t_1 = 1, \dots$).
We then randomly embed in the curve the same number (8 or 10) of occurrences of 4 distinct motifs of length $4T_{knot} = 40$ (i.e.~, $\ell = 41$ points). 
Coefficients defining both the curve and the motifs are randomly generated from a $Beta(-0.45, 0.45)$ and then rescaled to $[-15,15]$. 
We incorporate into every motif occurrence a vertical shift drawn uniformly from $[-10,10]$ as well as noise -- adding to the coefficients independent draws from $\mathcal{N}(0,\,\sigma^{2})$. 
We note that this way of generating motifs within a curve does {\em not} match the additive model of (\ref{eq:additive_model}); thus, in our simulations we are challenging {\em funBIalign} with motifs that may be harder for it to identify. 
We also note that the curve background can, by chance, comprise segments that resemble one of the motifs; that is, extra portions that were not intentionally embedded but do follow the motif pattern. 
Moreover, the background may reveal entirely different and distinguishable motifs; that is, patterns that, while not intentionally inserted in multiple occurrences, happen to repeat themselves along the curve. 
Extra portions or additional motifs emerging from the background introduce an added complexity in interpreting simulation results (see below).

We construct a total of $100$ simulations. We consider $10$ alternative sets of $4$ motifs; $2$ alternative numbers of occurrences ($8$ or $10$) which, for simplicity, are the same for all motifs; and $4$ alternative levels of noise, expressed by $\sigma=$ $0.1$, $0.5$, $1$ or $2$. 
In a first batch of $10 \times 2 \times 4 = 80$ simulations, all motifs share the same $\sigma$, and we use all $\sigma$ values in turn. In a second batch of $10 \times 2 = 20$ simulations, each motif is attributed a different $\sigma$. 
%
We run \emph{funBIalign} setting $\ell = 41$ (the true length of the motifs) and using different minimum cardinalities $n_{min}= 5, 6, 7, 8$ (progressively closer to the true number of occurrences, $8$ or $10$). 
We post-process the candidate motifs produced in each run, ranking them according to two criteria 
-- the adjusted fMSR and the rank sum -- and identifying results which are most similar to the intentionally embedded motifs. 
Here we focus on those; that is, on the detection of occurrences of the motifs we embedded in the curve, plus potential {\em extra} portions of the background that the algorithm associates with such motifs. 
{\em funBIalign} may ``discover'' in the background motifs other than those we created, but we will ignore them in the discussion to follow. 
We also restrict the main text presentation to the $40$ simulations where motifs have $8$ occurrences and shared noise levels. 
Results for simulations where motifs have $10$ occurrences and shared noise levels, or where motifs have $8$ or $10$ occurrences and different noise levels, are entirely consistent with those presented here and are provided in Sections \ref{sec:same} and \ref{sec:different}.

\begin{figure}[!b]
\includegraphics[width=\linewidth]{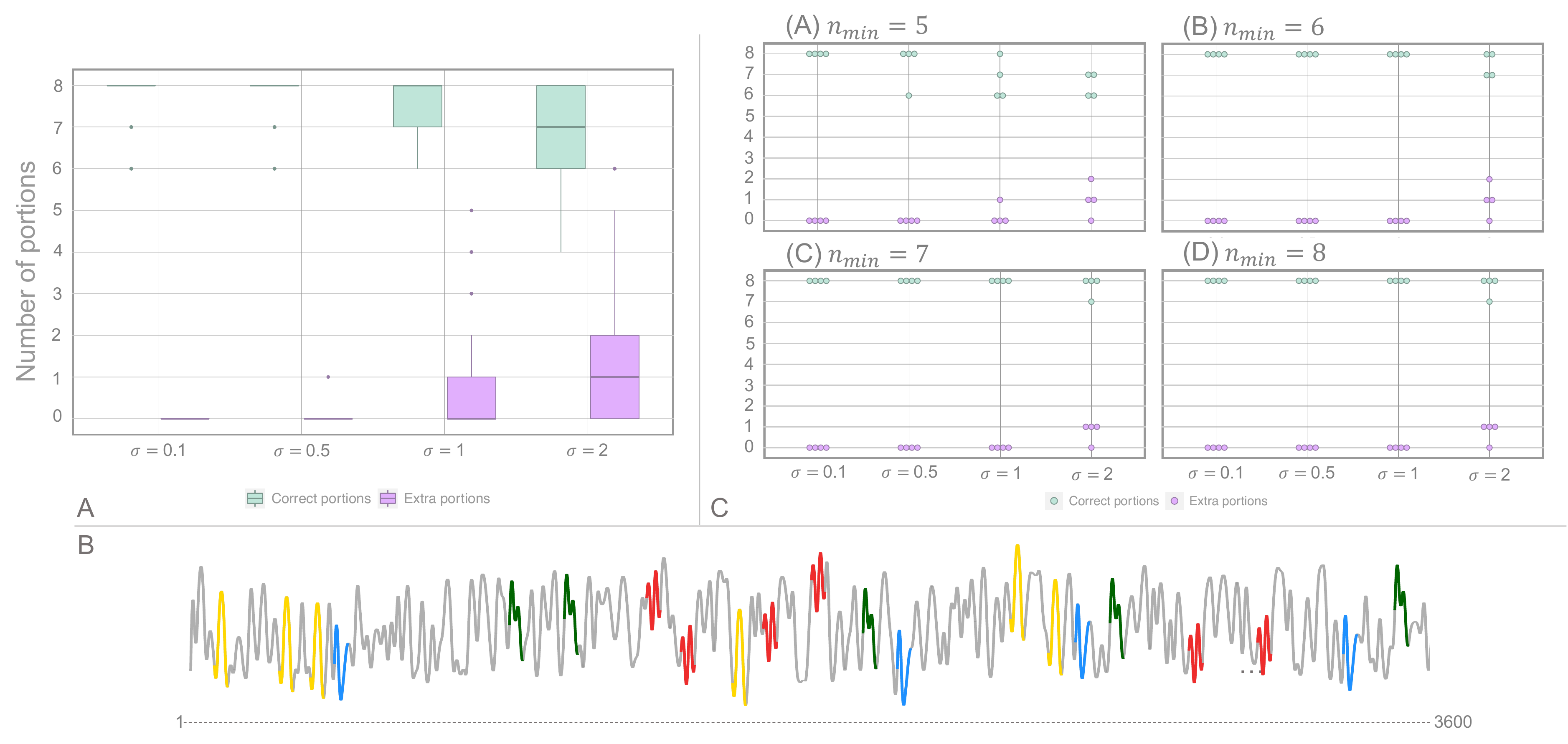}
\caption{
    \textbf{A.} Performance of \emph{funBIalign} for all simulations where motifs have $8$ occurrences and shared noise levels. 
    The algorithm is run with minimum cardinalities $n_{min}=5,6,7,8$ and results are pooled. 
    For each $\sigma$, the green and pink boxplots (left and right) represent correctly identified portions and extra portions, respectively. 
    \textbf{B.} First half of the curve for a simulation employing motif set n.~7 and $\sigma = 0.1$. Occurrences of the motifs are color-coded.
    \textbf{C.} Performance of {\em funBIalign} for simulations employing motif set n.~7, shown separately for runs with varying $n_{min}$. Green and pink jittered dots represent correctly identified portions and extra portions, respectively.
}
\label{fig:samemotif_summary_fig}
\end{figure}

\begin{figure}[!b]
\centering
\includegraphics[width=0.95\linewidth]{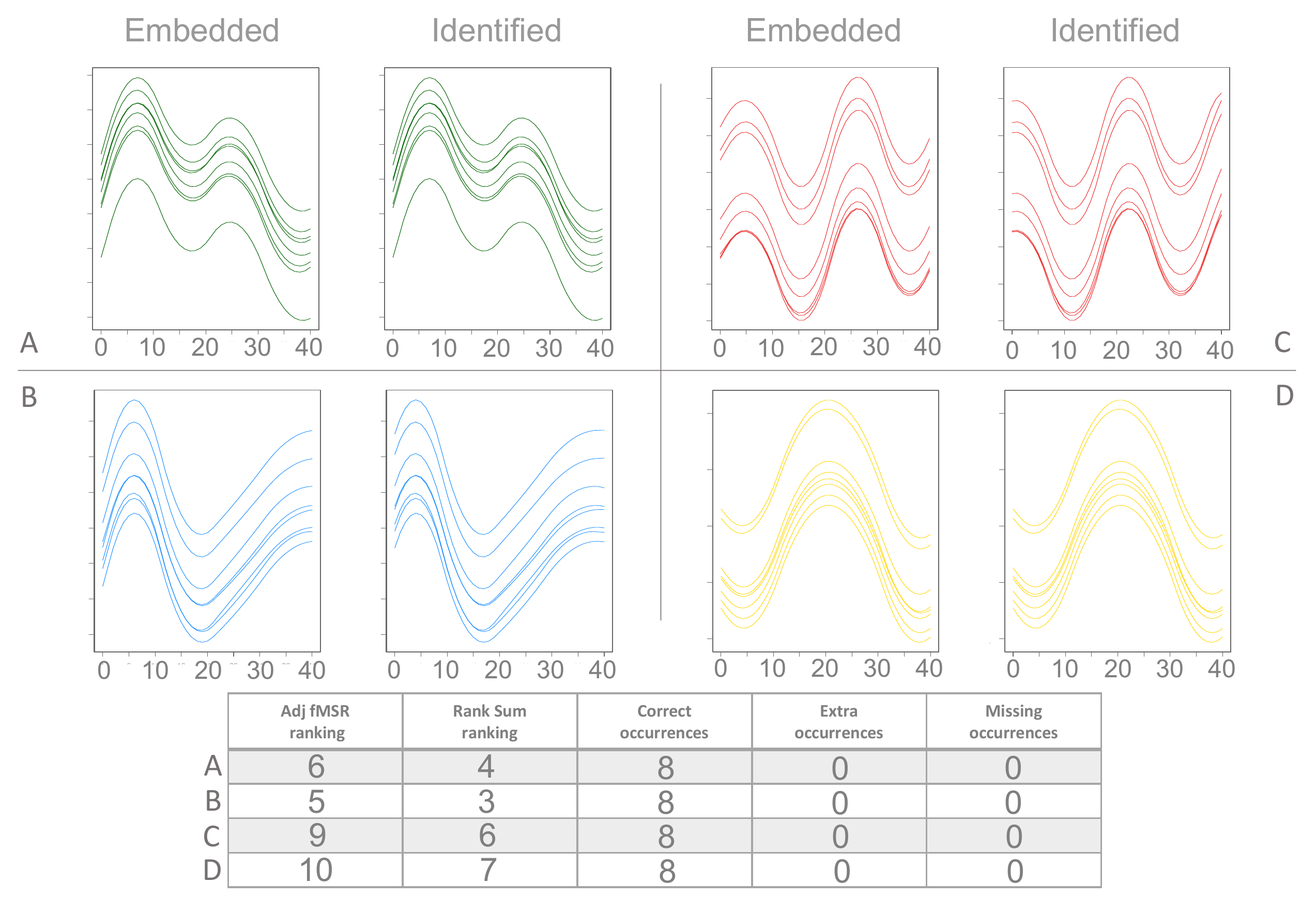}
\caption{
    Motif identification for the simulation employing motif set n.~7 and $\sigma=0.5$. {\em funBIalign} is run with $n_{min}=6$. 
    For each of the $4$ motifs, color-coded in green, red, blue and yellow, left and right panels show the $8$ occurrences embedded in the curve and the most similar portions identified by the algorithm, respectively. 
    The table provides rankings and numbers of correctly identified, extra and missing occurrences for each motif. With low noise, no embedded occurrences are missed, and no extra occurrences are identified.
}
\label{fig:simulation7_sd0.5_card6_details}
\end{figure}

\begin{figure}[!b]
\centering
\includegraphics[width=0.95\linewidth]{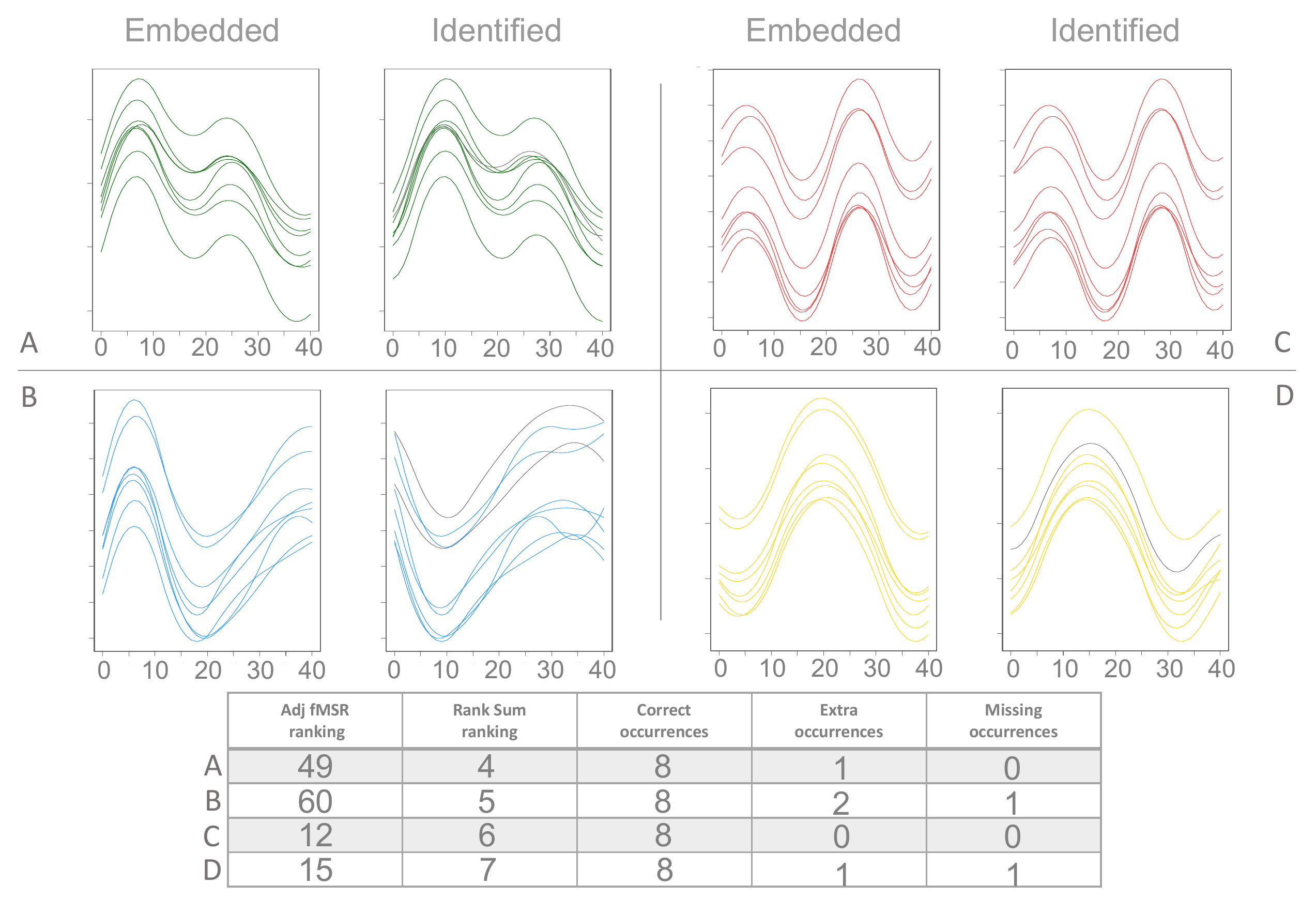}
\caption{
    Motif identification for the simulation employing motif set n.~7 and $\sigma=2$. {\em funBIalign} is run with $n_{min}=6$. 
    See legend for Fig.~\ref{fig:simulation7_sd0.5_card6_details}.
    With high noise, $2$ embedded occurrences are missed, and $4$ extra occurrences are identified (represented in gray among the identified portions panels). 
}
\label{fig:simulation7_sd2_card6_details}
\end{figure}

Fig.~\ref{fig:samemotif_summary_fig}A summarizes performance for the $40$ simulations,
pooling results across algorithm runs with different 
$n_{min}$'s. 
For every level of $\sigma$ we display two boxplots, each comprising $4 \times 10 \times 4 = 160$ values: for each of the $4$ motifs in each of the $10$ sets and across the $4$ runs, we count the number of portions correctly identified (left boxplot) and the number of extra portions (right boxplot).
We can see that {\it funBIalign} is quite effective in identifying embedded motif occurrences. 
However, as expected, as the level of noise increases some embedded portions are missed, and some extra portions are found -- though these are usually very similar to the embedded portions (see Fig.~\ref{fig:simulation7_sd2_card6_details} below).  
Similar results, again pooling across runs with different $n_{min}$'s, but separating each of the $10$ alternative motif sets, are provided in Fig.~\ref{fig:simulations_performance_8}. 
Fig.~\ref{fig:samemotif_summary_fig}C shows results (identification of true occurrences and of extra portions) for one motif set, n.~7, but separately for the $4$ runs with varying $n_{min}$.
The minimum cardinality used in the algorithm can indeed impact performance, especially through the ranking of the candidate motifs (see Figs.~\ref{fig:cardinality_effect_performance_8}-\ref{fig:cardinality_effect_ranking_8}).
When $n_{min}$ is much lower than the true number of occurrences, rank sum is preferable to adjusted fMSR as a ranking criterion, because it tends to privilege results more similar to the embedded motifs.
On the other hand, when $n_{min}$ is close to the true number of occurrences, best results do not necessarily have highest rank sums.
This fact can hinder their identification.
In addition, as expected, when $n_{min}$ is too low we miss some occurrences and, as $n_{min}$ gets higher, we identify more extra portions.
%
Fig.~\ref{fig:samemotif_summary_fig}B shows the first half of the curve for the simulation using motif set n.~7, with noise level $\sigma = 0.1$, color-coding motif occurrences interspersed across the curve.
Fig.~\ref{fig:simulation7_sd0.5_card6_details} provides details on the motifs identified in the simulation with noise level $\sigma = 0.5$, and Fig.~\ref{fig:simulation7_sd2_card6_details} on those 
corresponding to $\sigma = 2$, running {\em funBIalign} with $n_{min}=6$ in both cases.
Performance is excellent, though noisier motifs, as expected, cause a slight deterioration. 
We also see how, at the same level of noise, motifs may be easier or harder to identify depending on their shapes.
Importantly, we note that when {\em funBIalign} identifies extra portions, i.e.~segments of the background that resemble a motif by chance, these are indeed very similar to the occurrences intentionally embedded in the curve (Fig.~\ref{fig:simulation7_sd2_card6_details}). In effect, they should be thought of as ``unplanned'' true positives, not as false positives.

\section{Comparisons with related methods}
\label{sec:comparison_main}

We compare \emph{funBIalign} with two different methods; namely, {\em probKMA} (\cite{cremona2022}), a functional motif discovery algorithm employing probabilistic $K$-mean with local alignment, and {\em SCRIMP-MP} (\cite{zhu2018matrix}), one of the most recent Matrix Profile (MP) algorithms for motif discovery in univariate time series.
We focus on the ability of the three methods to correctly identify embedded motifs, but refrain from conducting a comparative analysis of their computational burdens; such an analysis is very hard to perform rigorously due to differences in coding languages employed, degree of code optimization, and overall structure of the pipelines being compares (e.g., number and nature of algorithmic and post-processing parameters to be fixed or tuned). 

\emph{funBIalign} and {\em probKMA} differ in various respects. 
\emph{funBIalign} employs agglomerative clustering with adjusted fMSR, which generates a complete hierarchy of all curve portions. In contrast, {\em probKMA} relies on local functional $K$-means with Sobolev distance. 
Moreover, while \emph{probKMA} can extend the length of motifs endogenously and discover motifs of varying and unknown sizes, starting from a user-defined set of minimum motif lengths, {\em funBIalign} requires the length of all motifs to be the same and fixed beforehand. 
However, being based on $K$-means, {\em probKMA} must be run several times with different initialization, which can add to the computational cost.
In addition, {\em probKMA} implements a more complex post-processing, involving several tuning parameters -- and this could make the method less attractive for non-specialized user.
Finally, \emph{funBIalign} has the benefit of being equally applicable to a single curve or to sets of curves, whereas {\em probKMA} is designed to operate on sets of curves; to use it in applications involving a single curve, this must be split at the outset -- which requires further arbitrary choices, and may potentially lead to the loss of interesting motifs.
We compare the performance of the two algorithms 
using two simulation settings for scenario (1) introduced in Section 4.2 of \citet{cremona2022}. 
In both, two motifs of length $61$, say A and B, occur each $12$ times across $20$ curves.
In particular, $12$ curves contain $1$ occurrence of a single motif (of A for $6$, and of B for $6$ curves), $4$ curves contain $2$ occurrences of a single motif (of A for $2$, and of B for $2$ curves), $2$ curves contain $1$ occurrence of both motifs, and $2$ curves contain no motif. 
The two settings differ in terms of length of the curves $L$ and of noise incorporated in the motifs; one 
has short curves and low noise ($L=200$ and $\sigma = 0.1$), and the other long curves and high noise ($L=500$ and $\sigma = 2$).
\begin{table}[!b]
\centering
\footnotesize
\begin{tabular}{ccccccc}
\multirow{2}{*}{Setting}
& \multirow{2}{*}{Motif}                        & \multirow{2}{*}{Portion}     & \multirow{2}{*}{\emph{probKMA}} & \multicolumn{3}{l}{\emph{funBIalign - $n_{min}$}} \\ \cline{5-7} 
                                                                                                       &                                               &                              &                             & $ = 8$    & $ = 10$   \\ \hline
\multicolumn{1}{c|}{\multirow{4}{*}{\begin{tabular}[c]{@{}l@{}}$L =200$\\ $\sigma =0.1$\end{tabular}}} & \multicolumn{1}{c|}{\multirow{2}{*}{Motif A}} & \multicolumn{1}{c|}{Correct} & \multicolumn{1}{c|}{12}         & 12       & 12       \\ \cline{3-7} 
\multicolumn{1}{c|}{}                                                                                  & \multicolumn{1}{c|}{}                         & \multicolumn{1}{c|}{Extra}   & \multicolumn{1}{c|}{0}           & 0        & 0        \\ \cline{2-7} 
\multicolumn{1}{c|}{}                                                                                  & \multicolumn{1}{c|}{\multirow{2}{*}{Motif B}} & \multicolumn{1}{c|}{Correct} & \multicolumn{1}{c|}{12}         & 12       & 12       \\ \cline{3-7} 
\multicolumn{1}{c|}{}                                                                                  & \multicolumn{1}{c|}{}                         & \multicolumn{1}{c|}{Extra}   & \multicolumn{1}{c|}{0}          & 0        & 0        \\ \hline
\multicolumn{1}{c|}{\multirow{4}{*}{\begin{tabular}[c]{@{}l@{}}$L =500$\\ $\sigma =2$\end{tabular}}}   & \multicolumn{1}{c|}{\multirow{2}{*}{Motif A}} & \multicolumn{1}{c|}{Correct} & \multicolumn{1}{c|}{11}         & 12       & 12       \\ \cline{3-7} 
\multicolumn{1}{c|}{}                                                                                  & \multicolumn{1}{c|}{}                         & \multicolumn{1}{c|}{Extra}   & \multicolumn{1}{c|}{2}           & 1       & 1        \\ \cline{2-7} 
\multicolumn{1}{c|}{}                                                                                  & \multicolumn{1}{c|}{\multirow{2}{*}{Motif B}} & \multicolumn{1}{c|}{Correct} & \multicolumn{1}{c|}{12}        & 10       & 12       \\ \cline{3-7} 
\multicolumn{1}{c|}{}                                                                                  & \multicolumn{1}{c|}{}                         & \multicolumn{1}{c|}{Extra}   & \multicolumn{1}{c|}{1}        & 0        & 3       \\ \hline
\end{tabular}
\caption{Comparison between \emph{funBIalign} and \emph{probKMA}. For \emph{probKMA}, median results across $10$ runs are reported.}
\label{table:probkma}
\end{table}
We run \emph{funBIalign} with $\ell = 61$ (the true length of the motifs) and $n_{min}=8, 10$, and post-process candidate motifs with the rank sum criterion. 
{\em probKMA} is run using $K = 2,3$, minimum motif length $v = 41, 51, 61$, and $20$ random initializations for each $(K,v)$ pair; results from the $120$ runs with different parameters/initializations are pooled following the motif discovery post-processing recommended in \cite{cremona2022}. The whole procedure is repeated $10$ times, and medians are taken over these $10$ repetitions when evalutating performance.
Results are summarized in Table \ref{table:probkma}.
As expected, the choice of $n_{min}$ has an impact on the performance of \emph{funBIalign}: smaller values can lead to missed portions, and larger values to extra portions.
However, with an appropriate $n_{min}$, \emph{funBIalign} can match and even exceed the performance of {\em probKMA}.

\begin{table}[!b]
\centering
\footnotesize
\resizebox{0.75\linewidth}{!}{%
\begin{tabular}{ccccccccccc}
                                                      & \multirow{2}{*}{Motif}                        & \multicolumn{4}{c}{\textbf{\emph{funBIalign} - $n_{min}$}}                                                                                                                                                                                   & \textbf{MP}                     & \multicolumn{4}{c}{$R$}                                                                                                                               \\ \cline{3-11} 
                                                      &                                               & $5$                                              & $6$                                              & $7$                                                & $8$                                                & $k_{neighbor}$                 & 3                                   & 10                                  & 25                                  & 50                                  \\ \cline{2-11} 
\multicolumn{1}{c|}{\multirow{12}{*}{$\sigma = 0.5$}} & \multicolumn{1}{c|}{\multirow{3}{*}{Motif 1}} & \multicolumn{1}{c|}{\multirow{3}{*}{\textbf{(8,0)}}} & \multicolumn{1}{c|}{\multirow{3}{*}{\textbf{(8,0)}}} & \multicolumn{1}{c|}{\multirow{3}{*}{\textbf{(8,0)}}} & \multicolumn{1}{c|}{\multirow{3}{*}{\textbf{(8,0)}}} & \multicolumn{1}{c|}{4}          & \multicolumn{1}{c|}{(6,0)}          & \multicolumn{1}{c|}{(6,0)}          & \multicolumn{1}{c|}{(6,0)}          & \multicolumn{1}{c|}{(6,0)}          \\ \cline{7-11} 
\multicolumn{1}{c|}{}                                 & \multicolumn{1}{c|}{}                         & \multicolumn{1}{c|}{}                                & \multicolumn{1}{c|}{}                                & \multicolumn{1}{c|}{}                                & \multicolumn{1}{c|}{}                                & \multicolumn{1}{c|}{6}          & \multicolumn{1}{c|}{(6,0)}          & \multicolumn{1}{c|}{\textbf{(8,0)}} & \multicolumn{1}{c|}{\textbf{(8,0)}} & \multicolumn{1}{c|}{\textbf{(8,0)}} \\ \cline{7-11} 
\multicolumn{1}{c|}{}                                 & \multicolumn{1}{c|}{}                         & \multicolumn{1}{c|}{}                                & \multicolumn{1}{c|}{}                                & \multicolumn{1}{c|}{}                                & \multicolumn{1}{c|}{}                                & \multicolumn{1}{c|}{8}          & \multicolumn{1}{c|}{(7,3)}          & \multicolumn{1}{c|}{\textbf{(8,0)}} & \multicolumn{1}{c|}{(8,2)}          & \multicolumn{1}{c|}{(8,2)}          \\ \cline{2-11} 
\multicolumn{1}{c|}{}                                 & \multicolumn{1}{c|}{\multirow{3}{*}{Motif 2}} & \multicolumn{1}{c|}{\multirow{3}{*}{\textbf{(8,0)}}} & \multicolumn{1}{c|}{\multirow{3}{*}{\textbf{(8,0)}}} & \multicolumn{1}{c|}{\multirow{3}{*}{\textbf{(8,0)}}} & \multicolumn{1}{c|}{\multirow{3}{*}{\textbf{(8,0)}}} & \multicolumn{1}{c|}{4}          & \multicolumn{1}{c|}{(6,0)}          & \multicolumn{1}{c|}{(6,0)}          & \multicolumn{1}{c|}{(6,0)}          & \multicolumn{1}{c|}{(6,0)}          \\ \cline{7-11} 
\multicolumn{1}{c|}{}                                 & \multicolumn{1}{c|}{}                         & \multicolumn{1}{c|}{}                                & \multicolumn{1}{c|}{}                                & \multicolumn{1}{c|}{}                                & \multicolumn{1}{c|}{}                                & \multicolumn{1}{c|}{6} & \multicolumn{1}{c|}{\textbf{(8,0)}} & \multicolumn{1}{c|}{\textbf{(8,0)}} & \multicolumn{1}{c|}{\textbf{(8,0)}} & \multicolumn{1}{c|}{\textbf{(8,0)}} \\ \cline{7-11} 
\multicolumn{1}{c|}{}                                 & \multicolumn{1}{c|}{}                         & \multicolumn{1}{c|}{}                                & \multicolumn{1}{c|}{}                                & \multicolumn{1}{c|}{}                                & \multicolumn{1}{c|}{}                                & \multicolumn{1}{c|}{8} & \multicolumn{1}{c|}{\textbf{(8,0)}} & \multicolumn{1}{c|}{\textbf{(8,0)}} & \multicolumn{1}{c|}{(8,2)} & \multicolumn{1}{c|}{(8,2)}          \\ \cline{2-11} 
\multicolumn{1}{c|}{}                                 & \multicolumn{1}{c|}{\multirow{3}{*}{Motif 3}} & \multicolumn{1}{c|}{\multirow{3}{*}{(6,0)}}          & \multicolumn{1}{c|}{\multirow{3}{*}{\textbf{(8,0)}}} & \multicolumn{1}{c|}{\multirow{3}{*}{\textbf{(8,0)}}} & \multicolumn{1}{c|}{\multirow{3}{*}{\textbf{(8,0)}}} & \multicolumn{1}{c|}{4}          & \multicolumn{1}{c|}{(6,0)}          & \multicolumn{1}{c|}{(6,0)}          & \multicolumn{1}{c|}{(6,0)}          & \multicolumn{1}{c|}{(6,0)}          \\ \cline{7-11} 
\multicolumn{1}{c|}{}                                 & \multicolumn{1}{c|}{}                         & \multicolumn{1}{c|}{}                                & \multicolumn{1}{c|}{}                                & \multicolumn{1}{c|}{}                                & \multicolumn{1}{c|}{}                                & \multicolumn{1}{c|}{6} & \multicolumn{1}{c|}{\textbf{(8,0)}} & \multicolumn{1}{c|}{\textbf{(8,0)}} & \multicolumn{1}{c|}{\textbf{(8,0)}} & \multicolumn{1}{c|}{\textbf{(8,0)}} \\ \cline{7-11} 
\multicolumn{1}{c|}{}                                 & \multicolumn{1}{c|}{}                         & \multicolumn{1}{c|}{}                                & \multicolumn{1}{c|}{}                                & \multicolumn{1}{c|}{}                                & \multicolumn{1}{c|}{}                                & \multicolumn{1}{c|}{8} & \multicolumn{1}{c|}{\textbf{(8,0)}} & \multicolumn{1}{c|}{\textbf{(8,0)}} & \multicolumn{1}{c|}{(8,2)}          & \multicolumn{1}{c|}{(8,2)}          \\ \cline{2-11} 
\multicolumn{1}{c|}{}                                 & \multicolumn{1}{c|}{\multirow{3}{*}{Motif 4}} & \multicolumn{1}{c|}{\multirow{3}{*}{\textbf{(8,0)}}} & \multicolumn{1}{c|}{\multirow{3}{*}{\textbf{(8,0)}}} & \multicolumn{1}{c|}{\multirow{3}{*}{\textbf{(8,0)}}} & \multicolumn{1}{c|}{\multirow{3}{*}{\textbf{(8,0)}}} & \multicolumn{1}{c|}{4}          & \multicolumn{1}{c|}{(5,0)}          & \multicolumn{1}{c|}{(6,0)}          & \multicolumn{1}{c|}{(6,0)}          & \multicolumn{1}{c|}{(6,0)}          \\ \cline{7-11} 
\multicolumn{1}{c|}{}                                 & \multicolumn{1}{c|}{}                         & \multicolumn{1}{c|}{}                                & \multicolumn{1}{c|}{}                                & \multicolumn{1}{c|}{}                                & \multicolumn{1}{c|}{}                                & \multicolumn{1}{c|}{6}          & \multicolumn{1}{c|}{(5,0)}          & \multicolumn{1}{c|}{\textbf{(8,0)}} & \multicolumn{1}{c|}{\textbf{(8,0)}} & \multicolumn{1}{c|}{\textbf{(8,0)}} \\ \cline{7-11} 
\multicolumn{1}{c|}{}                                 & \multicolumn{1}{c|}{}                         & \multicolumn{1}{c|}{}                                & \multicolumn{1}{c|}{}                                & \multicolumn{1}{c|}{}                                & \multicolumn{1}{c|}{}                                & \multicolumn{1}{c|}{8}          & \multicolumn{1}{c|}{(5,0)}          & \multicolumn{1}{c|}{(8,1)} & \multicolumn{1}{c|}{(8,2)}          & \multicolumn{1}{c|}{(8,2)}          \\ \hline
\multicolumn{1}{c|}{\multirow{12}{*}{$\sigma = 2$}}   & \multicolumn{1}{c|}{\multirow{3}{*}{Motif 1}} & \multicolumn{1}{c|}{\multirow{3}{*}{(7,0)}}          & \multicolumn{1}{c|}{\multirow{3}{*}{(8,1)}}          & \multicolumn{1}{c|}{\multirow{3}{*}{(8,1)}}          & \multicolumn{1}{c|}{\multirow{3}{*}{(8,1)}}          & \multicolumn{1}{c|}{4}          & \multicolumn{1}{c|}{(3,0)}          & \multicolumn{1}{c|}{(6,0)}          & \multicolumn{1}{c|}{(6,0)}          & \multicolumn{1}{c|}{(6,0)}          \\ \cline{7-11} 
\multicolumn{1}{c|}{}                                 & \multicolumn{1}{c|}{}                         & \multicolumn{1}{c|}{}                                & \multicolumn{1}{c|}{}                                & \multicolumn{1}{c|}{}                                & \multicolumn{1}{c|}{}                                & \multicolumn{1}{c|}{6}          & \multicolumn{1}{c|}{(3,0)}          & \multicolumn{1}{c|}{\textbf{(8,0)}} & \multicolumn{1}{c|}{\textbf{(8,0)}} & \multicolumn{1}{c|}{\textbf{(8,0)}} \\ \cline{7-11} 
\multicolumn{1}{c|}{}                                 & \multicolumn{1}{c|}{}                         & \multicolumn{1}{c|}{}                                & \multicolumn{1}{c|}{}                                & \multicolumn{1}{c|}{}                                & \multicolumn{1}{c|}{}                                & \multicolumn{1}{c|}{8}          & \multicolumn{1}{c|}{(3,0)}          & \multicolumn{1}{c|}{(8,2)}          & \multicolumn{1}{c|}{(8,2)}          & \multicolumn{1}{c|}{(8,2)}          \\ \cline{2-11} 
\multicolumn{1}{c|}{}                                 & \multicolumn{1}{c|}{\multirow{3}{*}{Motif 2}} & \multicolumn{1}{c|}{\multirow{3}{*}{(7,2)}}          & \multicolumn{1}{c|}{\multirow{3}{*}{(7,2)}}          & \multicolumn{1}{c|}{\multirow{3}{*}{(8,1)}}          & \multicolumn{1}{c|}{\multirow{3}{*}{(8,1)}}          & \multicolumn{1}{c|}{4}          & \multicolumn{1}{c|}{(5,1)}          & \multicolumn{1}{c|}{(4,2)}          & \multicolumn{1}{c|}{(4,2)}          & \multicolumn{1}{c|}{(4,2)}          \\ \cline{7-11} 
\multicolumn{1}{c|}{}                                 & \multicolumn{1}{c|}{}                         & \multicolumn{1}{c|}{}                                & \multicolumn{1}{c|}{}                                & \multicolumn{1}{c|}{}                                & \multicolumn{1}{c|}{}                                & \multicolumn{1}{c|}{6}          & \multicolumn{1}{c|}{(6,2)}          & \multicolumn{1}{c|}{(5,3)}          & \multicolumn{1}{c|}{(5,3)}          & \multicolumn{1}{c|}{(5,3)}          \\ \cline{7-11} 
\multicolumn{1}{c|}{}                                 & \multicolumn{1}{c|}{}                         & \multicolumn{1}{c|}{}                                & \multicolumn{1}{c|}{}                                & \multicolumn{1}{c|}{}                                & \multicolumn{1}{c|}{}                                & \multicolumn{1}{c|}{8}          & \multicolumn{1}{c|}{(7,3)}          & \multicolumn{1}{c|}{(6,4)}          & \multicolumn{1}{c|}{(6,4)}          & \multicolumn{1}{c|}{(6,4)}          \\ \cline{2-11} 
\multicolumn{1}{c|}{}                                 & \multicolumn{1}{c|}{\multirow{3}{*}{Motif 3}} & \multicolumn{1}{c|}{\multirow{3}{*}{(6,1)}}          & \multicolumn{1}{c|}{\multirow{3}{*}{\textbf{(8,0)} }}          & \multicolumn{1}{c|}{\multirow{3}{*}{\textbf{(8,0)}}}          & \multicolumn{1}{c|}{\multirow{3}{*}{\textbf{(8,0)}}}          & \multicolumn{1}{c|}{4}          & \multicolumn{1}{c|}{(5,1)}          & \multicolumn{1}{c|}{(5,1)}          & \multicolumn{1}{c|}{(5,1)}          & \multicolumn{1}{c|}{(5,1)}          \\ \cline{7-11} 
\multicolumn{1}{c|}{}                                 & \multicolumn{1}{c|}{}                         & \multicolumn{1}{c|}{}                                & \multicolumn{1}{c|}{}                                & \multicolumn{1}{c|}{}                                & \multicolumn{1}{c|}{}                                & \multicolumn{1}{c|}{6}          & \multicolumn{1}{c|}{(5,1)}          & \multicolumn{1}{c|}{(6,2)}          & \multicolumn{1}{c|}{(6,2)}          & \multicolumn{1}{c|}{(6,2)}          \\ \cline{7-11} 
\multicolumn{1}{c|}{}                                 & \multicolumn{1}{c|}{}                         & \multicolumn{1}{c|}{}                                & \multicolumn{1}{c|}{}                                & \multicolumn{1}{c|}{}                                & \multicolumn{1}{c|}{}                                & \multicolumn{1}{c|}{8}          & \multicolumn{1}{c|}{(5,1)}          & \multicolumn{1}{c|}{(7,3)}          & \multicolumn{1}{c|}{(7,3)}          & \multicolumn{1}{c|}{(7,3)}          \\ \cline{2-11} 
\multicolumn{1}{c|}{}                                 & \multicolumn{1}{c|}{\multirow{3}{*}{Motif 4}} & \multicolumn{1}{c|}{\multirow{3}{*}{(6,1)}}          & \multicolumn{1}{c|}{\multirow{3}{*}{(7,1)}}          & \multicolumn{1}{c|}{\multirow{3}{*}{(7,1)}}          & \multicolumn{1}{c|}{\multirow{3}{*}{(7,1)}}          & \multicolumn{1}{c|}{4}          & \multicolumn{1}{c|}{(5,1)}          & \multicolumn{1}{c|}{(5,1)}          & \multicolumn{1}{c|}{(5,1)}          & \multicolumn{1}{c|}{(5,1)}          \\ \cline{7-11} 
\multicolumn{1}{c|}{}                                 & \multicolumn{1}{c|}{}                         & \multicolumn{1}{c|}{}                                & \multicolumn{1}{c|}{}                                & \multicolumn{1}{c|}{}                                & \multicolumn{1}{c|}{}                                & \multicolumn{1}{c|}{6}          & \multicolumn{1}{c|}{(5,3)}          & \multicolumn{1}{c|}{(5,3)}          & \multicolumn{1}{c|}{(5,3)}          & \multicolumn{1}{c|}{(5,3)}          \\ \cline{7-11} 
\multicolumn{1}{c|}{}                                 & \multicolumn{1}{c|}{}                         & \multicolumn{1}{c|}{}                                & \multicolumn{1}{c|}{}                                & \multicolumn{1}{c|}{}                                & \multicolumn{1}{c|}{}                                & \multicolumn{1}{c|}{8}          & \multicolumn{1}{c|}{(6,4)}          & \multicolumn{1}{c|}{(7,3)}          & \multicolumn{1}{c|}{(7,3)}          & \multicolumn{1}{c|}{(7,3)}          \\ \cline{2-11} 
\end{tabular}%
}
\caption{
    Comparison between \emph{funBIalign} and {\em SCRIMP-MP}. Cases in which all occurrences are correctly identified without the addition of any extra portions are in bold. 
}
\label{table:MP}
\end{table}

We next consider {\em SCRIMP-MP}, which identifies motifs with a $k$ Nearest Neighbours (kNN) routine. 
This method employs a similarity measure based on the z-normalized Euclidean distance.
In contrast to \emph{probKMA}, {\em SCRIMP-MP} is designed to operate with a single time series; to use it in applications involving a set of curves, those need to be joined -- which can be problematic.  
{\em SCRIMP-MP} also requires the user to fix several parameters {\em ex ante}; namely, the $u$ number of motifs to be discovered, their length $\ell$, a motif radius $R$
(i.e.~, the distance within which two portions of the time series are identified as belonging to the same motif), and a maximum number $k_{neighbor}$ of neighbors to consider.
Specifically, the algorithm will starts from pairs of most similar portions, to which other portions of the curve are added only if they are within a distance $R$ from the starting pair, and only up to a maximum of $k_{neighbor}$. 
We compare the performance of \emph{funBIalign} and {\em SCRIMP-MP} using two curves from our own simulation study in Section~4, namely the ones employing motifs set n.~7 with $8$ occurrences each, with shared noise level $\sigma = 0.5$ and $\sigma =2$, respectively (see Figures~\ref{fig:simulation7_sd0.5_card6_details} and \ref{fig:simulation7_sd2_card6_details}).
We run {\em funBIalign} with $\ell = 41$ (the true length of the motifs) and $n_{min}=5,6,7,8$. 
We use {\em SCRIMP-MP}, as implemented in the R library \verb|tsmp|, to identify a maximum of $u=50$ motifs with $\ell = 41$. All possible combinations of $R=3,10,25,50$ and $k_{neighbor}=4,6,8$ are tested. 
Table \ref{table:MP} summarizes the results most similar to the embedded motifs.
Also here the choice of $n_{min}$ has an impact on the performance of {\em funBIalign}, but 
$R$ and $k_{neighbor}$ have a yet stronger impact on the performance of {\em SCRIMP-MP}, which struggles more -- especially with higher noise. 
In addition, tuning these two parameters is not trivial, because intuition and knowledge about them is unlikely to be available {\em a priori}.
A similar comparison using motifs with different noise levels is presented in Section~\ref{sec:comparison_MP}.

\section{Case studies}
\label{sec:case}

In this section, we assess the ability of {\em funBIalign} to discover functional motifs in real data sets through two case studies. The first concerns food price inflation over a period of around 60 years, and the second temperature changes over a period of around 30 years. In both cases we utilize monthly measurements across the world provided by FAO\footnote{Data used in these cases studies can be downloaded at \url{https://www.fao.org/faostat/en/#data/
ET}}. 

\subsection{Case study 1: Food price inflation}
The FAOSTAT data used in this case study comprise monthly food price inflation measurements from January 2001 to June 2022 (a total of 258 measurements) for different countries and geographical regions (details are available in the repository metadata section, and geographical regions, as defined by the United Nations, are in Section~\ref{sec:casestudies_inflation}). 
We first seek motifs in a single, world-wide food price inflation curve, and then seek motifs in the curves for 19 distinct geographical regions. 
Before 
running \emph{funBIalign}, we smooth the data using local polynomials with Gaussian kernels 
(\verb|locpoly| function of the R package \verb|KernSmooth|, \cite{wand2006kernsmooth}); a bandwidth parameter equal to 1.5 seems appropriate to avoid over-smoothing possibly interesting peaks in the data. 

\begin{figure}[!t]
\includegraphics[width=\linewidth]{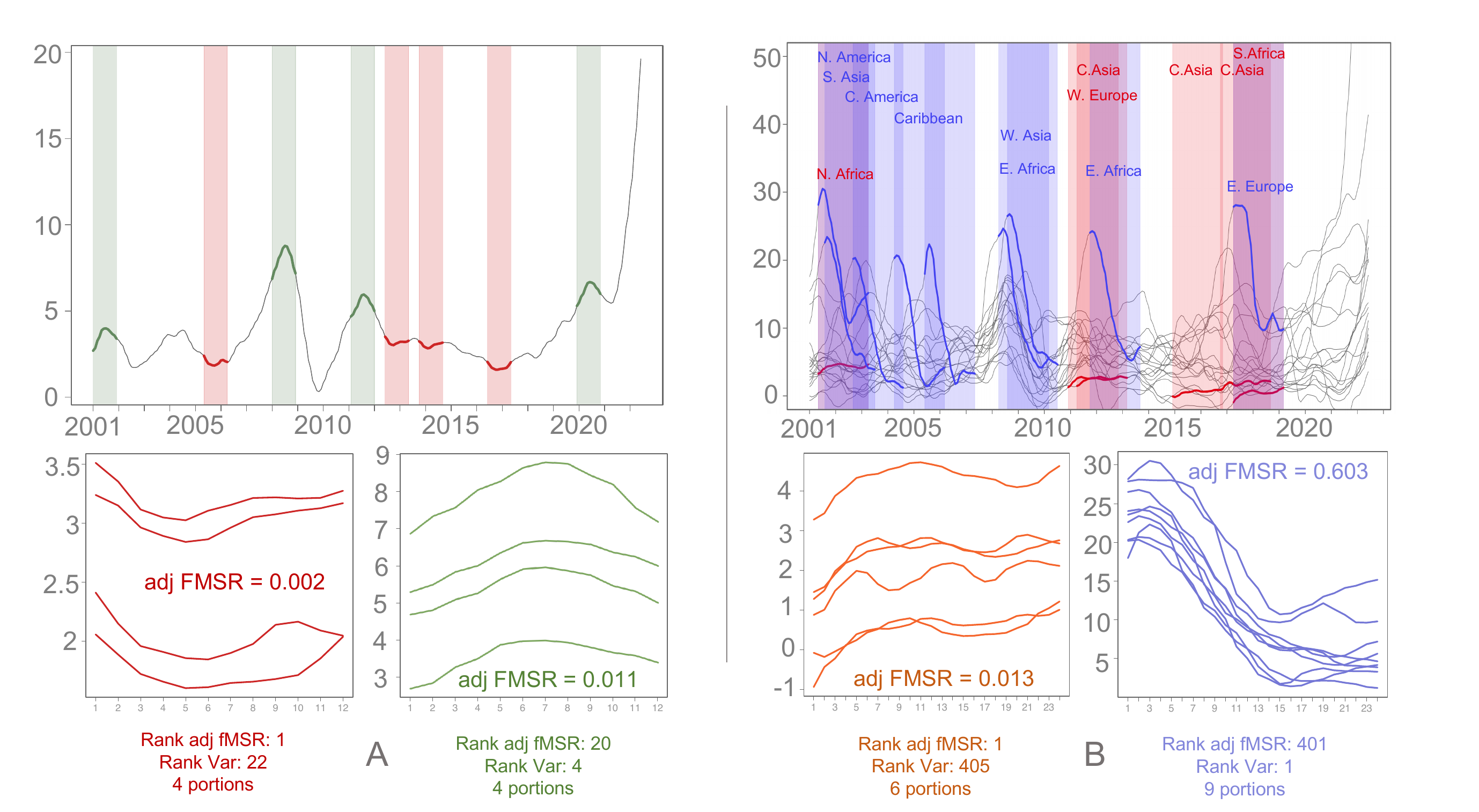}
\caption{
    Two of the motifs identified by \emph{funBIalign} in the world-wide food price inflation curve (panel A) and two of the motifs identified in the curves for 19 geographical regions (panel B). 
    In both panels, we show the top-ranking adjusted fMSR motif (bottom left plot) along with a high-ranking motif in terms of variance (bottom right plot) -- which, due to the criterion definition, may rank relatively low in terms of fMSR.
    In the top plot, N, S, E, W, and C stand for Northern, Southern, Eastern, Western, and Central, respectively. 
}
\label{fig:inflation}
\end{figure}

For the world-wide curve, we seek annual patterns setting $\ell = 12$ (months), and fix the minimum cardinality to  $n_{min} = 4$; a small $n_{min}$ is appropriate since we are considering a total of only $258$ measurements. \emph{funBIalign} identifies $30$ motifs capturing various types of shapes. 
Two, depicting ``valleys'' and ``peaks'' are displayed in Fig.~\ref{fig:inflation}A. Notably, the ``peak'' motif has occurrences corresponding to well-known economic crises; the 2001 recession, the global financial crisis of 2007-2008, and the 2020 COVID-19 recession. 
For the 19 regional curves, which comprise overall many more measurements, we seek longer, biannual patterns setting $\ell = 24$ (months) and fix a larger minimum cardinality $n_{min} = 6$. 
Here \emph{funBIalign} identifies $415$ motifs.
To sieve through such a large output, we suggest that users post-process and rank results based on various approaches. 
For instance, if one is interested in cohesive results regardless of how frequently a motif recurs, the adjusted fMSR criterion is the best choice. In contrast, if cardinality is important, it may be preferable to utilize the rank sum criterion. 
We note that, due to the definition of fMSR, motifs with lower variance - which look constant - tend to rank higher, a fact that could overshadow some patterns. 
For instance, among the $415$ motifs identified in the $19$ regional curves, some rather interesting high variance motifs are at the bottom of the fMSR ranking. 
Two motifs are 
shown in Fig.~\ref{fig:inflation}B. One, occurring 6 times, is the top-ranking in fMSR; it depicts a ``mild ascent'', with low variance. The other, occurring 9 times, is the top-ranking in terms of variance; it depicts a ``peak followed by a valley'' and ranks only 401$^\text{st}$ in fMSR.
We end this section remarking on the very steep ascent in food price inflation at the end of the time domain covered by the data (2021 and first half of 2022). This can be noticed in the world-wide curve (Fig.~\ref{fig:inflation}A), and appears to be especially driven by some of the 19 regional curves (Fig.~\ref{fig:inflation}B) -- in particular that of Western Asia, which raises to $\approx 100$ and is cut in the plot. Digging deeper, the rise in the Western Asia curve seems itself driven by countries such as Lebanon and the Syrian Arab Republic, which were experiencing conflicts and thus likely additional inflation pressures on top of world-wide COVID-19 related trends.

\subsection{Case study 2: Temperature changes}

Data used in this case study comprise monthly measurements of temperature changes with respect to a baseline climatology corresponding to the average temperature in the period 1951-1980 for different countries and geographical regions (further details can be found in the repository metadata section). 
We focus again on the $19$ geographical regions defined by the United Nations, and on a period from 1961 to 2021, for a total of $732$ monthly measurements (the measurements for a region are obtained by averaging those for the countries belonging it).
Also in this case study, before running \emph{funBIalign}, we smooth the data -- but this time we use cubic smoothing B-splines with knots at each month and roughness penalty on the curve second derivative. The smoothing parameter is selected minimizing the average generalized cross-validation error across curves.
%
\begin{figure}[!b]
\includegraphics[width=\linewidth]{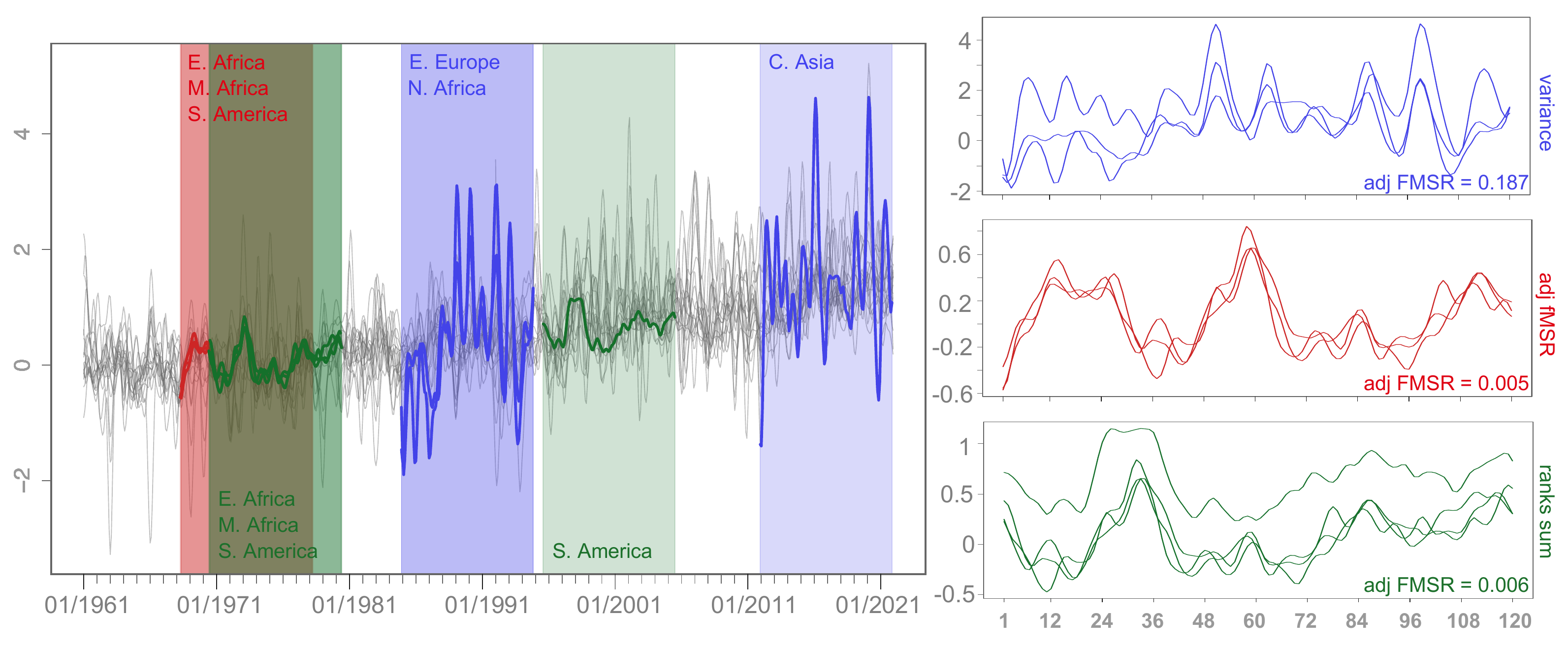}
\caption{
    Three motifs identified by {\em funBIalign} in the temperature change curves for $19$ geographical regions. 
    We show the top-ranking variance (top right -- 3 portions), adjusted fMSR (middle right -- 3 portions) and rank sum (bottom right -- 4 portions) motifs. 
    In the left plot, N,S,E,W,C and M stand for Northern, Southern, Eastern, Western, Central, and Middle respectively.
}
\label{fig:tempchange}
\end{figure}

In this analysis we seek long, 10-year patterns setting $\ell = 120$ (months). Considering the relative size of motifs to time period covered by the data, we fix again a small minimum cardinality $n_{min} = 3$. 
Also here \emph{funBIalign} identifies a very large number of motifs; $2367$.
We post-process and rank them using three different approaches: adjusted fMSR, rank sum, and variance. 
Top-ranking motifs based on each and the 19 curves are shown in Fig.~\ref{fig:tempchange}. 
Notably, a motif can occur simultaneously in multiple regions, and then separately in others -- e.g., the top variance motif, with peaks as much as 4°C above baseline, characterizes East Europe and North Africa from the mid '80s to the mid '90s, and then Central Asia almost 30 years later. 
Also notably, occurrences of different motifs can overlap -- e.g., occurrences of the top fMSR and the top rank sum motifs in East Africa, North Africa and South America in the '70s present a large overlap, almost extending one another along the time domain.
We end pointing out the increasing upward departures from baseline climatology shown by temperatures in all $19$ regions over the $60$ years covered by the data.

\section{Conclusions}

We contribute to the recent literature on functional motif discovery a definition of functional motifs based on an explicit additive model, and an algorithm designed to discover such motifs; \emph{funBIalign}. 
Related to our additive model, we also introduce the adjusted functional Mean Squared Residue (fMSR) score. 
The fMSR is a functional extension of the MSR score widely used in the multivariate biclustering literature. 
Building upon our own past work \citep{di2020bias}, we prove it to be biased towards motifs that occur less often in the data, and formulate a de-biasing adjustment. 
\emph{funBIalign} is a very flexible multi-step algorithm which requires only two, easy to interpret input parameters: the length $\ell$ and the minimum number of occurrences $n_{min}$ of the motifs to be discovered.
It uses agglomerative clustering to produce a hierarchy capturing the relationships among curve portions, followed by a dynamic cutting procedure to identify the most interesting candidate motifs based on such hierarchy, and by a post-processing step to eliminate redundant results.

\emph{funBIalign} shows very good performance both in extensive simulations and in two real-data case studies. 
Commenting on our simulation study, we noted how the algorithm can identify extra instances of a given motif, or even entirely different motifs, that were not intentionally embedded in the simulated curves. While this is an explainable bi-product of the procedure used to create our simulated curves (certain patterns may occur and recur by chance in the curve background), it points to the need for a more rigorous statistical treatment of motif discovery. In particular, in future work, we plan to develop a measure of significance to be used in conjunction with the adjusted fMSR for motif discovery. 

Parsing through all curve portions of a given length allows {\em funBIalign} to tackle applications where classical functional alignment fails or may be inadequate, such as identifying motifs embedded consecutively, or appearing only in one curve. 
However, this comes at a cost; storing the dissimilarity matrix $D_{fMRS}$ may necessitate a massive amount of memory when analyzing large functional data sets.
Nonetheless, we note that this matrix is only calculated once, and in our simulations (each involving one curve encompassing $7001$ measurements) and case studies (involving a maximum of $19$ curves of $732$ measurements) memory did not pose a challenge;
{\em funBIalign} run in under 1 minute without RAM problems on a local machine (64GB of memory, 8 performance cores and 2 efficiency cores).
We do believe that running time could be further improved leveraging parallelization.
In future work we plan to optimize {\em funBIalign} to allow it to scale efficiently also on much larger datasets.
Current R code for the algorithm is available in the 
GitHub repository \url{https://github.com/JacopoDior/funBIalign} and an R package is in preparation.

\printbibliography[heading=subbibintoc]

\end{document}


\def\spacingset#1{\renewcommand{\baselinestretch}%
{#1}\small\normalsize} \spacingset{1}


\if0\blind
{
  \title{\bf \emph{funBIalign}: a hierachical algorithm for functional motif discovery based on mean squared residue score - Supplementary Material}
  \author{Jacopo Di Iorio \hspace{.2cm}\\
    Dept. of Statistics, Penn State University\\
    \\
    Marzia A. Cremona  \hspace{.2cm}\\
    Dept. of Operations and Decision Systems, Université Laval \\
    \\
    Francesca Chiaromonte \hspace{.2cm}\\
    Dept. of Statistics, Penn State University}
  \maketitle
} \fi

\if1\blind
{
  \bigskip
  \bigskip
  \bigskip
  \begin{center}
    {\LARGE\bf Title}
\end{center}
  \medskip
} \fi

\bigskip

\section{Proof of Theorem 1}
\label{sec:s1_proof}

\begin{proof}
Given a functional motif $Q$ composed of $n_Q$ portions, the functional mean squared residue score is defined as 
\begin{equation*}
    H(Q) = \frac{1}{n_Q}\frac{1}{l}\sum_{k=1}^{n_Q}\int_{0}^{l}\left(p_{k}(t) - \overline{p}_{k} - \overline{p}(t) + \overline{p}\right)^{2}dt,
\end{equation*}
where we have
\begin{equation*}
    \overline{p}_k = \frac{1}{l}\int_{0}^{l}p_{k}(t)dt,
\end{equation*}
\begin{equation*}
    \overline{p}(t) = \frac{1}{n_Q}\sum_{k=1}^{n_Q}p_{k}(t),
\end{equation*}
\begin{equation*}
    \overline{p} = \frac{1}{n_Q}\frac{1}{l}\sum_{k = 1}^{n_Q}\int_{0}^{l}p_{k}(t)dt.
\end{equation*}
We define $g_{k}(t) := p_{k}(t) - \overline{p}_k - \overline{p}(t) + \overline{p}$, so that $H(Q) = \frac{1}{n_Q}\frac{1}{l}\sum_{k=1}^{n_Q}\int_{0}^{l} g_{k}^{2}(t) dt$.
We have that:
\begin{align*}
    g_{k}(t) 
    &= p_k(t) - \overline{p}_{k} - \frac{1}{n_Q} \sum_{j=1}^{n_Q}p_j(t) + \frac{1}{n_Q} \sum_{j =1}^{n_Q}\overline{p}_{j} \\
    &= \frac{n_Q-1}{n_Q} p_k(t) - \frac{n_Q-1}{n_Q} \overline{p}_k - \frac{1}{n_Q} \sum_{j \neq k} p_j(t) + \frac{1}{n_Q}\sum_{j \neq k}\overline{p}_j \\
    &= A + B + C + D.
\end{align*}
Then $g_{k}^2(t)$ is:
\begin{equation*}
    g_{k}^2(t) = A^2 + B^2 + C^2 + D^2 + 2AB + 2AC + 2AD + 2BC + 2BD + 2CD
\end{equation*}
where
\begin{equation*}
    A^2 = \frac{(n_Q-1)^2}{n_Q^2}p_k^2(t),
\end{equation*}
\begin{equation*}
    B^2 = \frac{(n_Q-1)^2}{n_Q^2}\overline{p}_k^2,
\end{equation*}
\begin{equation*}
    C^2 = \frac{1}{n_Q^2}\sum_{j \neq k} p_j^2(t) + \frac{1}{n_Q^2} \sum_{j \neq k}\left(p_j(t) \sum_{i \neq \{j,k\}} p_i(t) \right),
\end{equation*}
\begin{equation*}
    D^2 = \frac{1}{n_Q^2}\sum_{j \neq k} \overline{p}_j^2 + \frac{1}{n_Q^2} \sum_{j \neq k}\left(\overline{p}_j \sum_{i \neq \{j,k\}} \overline{p_i}\right),
\end{equation*}
\begin{equation*}
    2AB = - 2 \frac{(n_Q-1)^2}{n_Q^2} p_k(t) \overline{p}_k,
\end{equation*}
\begin{equation*}
    2AC = -2 \frac{n_Q-1}{n_Q^2}p_k(t) \sum_{j \neq k} p_j(t),
\end{equation*}
\begin{equation*}
    2AD = 2 \frac{n_Q-1}{n_Q^2} p_k(t) \sum_{j \neq k} \overline{p}_j,
\end{equation*}
\begin{equation*}
    2BC = 2 \frac{n_Q-1}{n_Q^2} \overline{p}_k \sum_{j \neq k} p_j(t),
\end{equation*}
\begin{equation*}
    2BD  = -2 \frac{n_Q-1}{n_Q^2} \overline{p}_k \sum_{j \neq k} \overline{p}_j,
\end{equation*}
\begin{equation*}
    2CD = -2 \frac{1}{n_Q^2} \left( \sum_{ j \neq k} p_j(t) \overline{p}_j \right).
\end{equation*}
Defining $z_k(t) := p_k^2(t)$, we can rewrite $A^2$ and $C^2$ in the following way:
\begin{equation*}
    A^2 = \frac{(n_Q-1)^2}{n_Q^2}z_k(t),
\end{equation*}
\begin{equation*}
    C^2 =  \frac{1}{n_Q^2}\sum_{j \neq k} z_j(t) + \frac{1}{n_Q^2} \sum_{j \neq k}\left(p_j(t) \sum_{i \neq \{j,k\}} p_i(t) \right).
\end{equation*}
Noting that $2\frac{1}{l} \int_0^l  BC dt + 2\frac{1}{l} \int_0^l  BD dt = 0$, we can then compute:
\begin{align*}
    \frac{1}{l} \int_0^l g_k^2(t)dt 
    =& \frac{1}{l} \int_0^l A^2 dt + \frac{1}{l} \int_0^l B^2 dt \\
    &+ \frac{1}{l} \int_0^l C^2 dt + \frac{1}{l} \int_0^l D^2 dt + \\
    &+ 2\frac{1}{l} \int_0^l  AB dt+  2\frac{1}{l} \int_0^l AC dt+  2\frac{1}{l} \int_0^l AD dt + 2\frac{1}{l} \int_0^l CD dt \\
    =& \frac{(n_Q-1)^2}{n_Q^2} \overline{z}_k + \frac{(n_Q-1)^2}{n_Q^2} \overline{p}_k^2 \\
    &+ \frac{1}{n_Q^2} \sum_{j \neq k} \overline{z}_j + \frac{1}{n_Q^2} \frac{1}{l} \int_0^l \sum_{j \neq k} \left( p_j(t) \sum_{i \neq \{j,k\}} p_i(t) \right)dt + \frac{1}{n_Q^2} \sum_{j \neq k} \overline{p}_j^2 +  \frac{1}{n_Q^2} \sum_{j \neq k} \left( \overline{p}_j \sum_{i \neq \{j,k\}} \overline{p}_i \right) + \\
    & -2 \frac{(n_Q-1)^2}{n_Q^2}\overline{p}_k^2  -2 \frac{n_Q-1}{n_Q^2} \frac{1}{l} \int_0^l \left( p_k(t) \sum_{j \neq k} p_j(t) \right)dt+  2 \frac{n_Q-1}{n_Q^2} \overline{p}_k \sum_{j \neq k} \overline{p}_j -2 \frac{1}{n_Q^2} \left( \sum_{j \neq k} \overline{p}_j \right)^2 .
    \end{align*}
%

We now compute the coefficients corresponding to each of the elements $\overline{z}_{\kappa}$, $\overline{p}_{\kappa}^2$, $\frac{1}{l} \int_0^l p_{\kappa}(t)p_{\iota}(t)dt$, and  $\overline{p}_{\kappa} \overline{p}_{\iota}$, with $\iota \neq \kappa$, in the H-score of a motif $Q$ composed of $n_Q$ portions, multiplied by $n_Q$:
\begin{align*}
    n_Q H(Q) &= \sum_{k=1}^{n_Q} \frac{1}{l}\int_{0}^{l} g_{k}^{2}(t) dt \\
    &= \sum_{\kappa=1}^{n_Q} \left[ h^Q_{\overline{z}_{\kappa}}\overline{z}_{\kappa} + h^Q_{\overline{p}_{\kappa}^2}\overline{p}_{\kappa}^2 + \sum_{\iota \neq \kappa} \left( h^Q_{\overline{p}_{\iota \kappa}}\frac{1}{l} \int_0^l p_{\kappa}(t)p_{\iota}(t)dt + h^Q_{\overline{p}_{\kappa} \overline{p}_{\iota}}\overline{p}_{\kappa} \overline{p}_{\iota} \right) \right].
\end{align*}
We start computing $h^Q_{\overline{z}_{\kappa}}$, the coefficient referring to $\overline{z}_{\kappa}$. This element appears twice:
in $\frac{(n_Q-1)^2}{n_Q^2} \overline{z}_k$ when $k = \kappa$, and in $\frac{1}{n_Q^2} \sum_{j \neq k} \overline{z}_j$  when $k \neq \kappa$ (that is, $n_Q-1$ times). Hence, we have
$$
    h^Q_{\overline{z}_{\kappa}} 
    = \frac{(n_Q-1)^2}{n_Q^2} + \frac{n_Q-1}{n_Q^2} 
    = \frac{n_Q-1}{n_Q}.
$$
To compute the coefficient $h^Q_{\overline{p}_{\kappa}^2}$ referring to the element $\overline{p}_{\kappa}^2$, we need to consider that the element appears four times:
in $\frac{(n_Q-1)^2}{n_Q^2} \overline{p}_k^2$ and in $-2 \frac{(n_Q-1)^2}{n_Q^2}\overline{p}_k^2$ when $k = \kappa$; in $\frac{1}{n_Q^2} \sum_{j \neq k} \overline{p}_j^2$ and in 
$-2 \frac{1}{n_Q^2} \left( \sum_{j \neq k} \overline{p}_j \right)^2$ when $ k \neq \kappa$ (that is, $n_Q-1$ times). Hence, we have
$$
    h^Q_{\overline{p}_{\kappa}^2} 
    = \frac{(n_Q-1)^2}{n_Q^2} -2\frac{(n_Q-1)^2}{n_Q^2} + \frac{n_Q-1}{n_Q^2} -2 \frac{n_Q-1}{n_Q^2}
    = - \frac{n_Q-1}{n_Q}.
$$
For the coefficient of the element $\frac{1}{l} \int_0^l p_{\kappa}(t)p_{\iota}(t)dt$ for $\iota \neq \kappa$, i.e. $h^Q_{\overline{p}_{\iota \kappa}}$, we need to consider the term $-2 \frac{n_Q-1}{n_Q^2} \frac{1}{l} \int_0^l \left( p_k(t) \sum_{j \neq k} p_j(t) \right)dt$ when $k = \kappa$ or $k = \iota$ (that is, $2$ times), and the term $\frac{1}{n_Q^2} \frac{1}{l} \int_0^l \sum_{j \neq k} \left( p_j(t) \sum_{i \neq \{j,k\}} p_i(t) \right)dt$ when $k \neq \{\kappa, \iota\}$ (that is, $2(n_Q-2)$ times). We obtain:
$$
    h^Q_{\overline{p}_{\iota \kappa}}
    = -2 \left( -2 \frac{n_Q-1}{n_Q^2} \right) + 2\frac{n_Q-2}{n_Q^2}
    = - \frac{2}{n_Q}.
$$
Finally, we compute $h^Q_{\overline{p}_{\kappa} \overline{p}_{\iota}}$ for $\iota \neq \kappa$, the coefficient corresponding to $\overline{p}_{\kappa} \overline{p}_{\iota}$. This element appears in the term $2 \frac{n_Q-1}{n_Q^2} \overline{p}_k \sum_{j \neq k} \overline{p}_j$ when $k \neq \kappa$ or $k \neq \iota$ (that is, $2$ times), in the term $\frac{1}{n_Q^2} \sum_{j \neq k} \left( \overline{p}_j \sum_{i \neq \{j,k\}} \overline{p}_i \right)$ when $k \neq \{\kappa, \iota\}$ (that is, $2(n_Q-2)$ times), and in the term $-2 \frac{1}{n_Q^2} \left( \sum_{j \neq k} \overline{p}_j \right)^2$ when $k \neq \{\kappa, \iota\}$ (that is, $2(n_Q-2)$ times). Hence, we have:
$$
    h^Q_{\overline{p}_{\kappa} \overline{p}_{\iota}}
    = 2 \left(2 \frac{n_Q-1}{n_Q^2}\right) + 2\frac{n_Q-2}{n_Q^2} + 2(n_Q-2)\left(-2 \frac{1}{n_Q^2} \right)
    = \frac{2}{n_Q}.
$$

We now consider $\overline{H}_{n}$ for $n=2,3,\dots, n_Q$, the average fMRS of all $\binom{n_Q}{n}$ sub-motifs of $Q$ obtained selecting exactly $n$ of the $n_Q$ portions $p_{k}(t)$ belonging to $Q$:
\begin{align*}
    \overline{H}_n &= \binom{n_Q}{n}^{-1} \sum_{\substack{R \subseteq Q \\ |R|=n}}H(R) \\
    &= \sum_{\kappa=1}^{n_Q} \left[ \overline{h}^n_{\overline{z}_{\kappa}}\overline{z}_{\kappa} + \overline{h}^n_{\overline{p}_{\kappa}^2}\overline{p}_{\kappa}^2 + \sum_{\iota \neq \kappa} \left( \overline{h}^n_{\overline{p}_{\iota \kappa}}\frac{1}{l} \int_0^l p_{\kappa}(t)p_{\iota}(t)dt + \overline{h}^n_{\overline{p}_{\kappa} \overline{p}_{\iota}}\overline{p}_{\kappa} \overline{p}_{\iota} \right) \right].
\end{align*}
%
We start computing $\overline{h}^n_{\overline{z}_{\kappa}}$, the coefficient corresponding to $\overline{z}_{\kappa}$. This element appears in the $H(R)$ of the sub-motifs comprising the portion $p_{\kappa}(t)$, that are exactly $\binom{n_Q-1}{n-1}$. Hence, we have:
$$
    \overline{h}^n_{\overline{z}_{\kappa}} 
    = \frac{\binom{n_Q-1}{n-1}}{\binom{n_Q}{n}} \frac{h^R_{\overline{z}_{\kappa}}}{n}
    = \frac{n-1}{nn_Q}.
$$
To compute $\overline{h}^n_{\overline{p}_{\kappa}^2}$, the coefficient of the element $\overline{p}_{\kappa}^2$, we need to consider the sub-motifs comprising the portion $p_{\kappa}(t)$. There are exactly $\binom{n_Q-1}{n-1}$ such motifs, so we obtain:
$$
    \overline{h}^n_{\overline{p}_{\kappa}^2}
    = \frac{\binom{n_Q-1}{n-1}}{\binom{n_Q}{n}} \frac{h^R_{\overline{p}_{\kappa}^2}}{n} 
    = -\frac{n-1}{nn_Q}
$$
To compute the coefficient $\overline{h}^n_{\overline{p}_{\iota \kappa}}$ of the element $\frac{1}{l} \int_0^l p_{\kappa}(t)p_{\iota}(t)dt$ with $\iota \neq \kappa$ we need to consider the $\binom{n_Q-2}{n-2}$ sub-motifs comprising both portions $p_{\kappa}(t)$ and $p_{\iota}(t)$, obtaining:
$$
    \overline{h}^n_{\overline{p}_{\iota \kappa}}
    = \frac{\binom{n_Q-2}{n-2}}{\binom{n_Q}{n}} \frac{h^R_{\overline{p}_{\iota \kappa}}}{n}
    = -2\frac{n-1}{n_Q(n_Q-1)n}.
$$
Finally, for the coefficient $\overline{h}^n_{\overline{p}_{\kappa} \overline{p}_{\iota}}$ corresponding to the element $\overline{p}_{\kappa} \overline{p}_{\iota}$ with $\iota \neq \kappa$, we need to consider the $\binom{n_Q-2}{n-2}$ sub-motifs comprising both portions $p_{\kappa}(t)$ and $p_{\iota}(t)$. Hence, we have:
$$
    \overline{h}^n_{\overline{p}_{\kappa} \overline{p}_{\iota}}
    = \frac{\binom{n_Q-2}{n-2}}{\binom{n_Q}{n}} \frac{h^R_{\overline{p}_{\kappa} \overline{p}_{\iota}}}{n}
    = 2\frac{n-1}{n_Q(n_Q-1)n}.
$$
Putting all the terms together, we obtain:
\begin{align*}
    \overline{H}_n 
    &= \frac{n-1}{nn_Q} \sum_{\kappa=1}^{n_Q} \left[ \overline{z}_{\kappa} + \overline{p}_{\kappa}^2 - \frac{2}{n_Q-1} \sum_{\iota \neq \kappa} \left( \frac{1}{l} \int_0^l p_{\kappa}(t)p_{\iota}(t)dt - \overline{p}_{\kappa} \overline{p}_{\iota} \right) \right].
\end{align*}
We are interested in the relationship between $\overline{H}_{n+1}$ and $\overline{H}_n$. We have:
\begin{align*}
    \overline{H}_{n+1} 
    &= \frac{n}{(n+1)n_Q} \sum_{\kappa=1}^{n_Q} \left[ \overline{z}_{\kappa} + \overline{p}_{\kappa}^2 - \frac{2}{n_Q-1} \sum_{\iota \neq \kappa} \left( \frac{1}{l} \int_0^l p_{\kappa}(t)p_{\iota}(t)dt - \overline{p}_{\kappa} \overline{p}_{\iota} \right) \right] \\
    &= \frac{n^2}{(n+1)(n-1)} \overline{H}_n,
\end{align*}
from which we obtain
\begin{equation*}
    \overline{H}_{n+1} = \overline{H}_{n}\frac{n^2}{n^2-1}.
\end{equation*}

\end{proof}

\section{A schematic of \emph{funBIalign} algorithm}
\begin{figure}[H]
\centering
\includegraphics[width=0.70\linewidth]{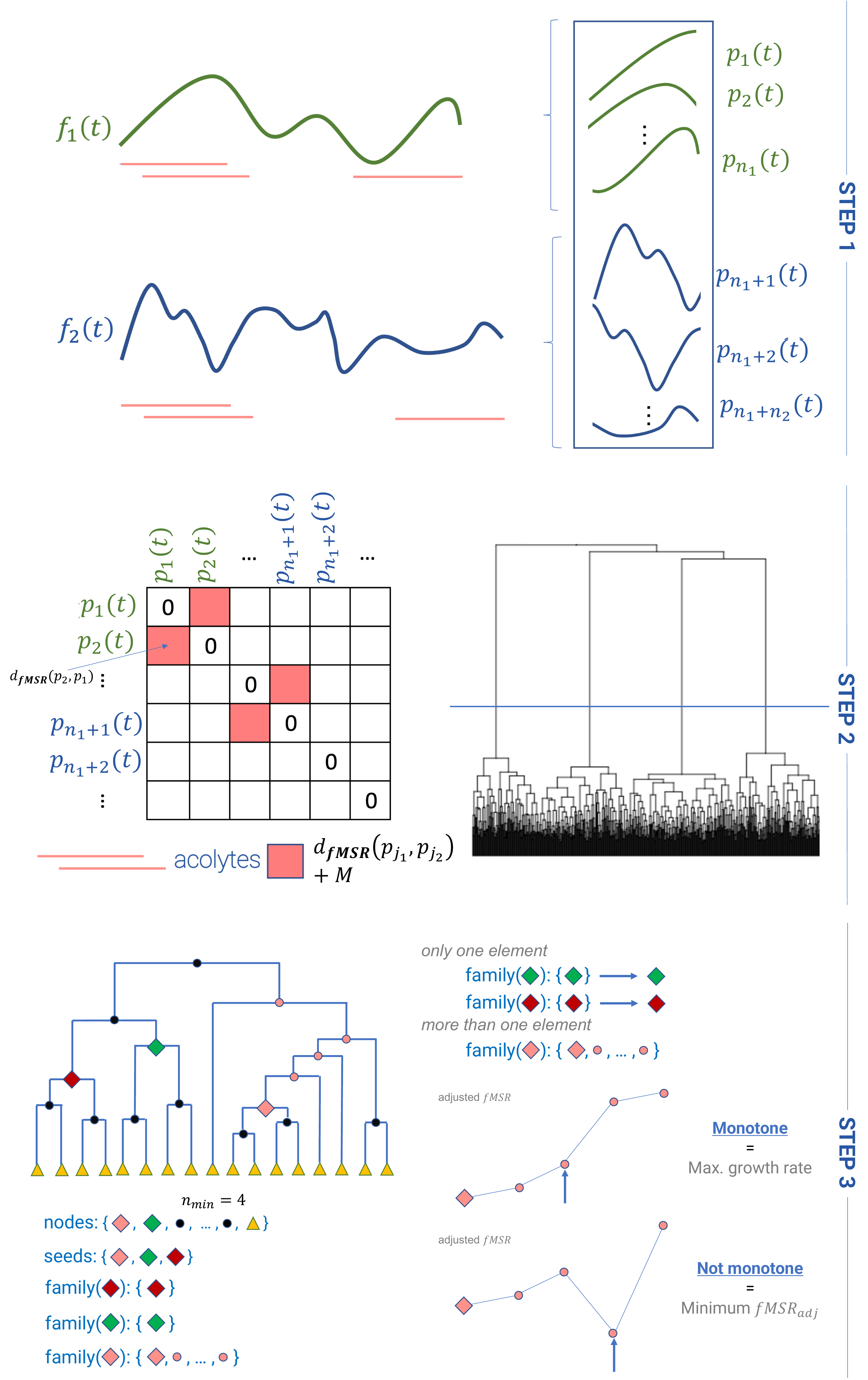}
\caption{
    A visual scheme of the three main steps of \emph{funBIalign} algorithm.
}
\label{sup:algo_scheme}
\end{figure}

\section{Motifs with shared noise level}
\label{sec:same}

\subsection{Motifs with 8 occurrences}
\label{sec:same8}

\begin{figure}[H]
\includegraphics[width=\linewidth, height=16cm]{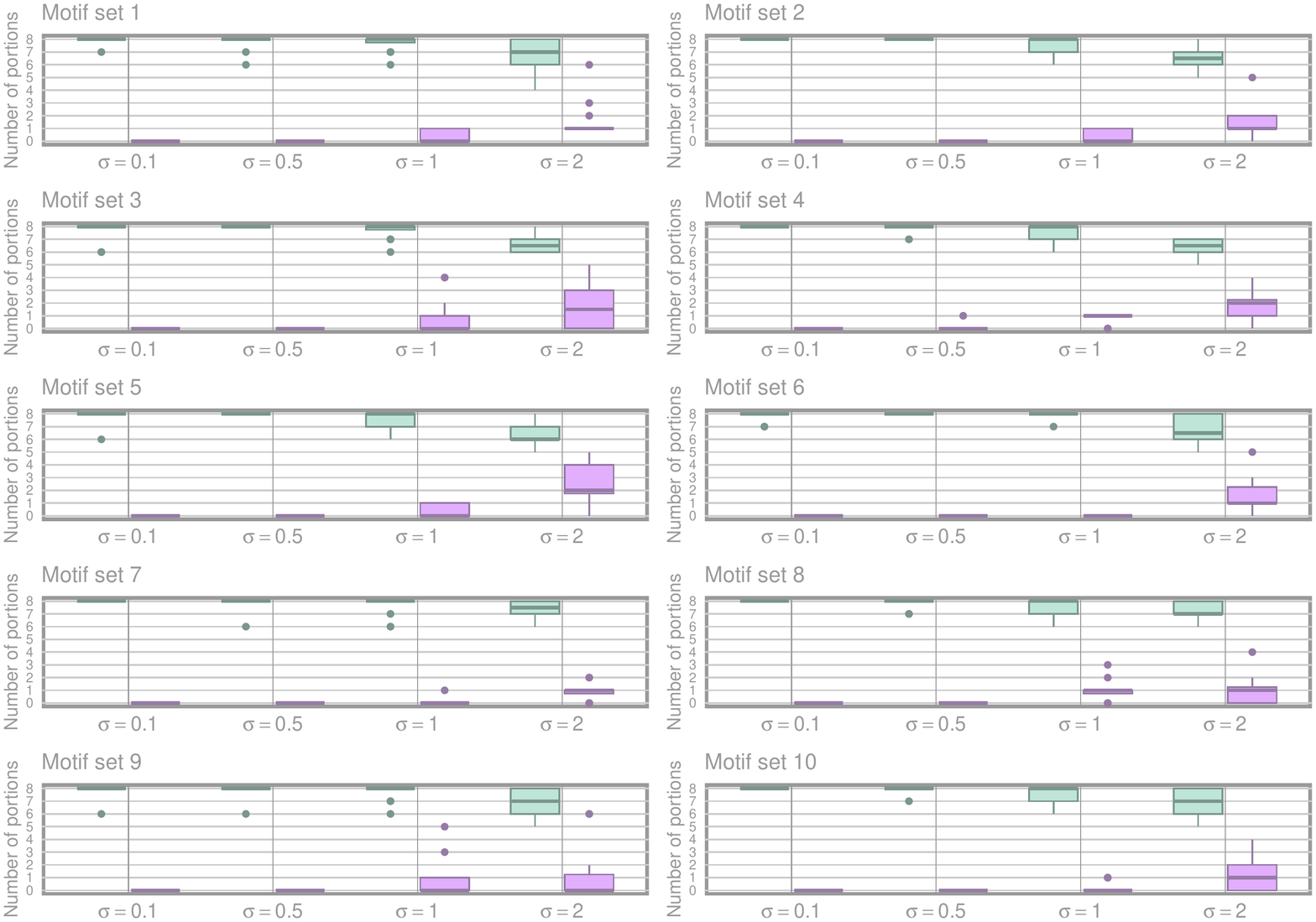}
\caption{
    Performance of \emph{funBIalign} for all simulations where motifs have $8$ occurrences and shared noise level. 
    Different panels show the performance pooling across runs of the algorithm with minimum cardinalities $n_{min} = 5, 6, 7, 8$, but separately for each of the 10 alternative motif sets. 
    For each $\sigma$, the green and pink boxplots (left and right) represent correctly identified portions and extra portions, respectively. 
}
\label{fig:simulations_performance_8}
\end{figure}

\begin{figure}[H]
\includegraphics[width=\linewidth]{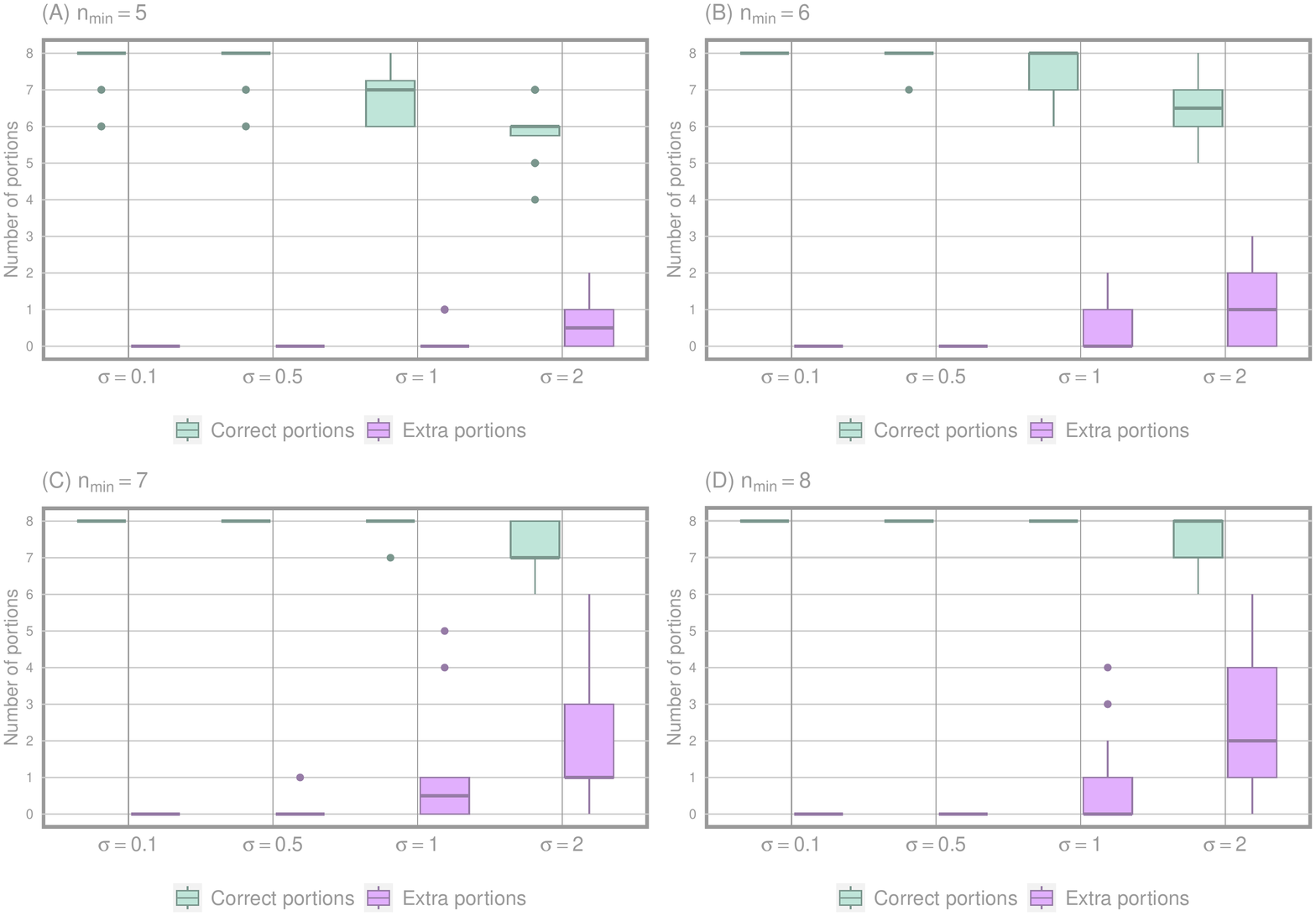}
\caption{
    Performance of \emph{funBIalign} for all simulations where motifs have $8$ occurrences and shared noise level. 
    The four panels show the performance of the algorithm run with minimum cardinalities $n_{min}=5,6,7,8$, respectively, but pooling across 10 alternative motif sets.
    For each $\sigma$, the green and pink boxplots (left and right) represent correctly identified portions and extra portions, respectively. 
}
\label{fig:cardinality_effect_performance_8}
\end{figure}

\begin{figure}[H]
\includegraphics[width=\linewidth]{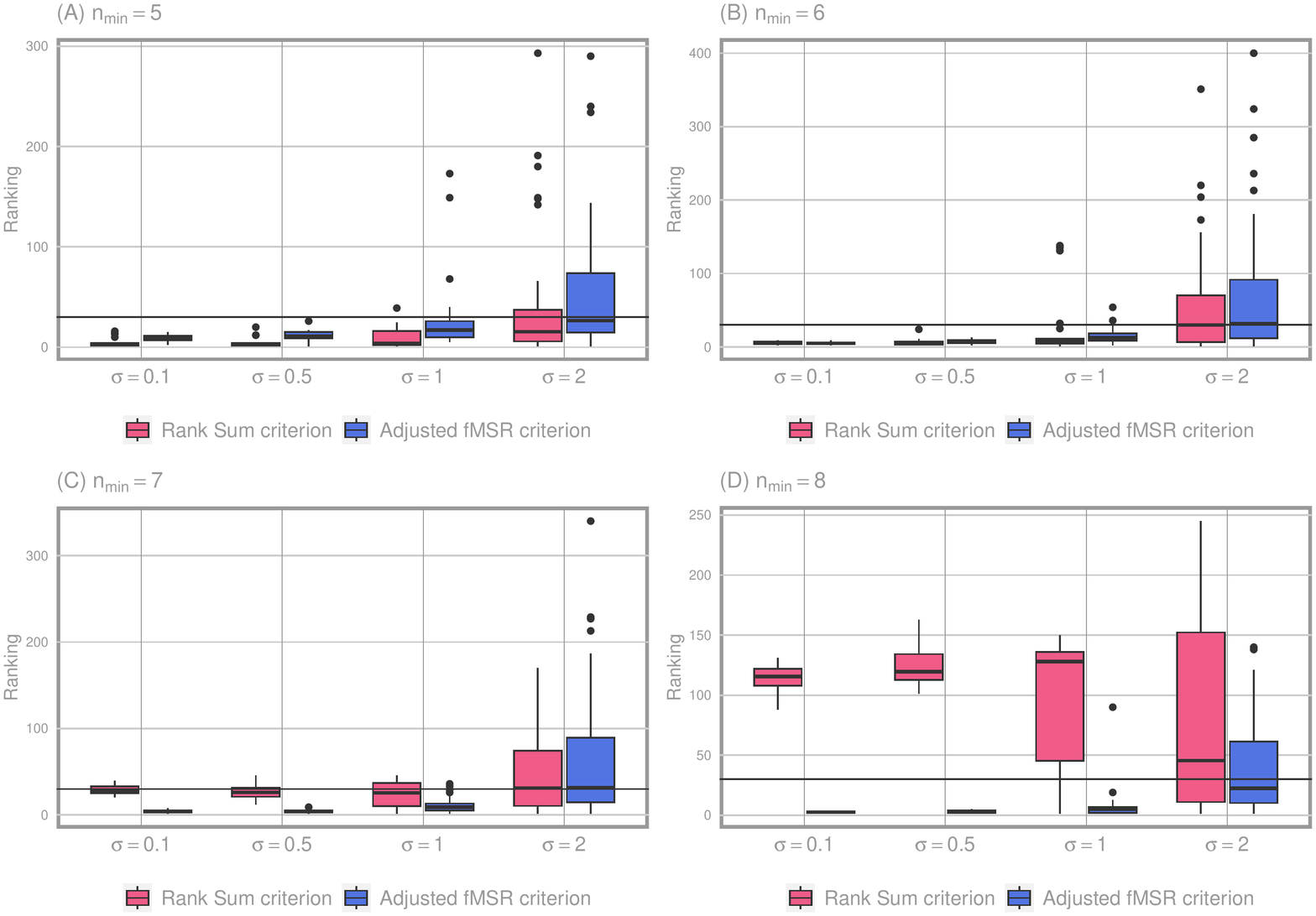}
\caption{
    Ranking of the results of \emph{funBIalign} which are most similar to the intentionally embedded motifs, for all simulations where motifs have $8$ occurrences and shared noise level. 
    The four panels show the ranking of the algorithm run with minimum cardinalities $n_{min}=5,6,7,8$, respectively, but pooling across 10 alternative motif sets.
    For each $\sigma$, the red and blue boxplots (left and right) represent ranking according to the rank sum and the adjusted fMSR criterion, respectively. 
    Horizontal line indicates rank 30.}
\label{fig:cardinality_effect_ranking_8}
\end{figure}

\subsection{Motifs with 10 occurrences}
\label{sec:same10}

\begin{figure}[H]
\includegraphics[width=\linewidth]{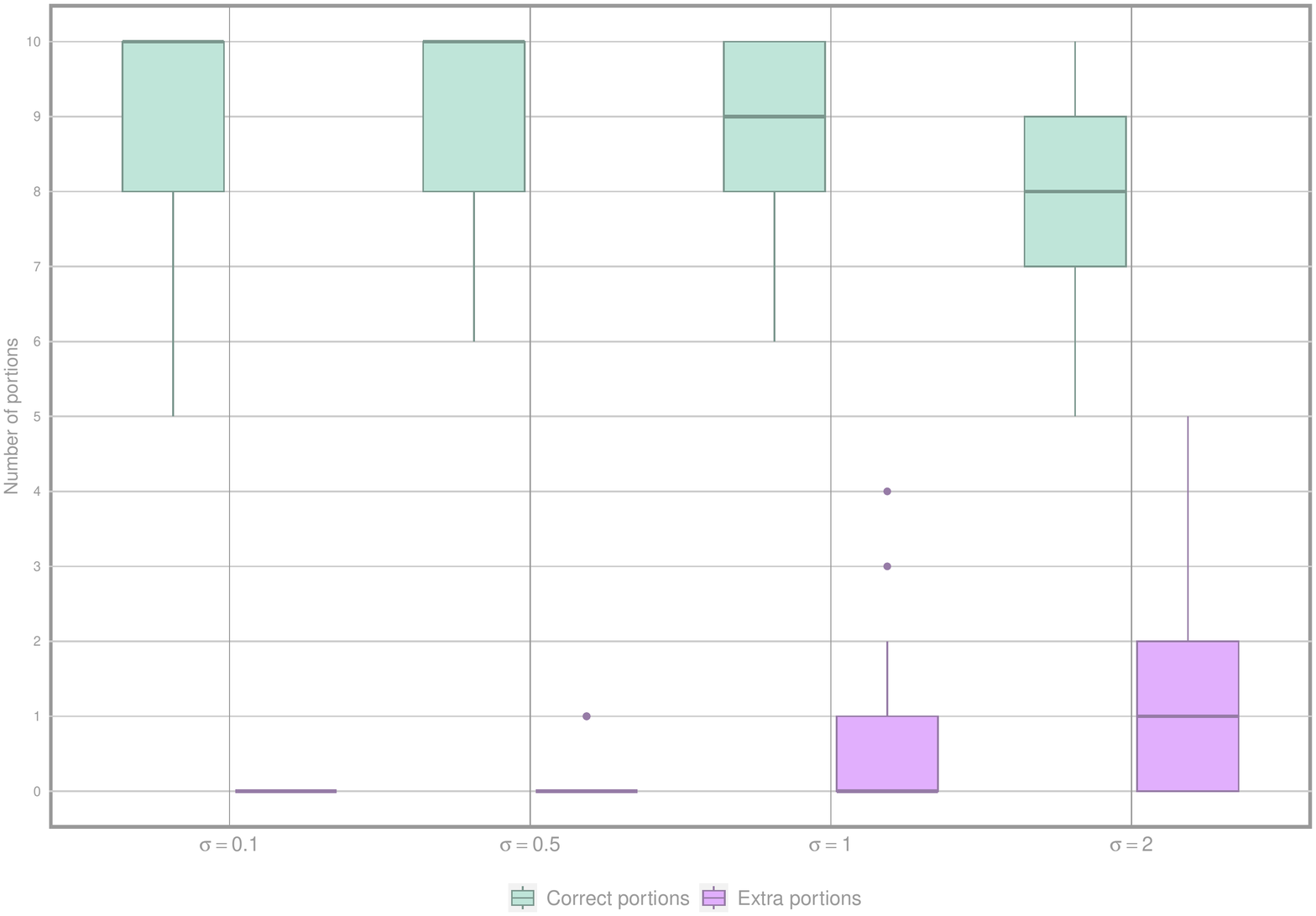}
\caption{
    Performance of \emph{funBIalign} for all simulations where motifs have $10$ occurrences and shared noise levels. 
    The algorithm is run with minimum cardinalities $n_{min}=5,6,7,8$ and results are pooled. 
    For each $\sigma$, the green and pink boxplots (left and right) represent correctly identified portions and extra portions, respectively.  
    See Figure 2.A.
}
\label{sup:general_performance_10}
\end{figure}

\begin{figure}[H]
\includegraphics[width=\linewidth, height = 15cm]{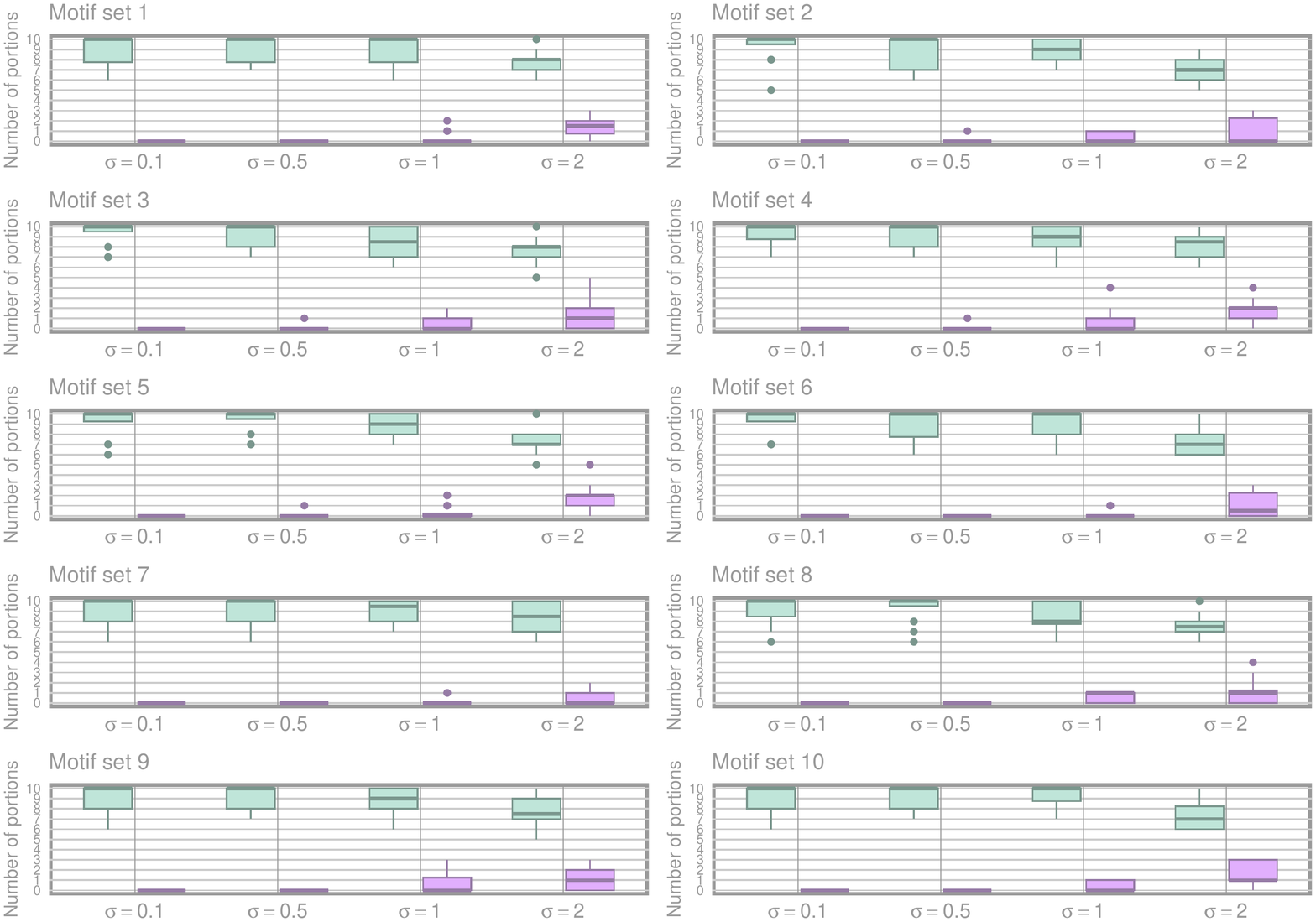}
\caption{
    Performance of \emph{funBIalign} for all simulations where motifs have $10$ occurrences and shared noise level. 
    Different panels show the performance pooling across runs of the algorithm with minimum cardinalities $n_{min} = 5, 6, 7, 8$, but separately for each of the 10 alternative motif sets. 
    For each $\sigma$, the green and pink boxplots (left and right) represent correctly identified portions and extra portions, respectively. 
}
\label{fig:simulations_performance_10}
\end{figure}

\begin{figure}[H]
\includegraphics[width=\linewidth]{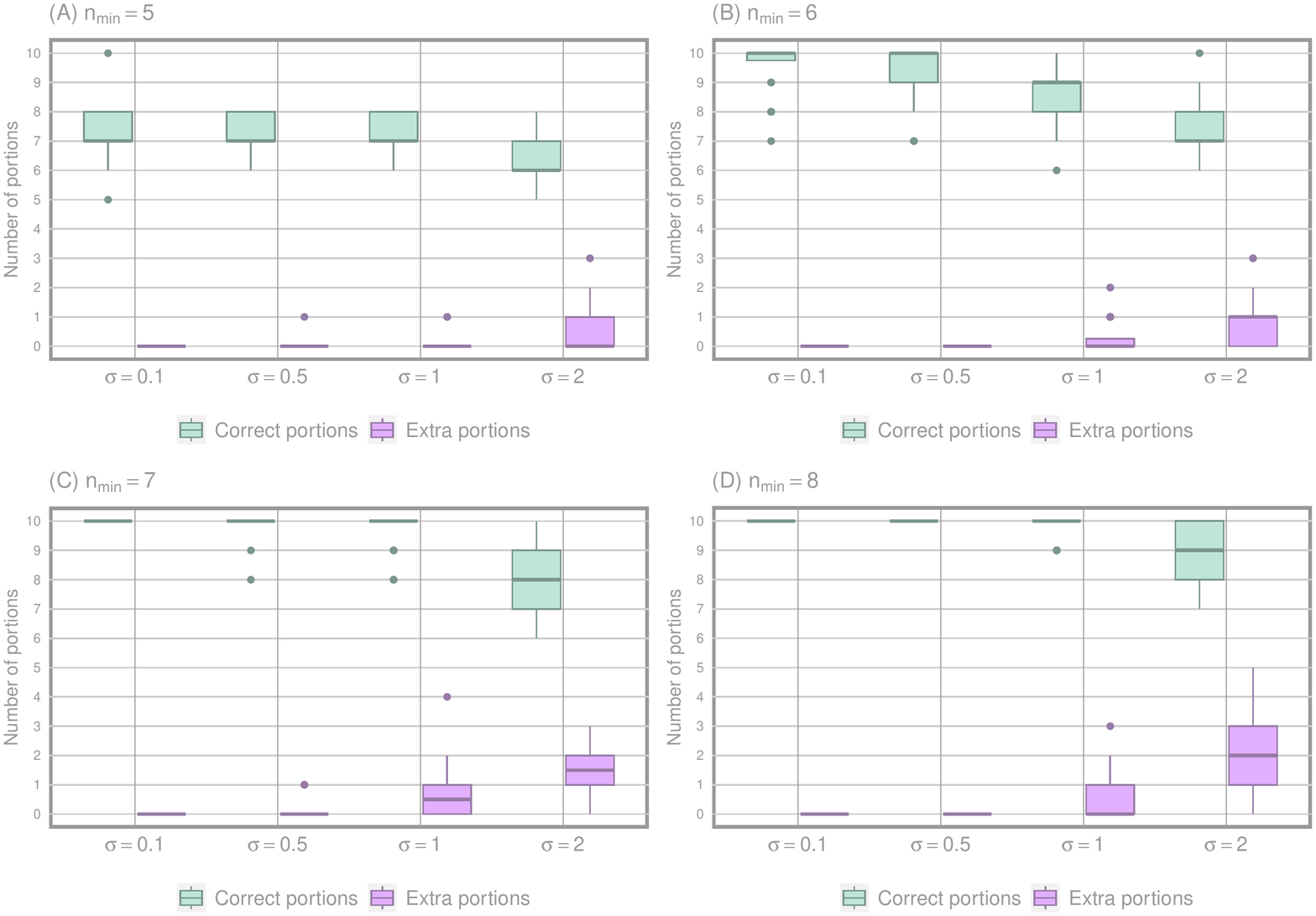}
\caption{
    Performance of \emph{funBIalign} for all simulations where motifs have $10$ occurrences and shared noise level. 
    The four panels show the performance of the algorithm run with minimum cardinalities $n_{min}=5,6,7,8$, respectively, but pooling across 10 alternative motif sets.
    For each $\sigma$, the green and pink boxplots (left and right) represent correctly identified portions and extra portions, respectively. 
}
\label{fig:cardinality_effect_performance_10}
\end{figure}

\begin{figure}[H]
\includegraphics[width=\linewidth]{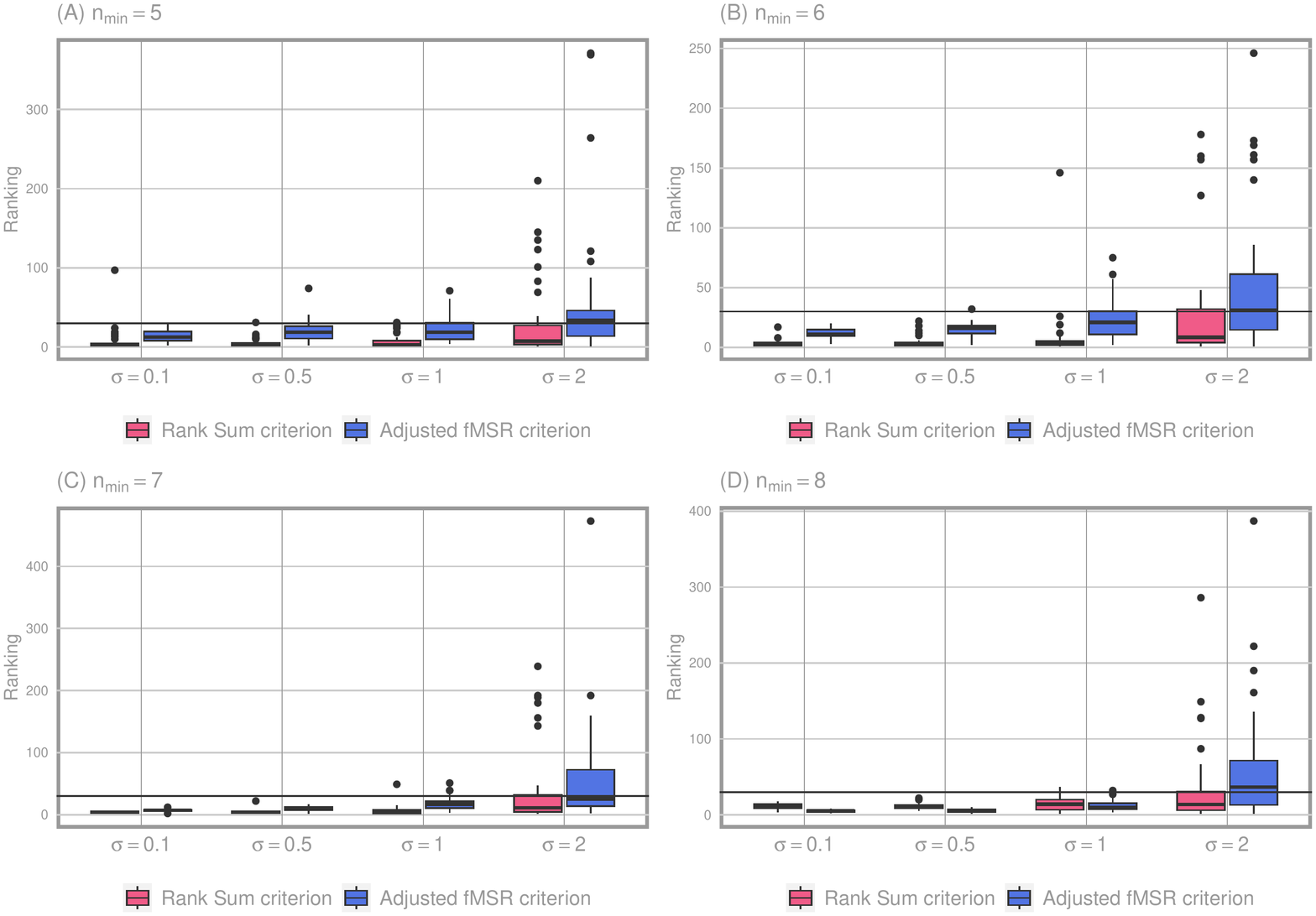}
\caption{
    Ranking of the results of \emph{funBIalign} which are most similar to the intentionally embedded motifs, for all simulations where motifs have $10$ occurrences and shared noise level. 
    The four panels show the ranking of the algorithm run with minimum cardinalities $n_{min}=5,6,7,8$, respectively, but pooling across 10 alternative motif sets.
    For each $\sigma$, the red and blue boxplots (left and right) represent ranking according to the rank sum and the adjusted fMSR criterion, respectively. 
    horizontal line indicates rank 30. 
}
\label{fig:cardinality_effect_ranking_10}
\end{figure}

\begin{figure}[H]
\includegraphics[width=\linewidth]{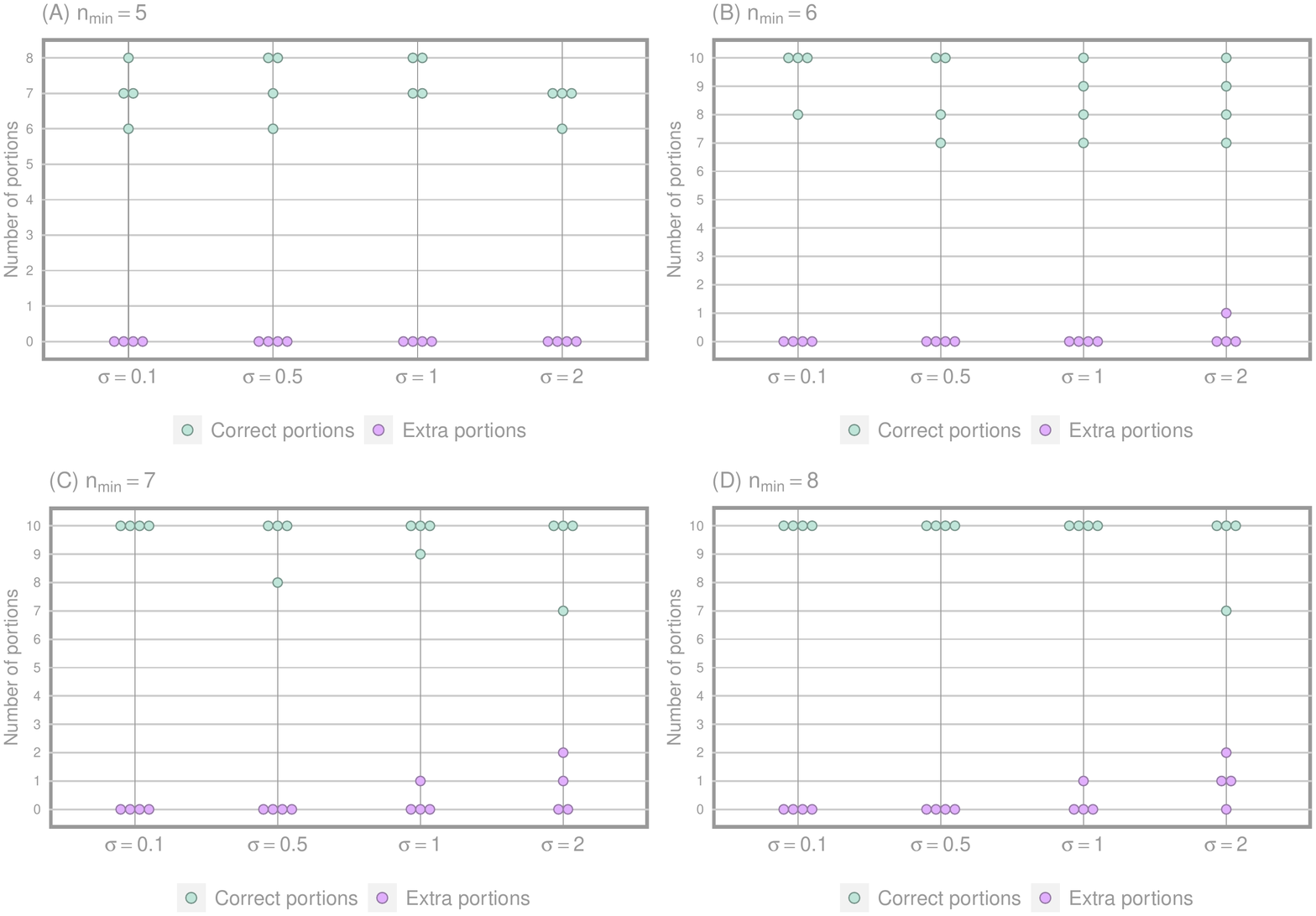}
\caption{
    Performance of {\em funBIalign} for simulations employing motif set n.~7 (where motifs have $10$ occurrences and shared noise level), shown separately for runs with varying $n_{min}$. 
    Green and pink jittered dots represent correctly identified portions and extra portions, respectively.
}
\label{fig:10portions_simulation_details}
\end{figure}

\begin{figure}[H]
\includegraphics[width=\linewidth]{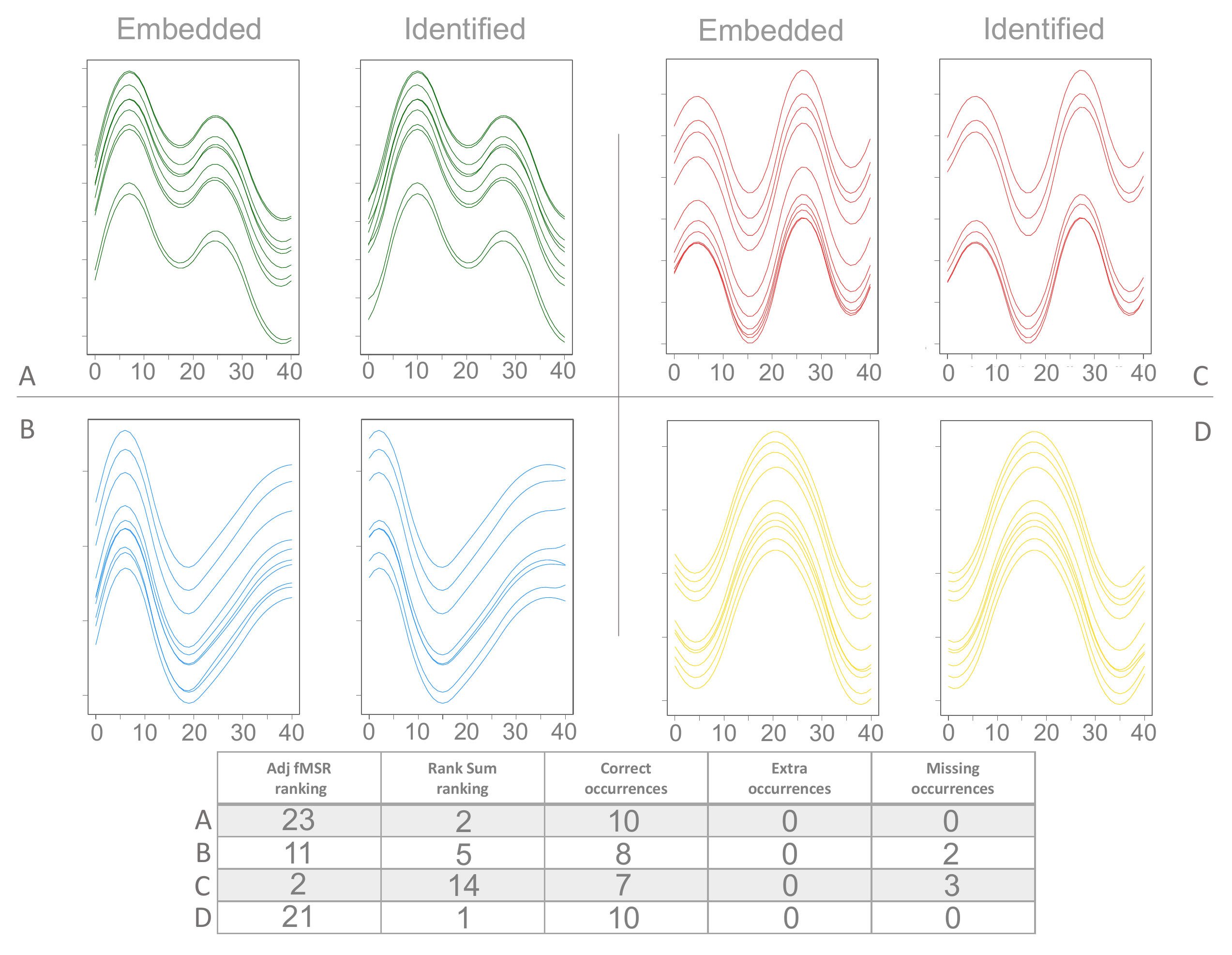}
\caption{
    Motif identification for the simulation employing motif set n.~7 and $\sigma=0.5$, where motifs have $10$ occurrences. {\em funBIalign} is run with $n_{min}=6$. 
    For each of the $4$ motifs, color-coded in green, red, blue and yellow, left and right panels show the $10$ occurrences embedded in the curve and the most similar portions identified by the algorithm, respectively. 
    The table provides rankings and numbers of correctly identified, extra and missing occurrences for each motif.
    No extra occurrences are identified, but $5$ embedded occurrences are missed.
}
\label{fig:portion10_simulation7_card6_sd05_details}
\end{figure}

\begin{figure}[H]
\includegraphics[width=\linewidth]{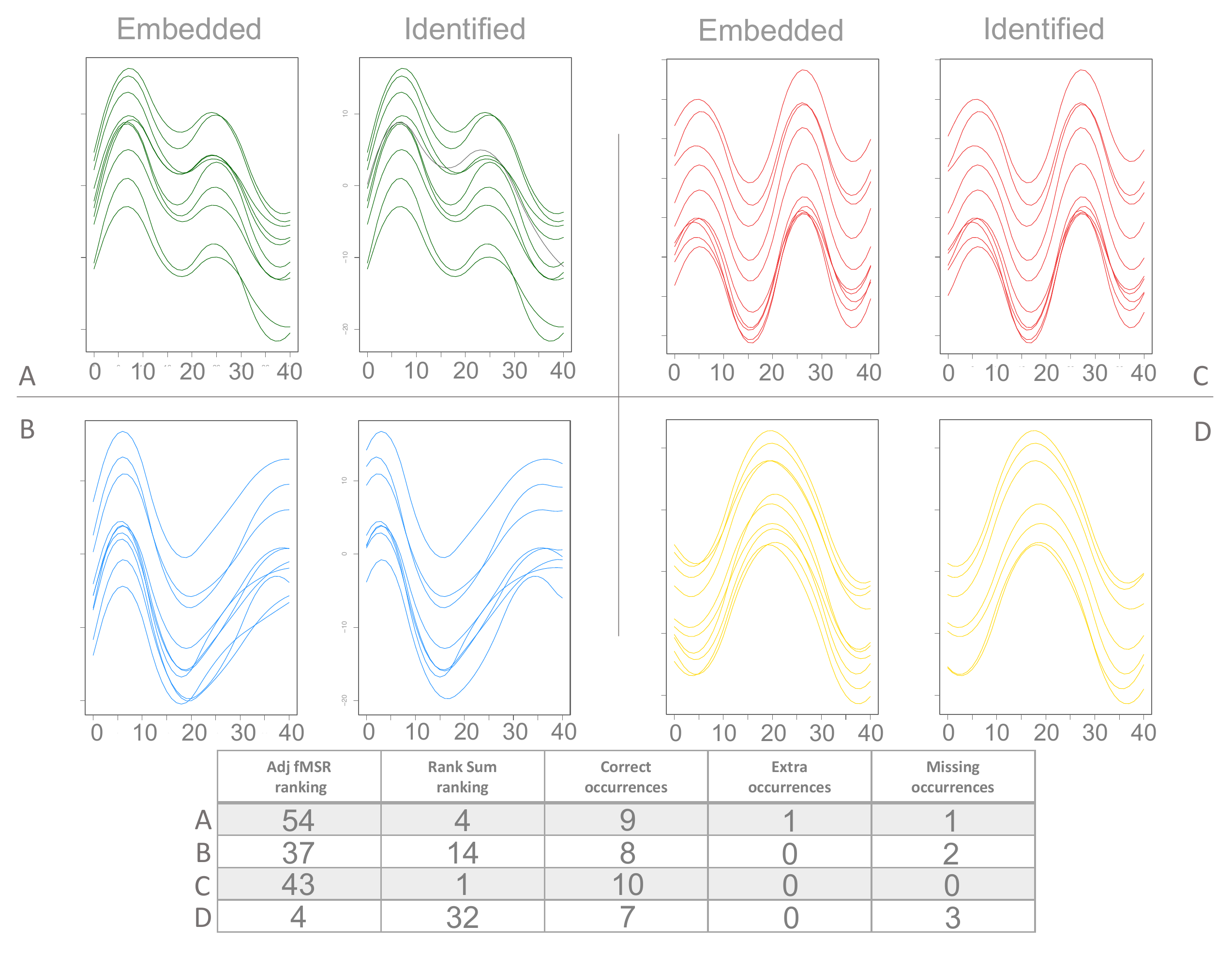}
\caption{
    Motif identification for the simulation employing motif set n.~7 and $\sigma=2$, where motifs have $10$ occurrences. {\em funBIalign} is run with $n_{min}=6$. 
    See legend for Figure~\ref{fig:portion10_simulation7_card6_sd05_details}.
    Extra occurrences are represented in gray among the identified portions panels. 
}
\label{fig:portion10_simulation7_card6_sd2_details}
\end{figure}

\section{Motifs with different noise levels}
\label{sec:different}

\subsection{Motifs with 8 occurrences}
\label{sec:different8}
\begin{figure}[H]
\includegraphics[width=\linewidth]{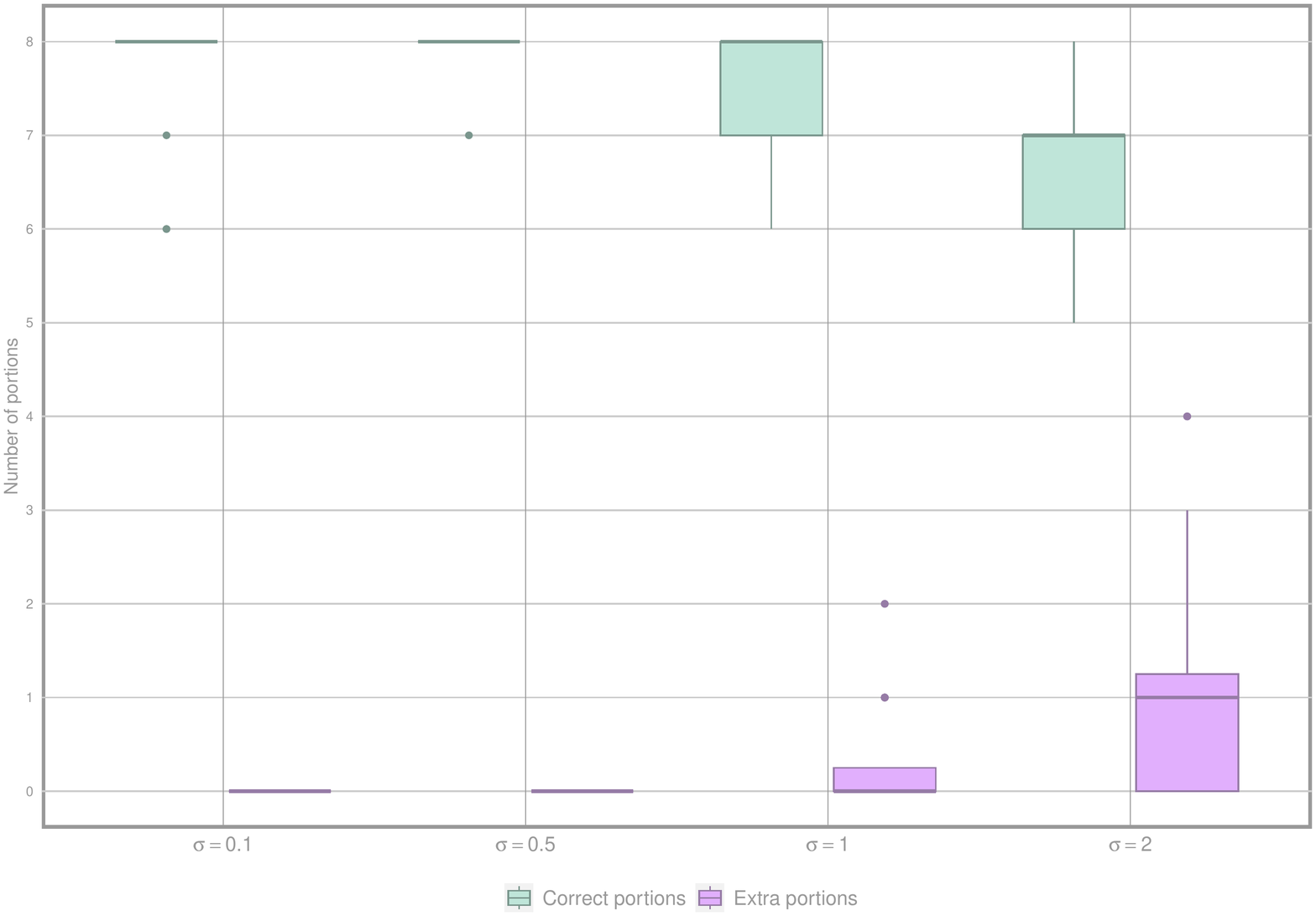}
\caption{
    Performance of \emph{funBIalign} for all simulations where motifs have $8$ occurrences and different noise levels. 
    The algorithm is run with minimum cardinalities $n_{min}=5,6,7,8$ and results are pooled. 
    For each $\sigma$, the green and pink boxplots (left and right) represent correctly identified portions and extra portions, respectively. See Figure 2.A.
}
\label{sup:hard_general_performance_8}
\end{figure}

\begin{figure}[H]
\includegraphics[width=\linewidth, height = 15cm]{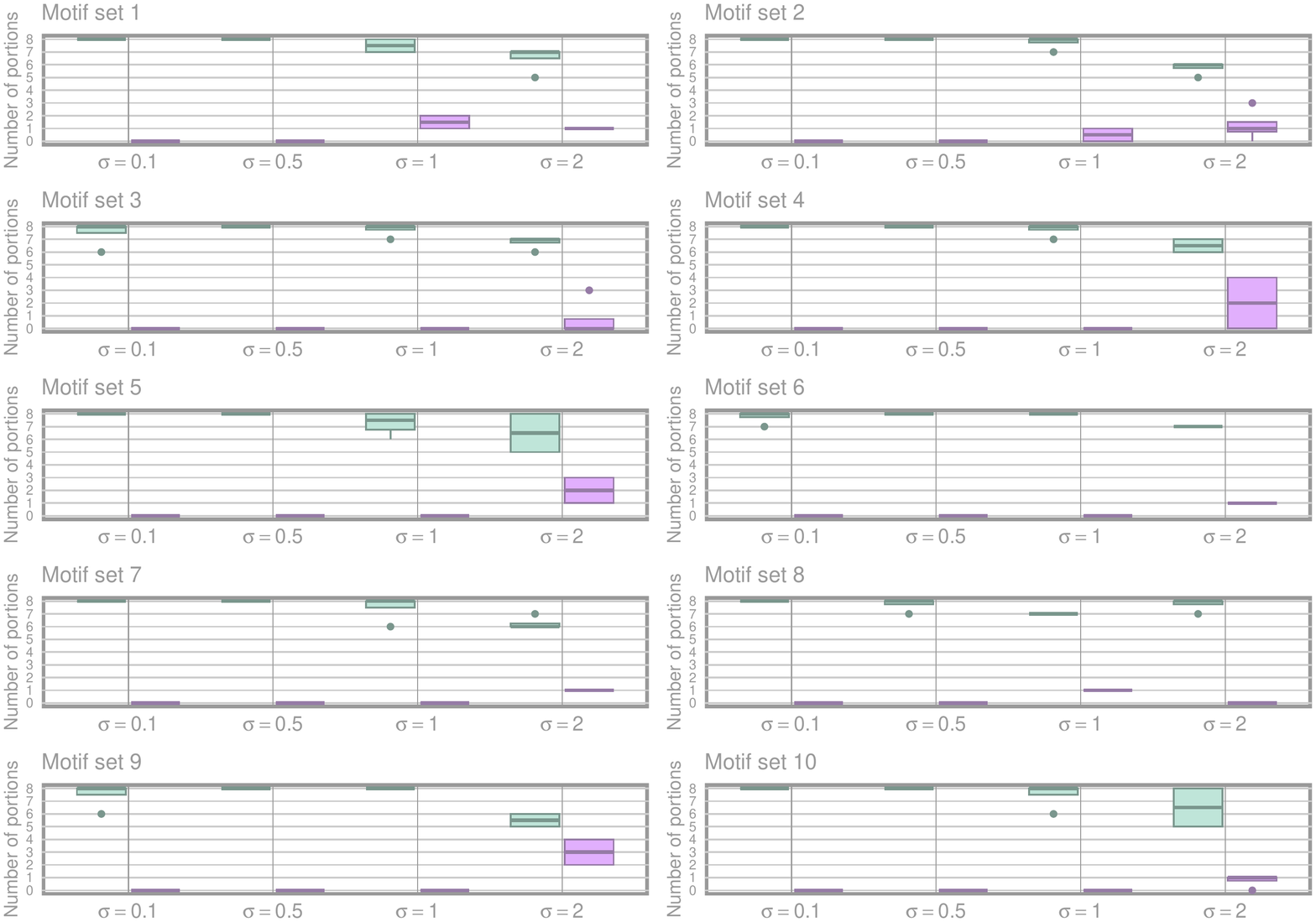}
\caption{ Performance of \emph{funBIalign} for all simulations where motifs have $8$ occurrences and different noise level. 
Different panels show the performance pooling across runs of the algorithm with minimum cardinalities $n_{min} = 5, 6, 7, 8$, but separately for each of the 10 alternative motif sets. 
For each $\sigma$, the green and pink boxplots (left and right) represent correctly identified portions and extra portions, respectively. }
\label{fig:hard_simulations_performance_8}
\end{figure}

\begin{figure}[H]
\includegraphics[width=\linewidth]{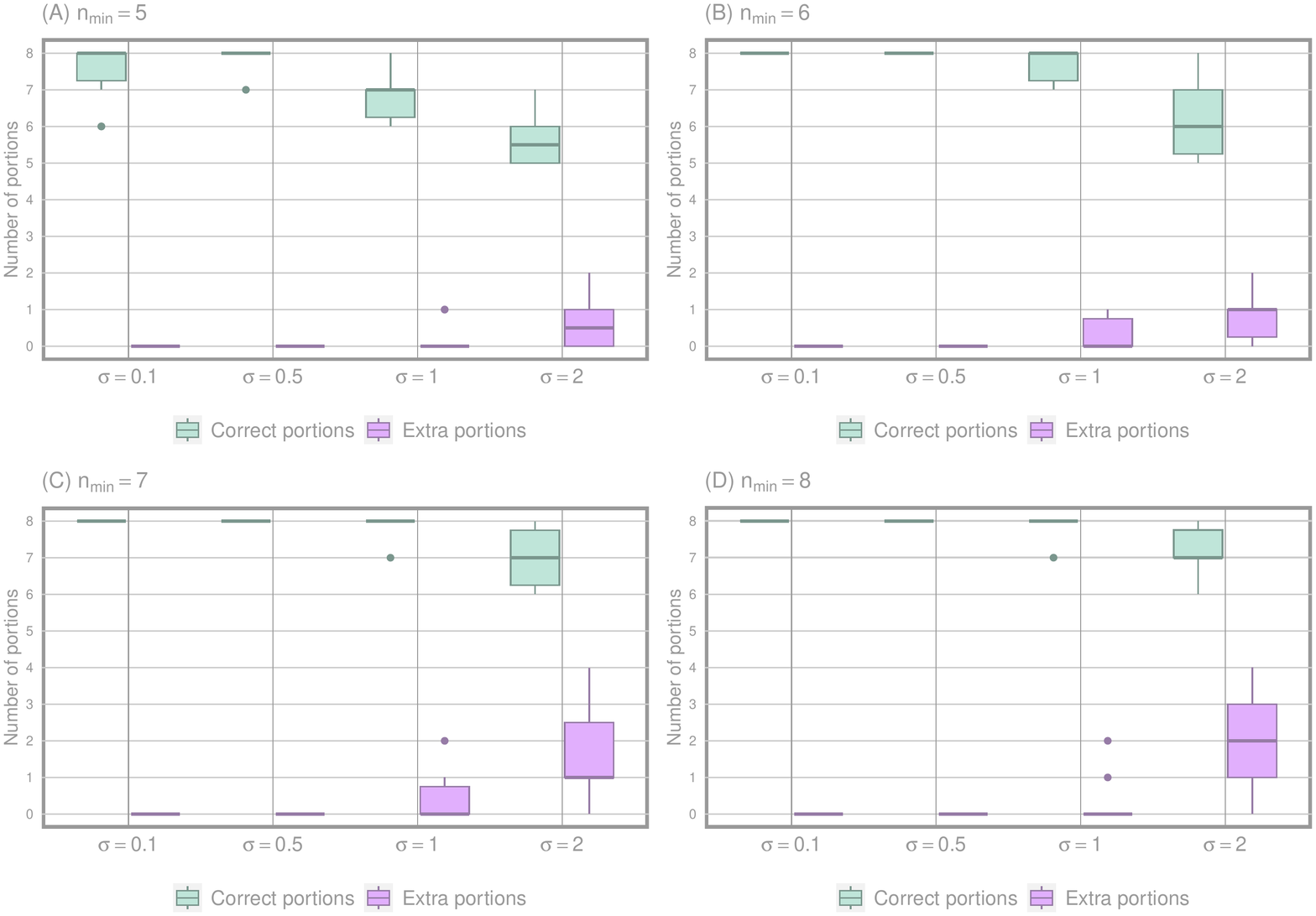}
\caption{
Performance of \emph{funBIalign} for all simulations where motifs have $8$ occurrences and different noise level. 
The four panels show the performance of the algorithm run with minimum cardinalities $n_{min}=5,6,7,8$, respectively, but pooling across 10 alternative motif sets.
For each $\sigma$, the green and pink boxplots (left and right) represent correctly identified portions and extra portions, respectively.}
\label{fig:hard_cardinality_effect_performance_8}
\end{figure}

\begin{figure}[H]
\includegraphics[width=\linewidth]{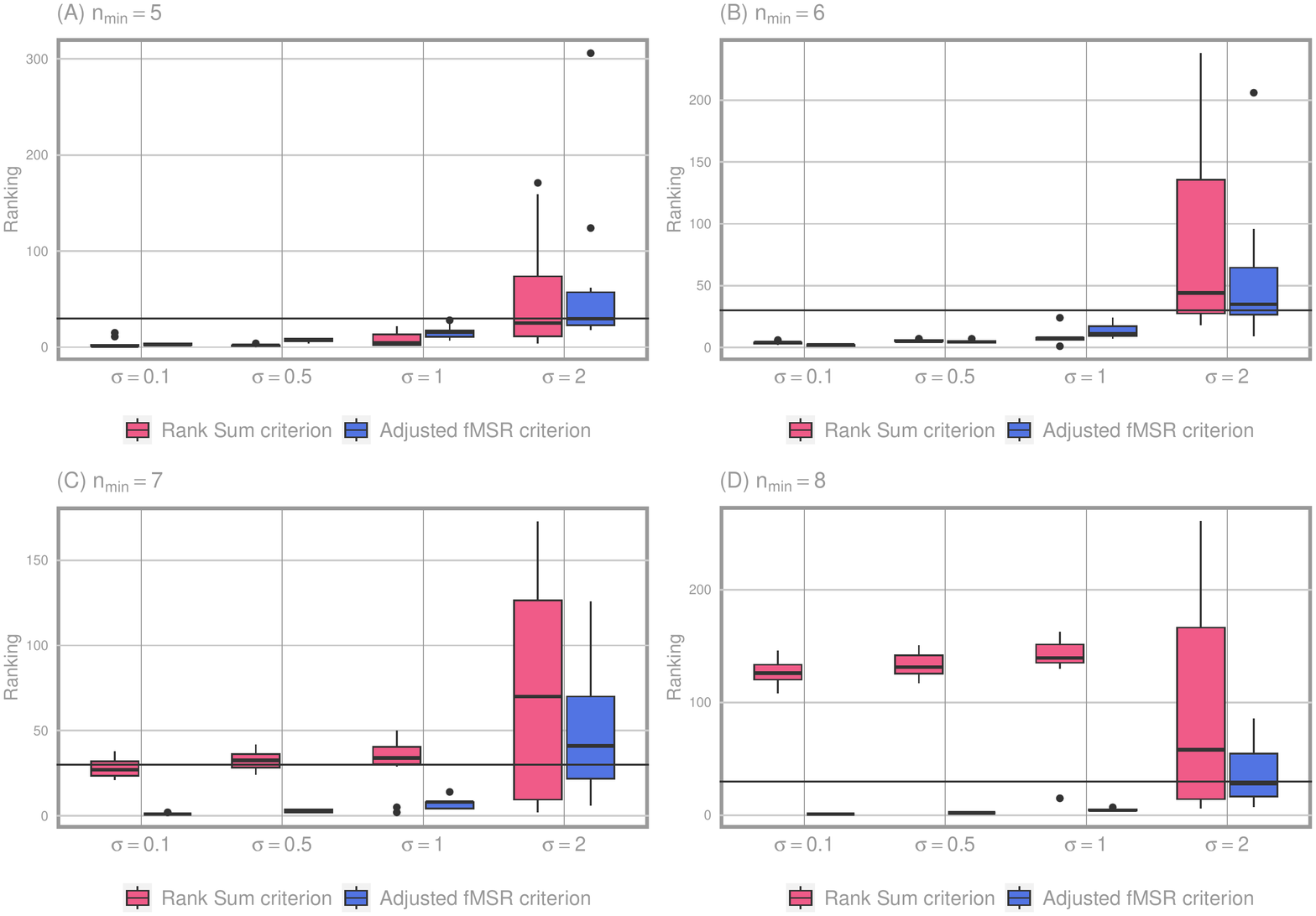}
\caption{Ranking of the results of \emph{funBIalign} which are most similar to the intentionally embedded motifs, for all simulations where motifs have $8$ occurrences and different noise level. 
The four panels show the ranking of the algorithm run with minimum cardinalities $n_{min}=5,6,7,8$, respectively, but pooling across 10 alternative motif sets.
For each $\sigma$, the red and blue boxplots (left and right) represent ranking according to the rank sum and the adjusted fMSR criterion, respectively. 
Horizontal line indicates rank 30.}
\label{fig:hard_cardinality_effect_ranking_8}
\end{figure}

\begin{figure}[H]
\includegraphics[width=\linewidth]{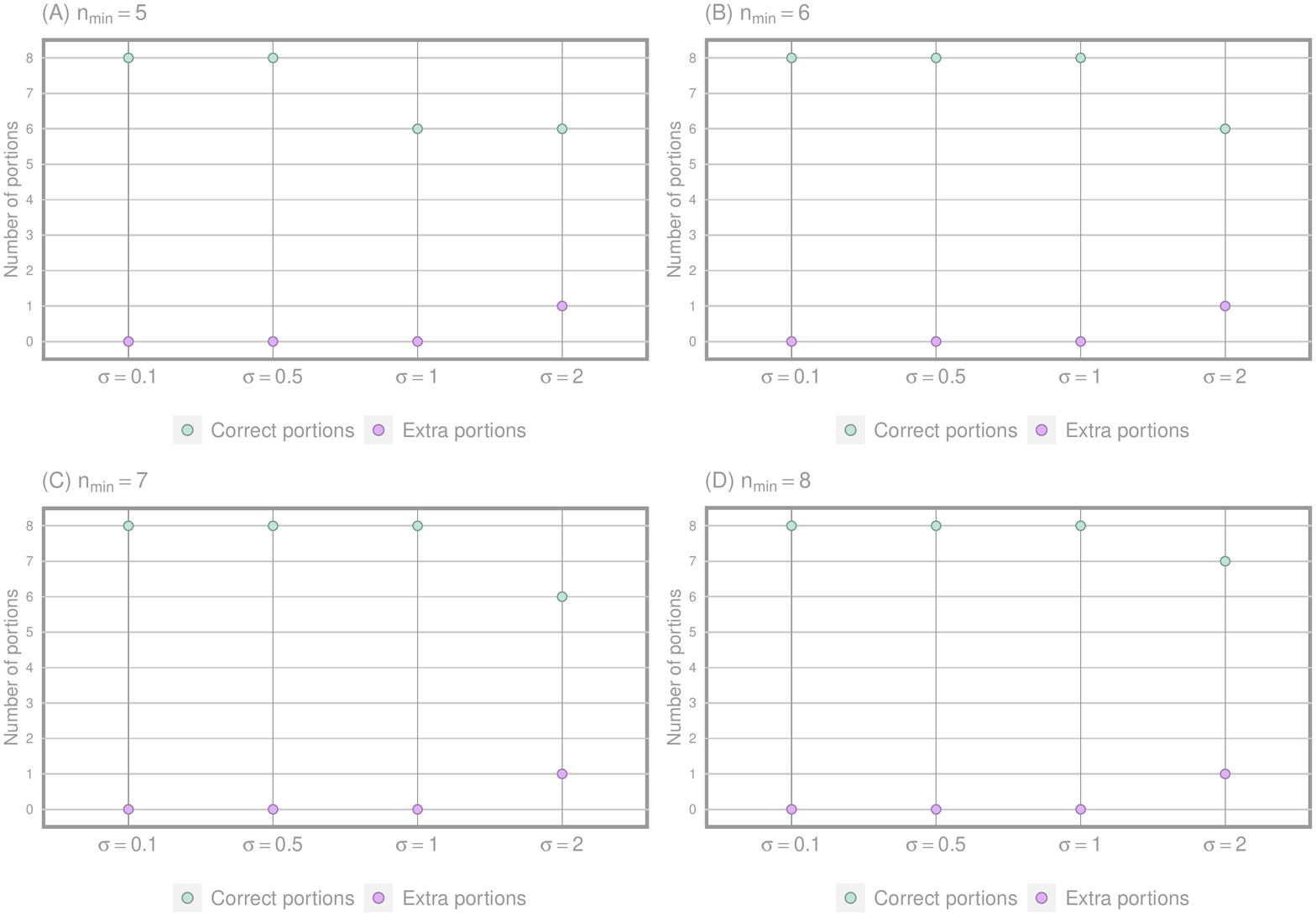}
\caption{ Performance of {\em funBIalign} for simulations employing motif set n.~7 (where motifs have $8$ occurrences and different noise level), shown separately for runs with varying $n_{min}$. Green and pink jittered dots represent correctly identified portions and extra portions, respectively.}
\label{fig:hard8_simulation5_details}
\end{figure}

\begin{figure}[H]
\includegraphics[width=\linewidth]{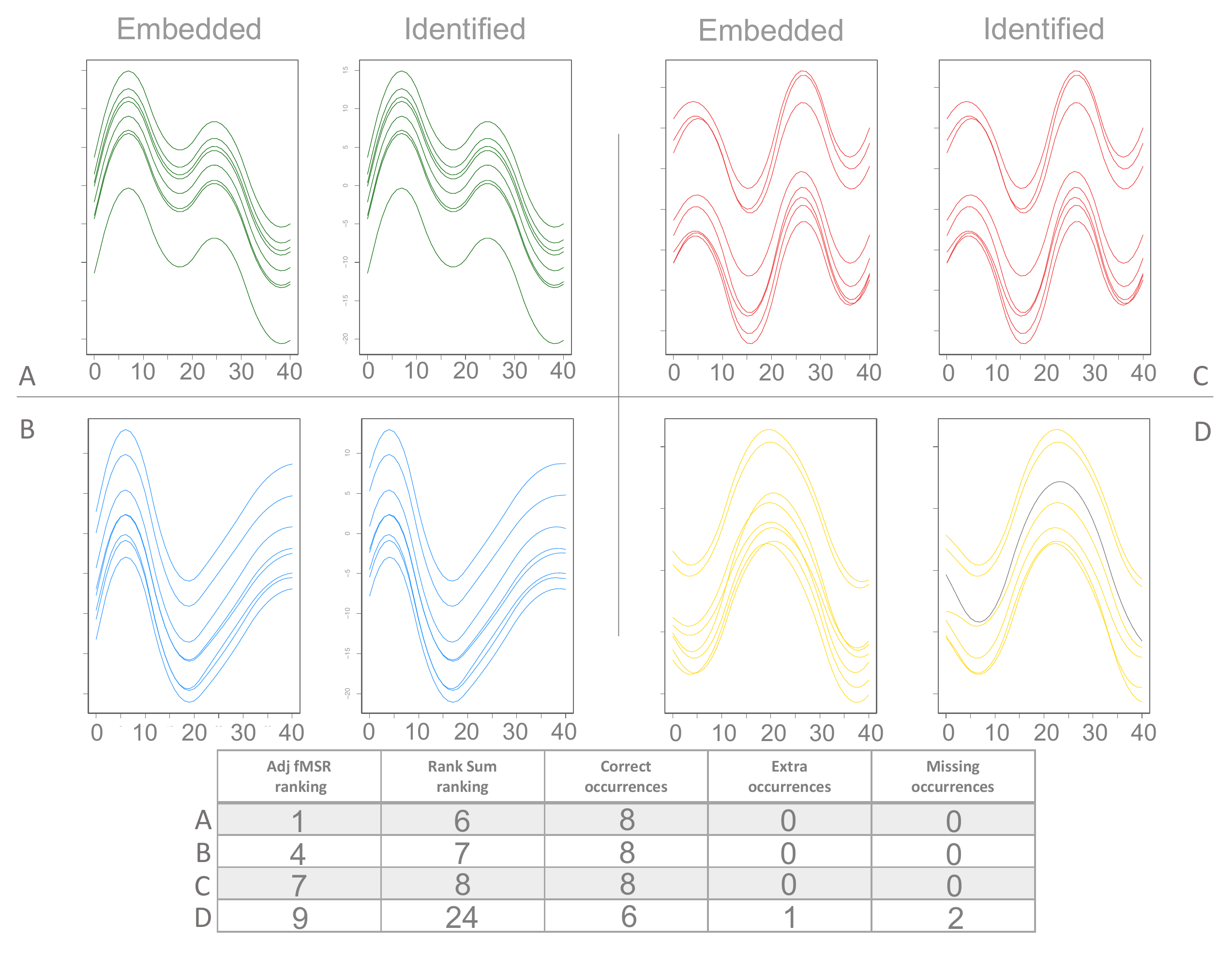}
\caption{Comparisons between the four embedded motifs (left) and the most similar ones identified by \emph{funBIalign} (right) for motif set n.~7 and different noise level (where motifs have $8$ occurrences). 
The algorithm is run with minimum cardinality $n_{min} = 6$.
The table reports the ranking according to the adjusted fMSR and the rank sum criteria, as well as details on the number of correct, extra, and missing occurrences. }
\label{fig:portion8_simulation7_card6_details}
\end{figure}

\subsection{Motifs with 10 occurrences}
\label{sec:different10}
\begin{figure}[H]
\includegraphics[width=\linewidth]{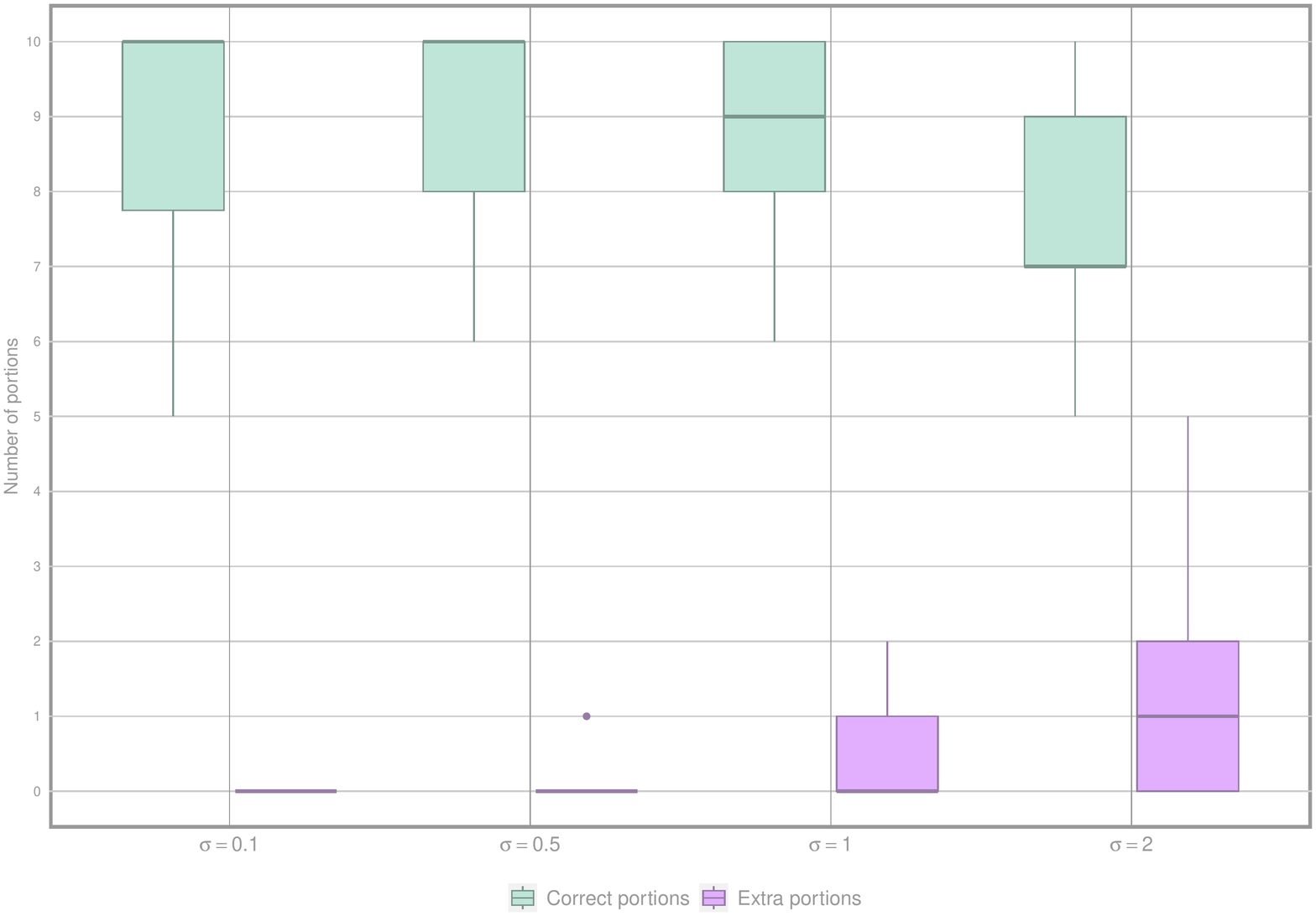}
\caption{Performance of \emph{funBIalign} for all simulations where motifs have $10$ occurrences and different noise levels. 
The algorithm is run with minimum cardinalities $n_{min}=5,6,7,8$ and results are pooled. 
For each $\sigma$, the green and pink boxplots (left and right) represent correctly identified portions and extra portions, respectively. See Figure 2.A.}
\label{sup:hard_general_performance_10}
\end{figure}

\begin{figure}[H]
\includegraphics[width=\linewidth]{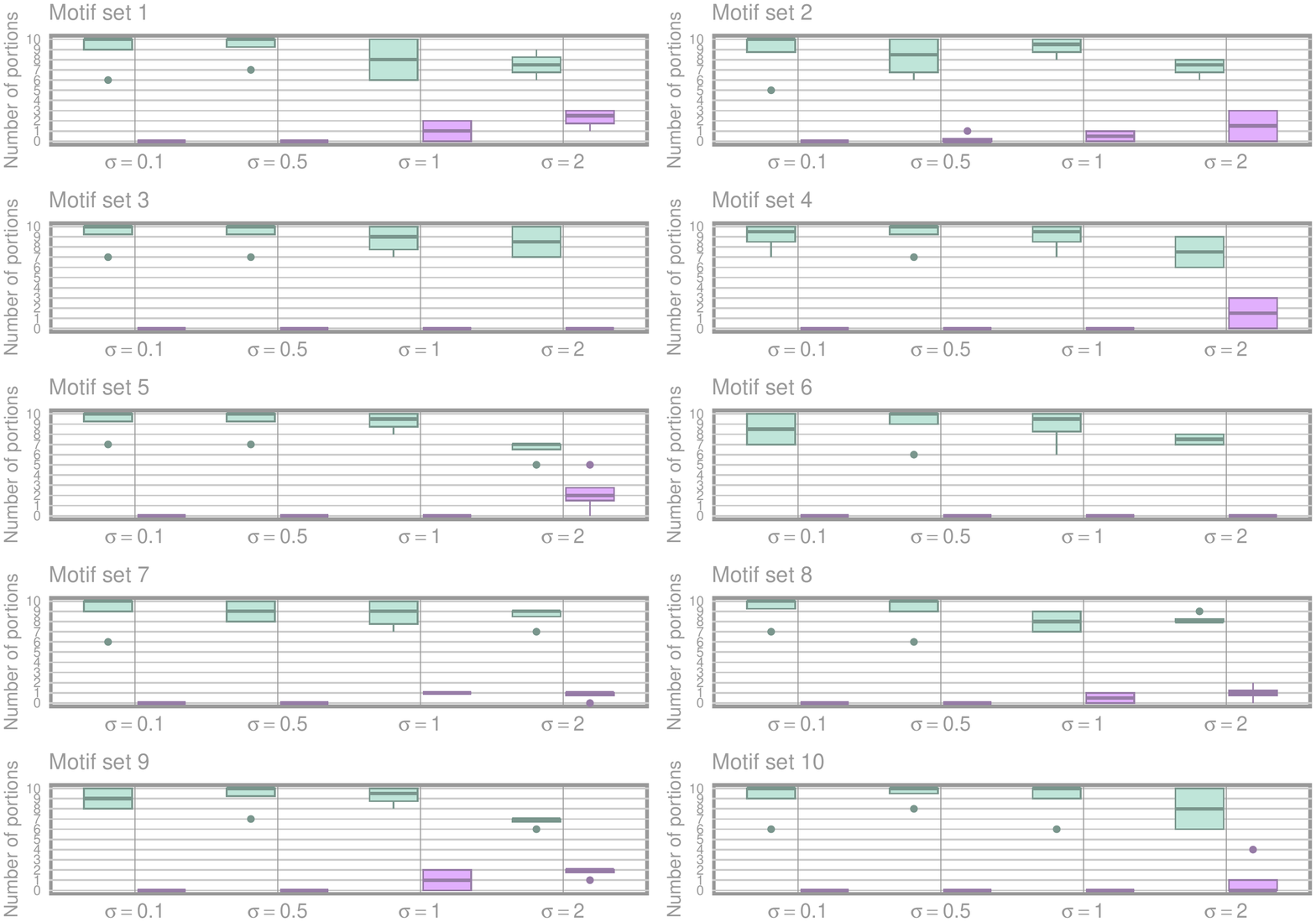}
\caption{Performance of \emph{funBIalign} for all simulations where motifs have $10$ occurrences and different noise level. 
Different panels show the performance pooling across runs of the algorithm with minimum cardinalities $n_{min} = 5, 6, 7, 8$, but separately for each of the 10 alternative motif sets. 
For each $\sigma$, the green and pink boxplots (left and right) represent correctly identified portions and extra portions, respectively. }
\label{fig:hard_simulations_performance_10}
\end{figure}

\begin{figure}[H]
\includegraphics[width=\linewidth]{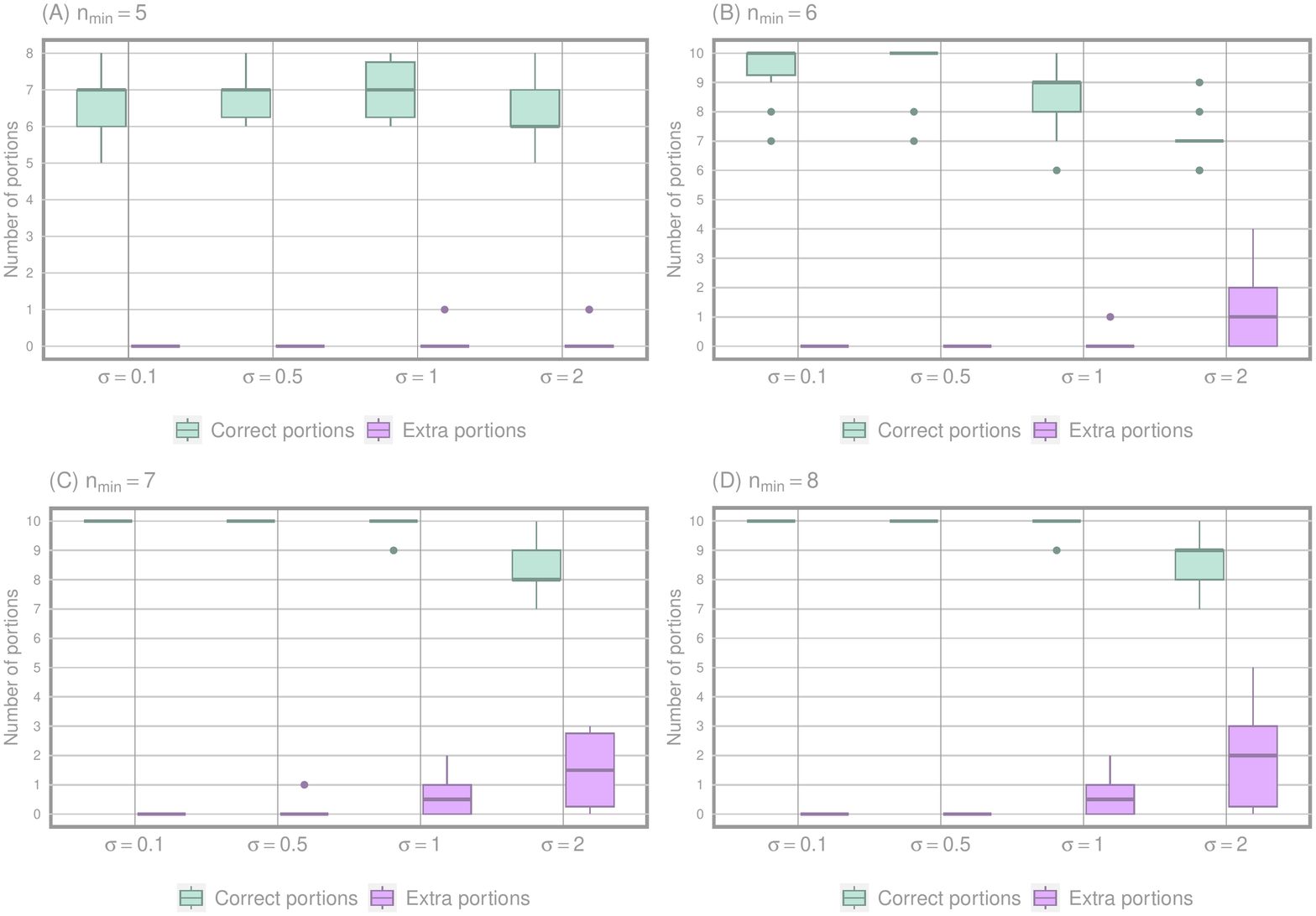}
\caption{Performance of \emph{funBIalign} for all simulations where motifs have $10$ occurrences and different noise level. 
The four panels show the performance of the algorithm run with minimum cardinalities $n_{min}=5,6,7,8$, respectively, but pooling across 10 alternative motif sets.
For each $\sigma$, the green and pink boxplots (left and right) represent correctly identified portions and extra portions, respectively.}
\label{fig:hard_cardinality_effect_performance_10}
\end{figure}

\begin{figure}[H]
\includegraphics[width=\linewidth]{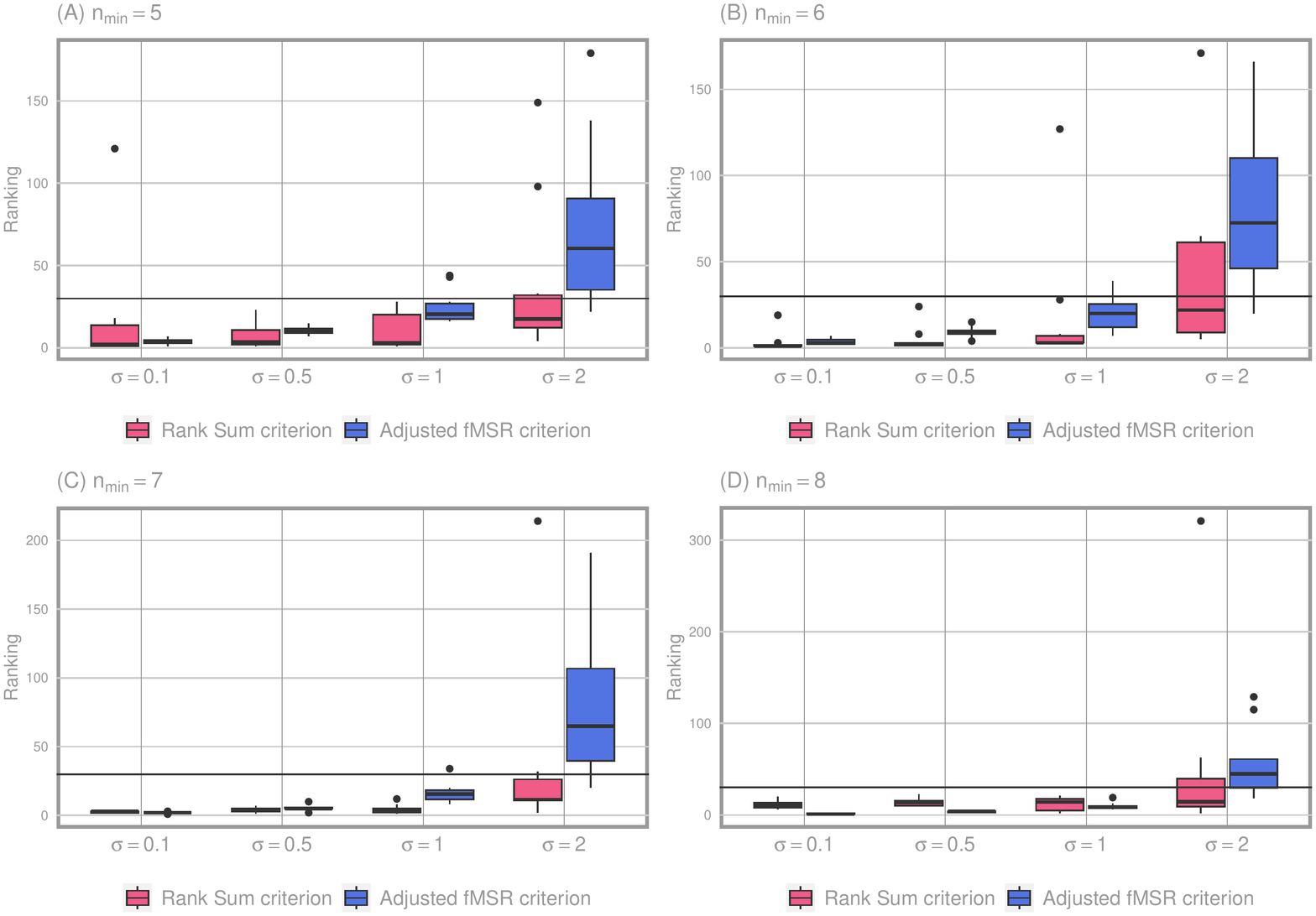}
\caption{Ranking of the results of \emph{funBIalign} which are most similar to the intentionally embedded motifs, for all simulations where motifs have $10$ occurrences and different noise level. 
The four panels show the ranking of the algorithm run with minimum cardinalities $n_{min}=5,6,7,8$, respectively, but pooling across 10 alternative motif sets.
For each $\sigma$, the red and blue boxplots (left and right) represent ranking according to the rank sum and the adjusted fMSR criterion, respectively. 
Horizontal line indicates rank 30.}
\label{fig:hard_cardinality_effect_ranking_10}
\end{figure}

\begin{figure}[H]
\includegraphics[width=\linewidth]{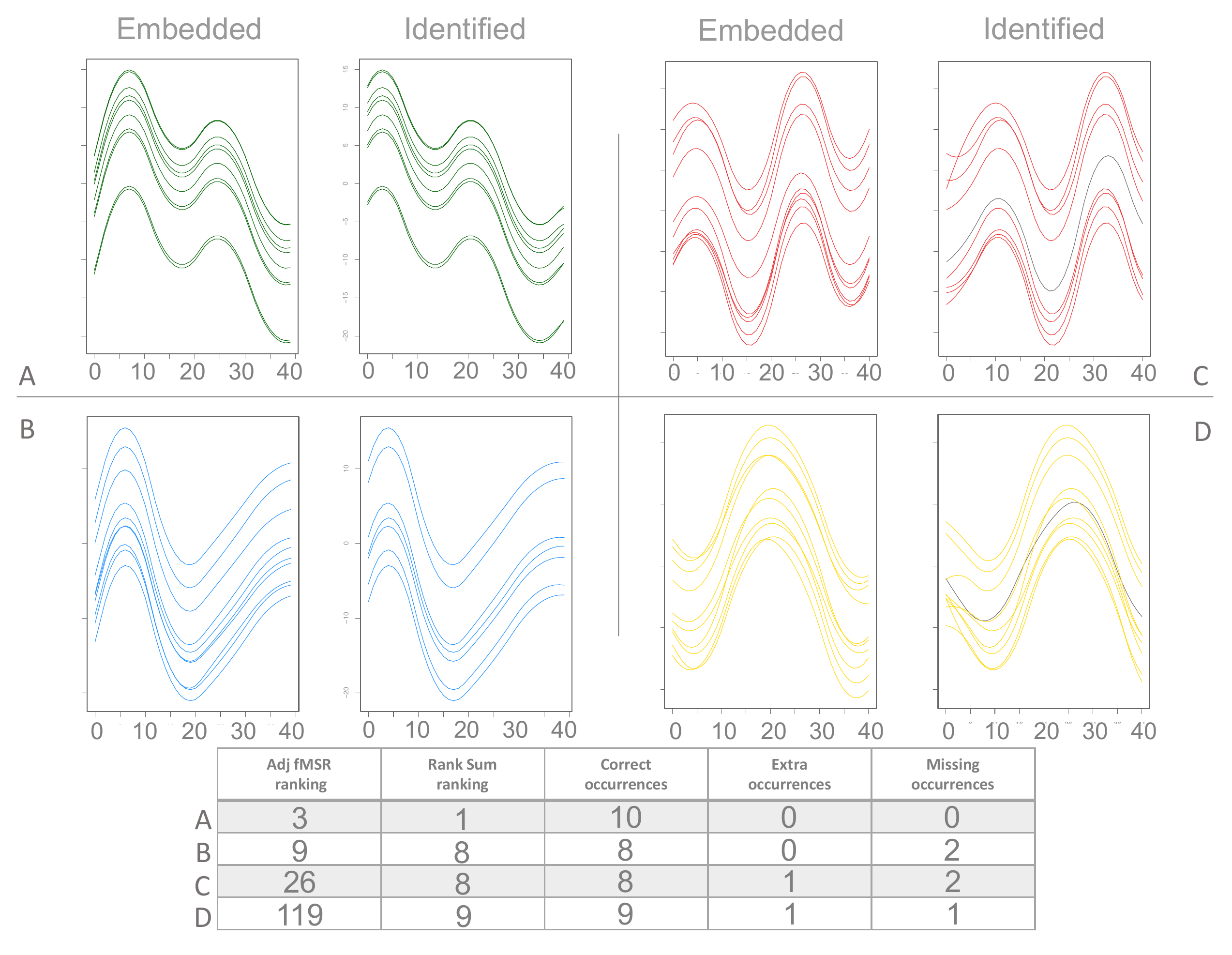}
\caption{Comparisons between the four embedded motifs (left) and the most similar ones identified by \emph{funBIalign} (right) for motif set n.~7 and different noise level (where motifs have $10$ occurrences). 
The algorithm is run with minimum cardinality $n_{min} = 6$.
The table reports the ranking according to the adjusted fMSR and the rank sum criteria, as well as details on the number of correct, extra, and missing occurrences. }
\label{fig:hard10_simulation7_details}
\end{figure}

\clearpage

\section{Comparison with {\em SCRIMP-MP}}
\label{sec:comparison_MP}

Table \ref{tab:MP_bis} presents the comparison between \emph{funBIalign} and \emph{SCRIMP-MP} in the case of motifs with $8$ occurrences each and different noise levels, using motif set n.~7 (see Section 4). 
As in the comparisons of Section 5, {\em funBIalign} is run with $\ell = 41$ (the true length of the motifs) and $n_{min}=5,6,7,8$, while {\em SCRIMP-MP} is run with $u=50$, $\ell = 41$ and all possible combinations of $R=3,10,25,50$ and $k_{neighbor}=4,6,8$. 
As expected, the identification of the motifs is harder for the motif with $\sigma = 2$ with both the methods. In general, \emph{funBIalign} performs better than \emph{SCRIMP-MP}.

\begin{table}[H]
\centering
\begin{tabular}{cccccccccc}
\footnotesize
\multirow{2}{*}{Motif}                                         & \multicolumn{4}{l}{\textbf{\emph{funBIalign} - $n_{min}$}}                                                                                                                                                                                   & \textbf{MP}            & \multicolumn{4}{c}{$R$}                                                                                                                               \\ \cline{2-10} 
                                                               & $5$                                              & $6$                                              & $7$                                                & $8$                                                & $k_{neighbor}$        & 3                                   & 10                                  & 25                                  & 50                                  \\ \hline
\multicolumn{1}{c|}{\multirow{3}{*}{Motif 1 ($\sigma = 0.1$)}} & \multicolumn{1}{c|}{\multirow{3}{*}{\textbf{(8,0)}}} & \multicolumn{1}{c|}{\multirow{3}{*}{\textbf{(8,0)}}} & \multicolumn{1}{c|}{\multirow{3}{*}{\textbf{(8,0)}}} & \multicolumn{1}{c|}{\multirow{3}{*}{\textbf{(8,0)}}} & \multicolumn{1}{c|}{4} & \multicolumn{1}{c|}{(6,0)}          & \multicolumn{1}{c|}{(6,0)}          & \multicolumn{1}{c|}{(6,0)}          & \multicolumn{1}{c|}{(6,0)}          \\ \cline{6-10} 
\multicolumn{1}{c|}{}                                          & \multicolumn{1}{c|}{}                                & \multicolumn{1}{c|}{}                                & \multicolumn{1}{c|}{}                                & \multicolumn{1}{c|}{}                                & \multicolumn{1}{c|}{6} & \multicolumn{1}{c|}{(7,0)}          & \multicolumn{1}{c|}{\textbf{(8,0)}} & \multicolumn{1}{c|}{\textbf{(8,0)}} & \multicolumn{1}{c|}{\textbf{(8,0)}} \\ \cline{6-10} 
\multicolumn{1}{c|}{}                                          & \multicolumn{1}{c|}{}                                & \multicolumn{1}{c|}{}                                & \multicolumn{1}{c|}{}                                & \multicolumn{1}{c|}{}                                & \multicolumn{1}{c|}{8} & \multicolumn{1}{c|}{(7,0)}          & \multicolumn{1}{c|}{\textbf{(8,0)}} & \multicolumn{1}{c|}{\textbf{(8,0)}} & \multicolumn{1}{c|}{\textbf{(8,0)}} \\ \hline
\multicolumn{1}{c|}{\multirow{3}{*}{Motif 2 ($\sigma = 0.5$)}} & \multicolumn{1}{c|}{\multirow{3}{*}{\textbf{(8,0)}}} & \multicolumn{1}{c|}{\multirow{3}{*}{\textbf{(8,0)}}} & \multicolumn{1}{c|}{\multirow{3}{*}{\textbf{(8,0)}}} & \multicolumn{1}{c|}{\multirow{3}{*}{\textbf{(8,0)}}} & \multicolumn{1}{c|}{4} & \multicolumn{1}{c|}{(6,0)}          & \multicolumn{1}{c|}{(6,0)}          & \multicolumn{1}{c|}{(6,0)}          & \multicolumn{1}{c|}{(6,0)}          \\ \cline{6-10} 
\multicolumn{1}{c|}{}                                          & \multicolumn{1}{c|}{}                                & \multicolumn{1}{c|}{}                                & \multicolumn{1}{c|}{}                                & \multicolumn{1}{c|}{}                                & \multicolumn{1}{c|}{6} & \multicolumn{1}{c|}{\textbf{(8,0)}} & \multicolumn{1}{c|}{\textbf{(8,0)}} & \multicolumn{1}{c|}{\textbf{(8,0)}} & \multicolumn{1}{c|}{\textbf{(8,0)}} \\ \cline{6-10} 
\multicolumn{1}{c|}{}                                          & \multicolumn{1}{c|}{}                                & \multicolumn{1}{c|}{}                                & \multicolumn{1}{c|}{}                                & \multicolumn{1}{c|}{}                                & \multicolumn{1}{c|}{8} & \multicolumn{1}{c|}{\textbf{(8,0)}} & \multicolumn{1}{c|}{\textbf{(8,0)}} & \multicolumn{1}{c|}{(8,2)}          & \multicolumn{1}{c|}{(8,2)}          \\ \hline
\multicolumn{1}{c|}{\multirow{3}{*}{Motif 3 ($\sigma = 1$)}}   & \multicolumn{1}{c|}{\multirow{3}{*}{(6,0)}}          & \multicolumn{1}{c|}{\multirow{3}{*}{\textbf{(8,0)}}} & \multicolumn{1}{c|}{\multirow{3}{*}{\textbf{(8,0)}}} & \multicolumn{1}{c|}{\multirow{3}{*}{\textbf{(8,0)}}} & \multicolumn{1}{c|}{4} & \multicolumn{1}{c|}{(6,0)}          & \multicolumn{1}{c|}{(6,0)}          & \multicolumn{1}{c|}{(6,0)}          & \multicolumn{1}{c|}{(6,0)}          \\ \cline{6-10} 
\multicolumn{1}{c|}{}                                          & \multicolumn{1}{c|}{}                                & \multicolumn{1}{c|}{}                                & \multicolumn{1}{c|}{}                                & \multicolumn{1}{c|}{}                                & \multicolumn{1}{c|}{6} & \multicolumn{1}{c|}{(7,1)}          & \multicolumn{1}{c|}{(7,1)}          & \multicolumn{1}{c|}{(7,1)}          & \multicolumn{1}{c|}{(7,1)}          \\ \cline{6-10} 
\multicolumn{1}{c|}{}                                          & \multicolumn{1}{c|}{}                                & \multicolumn{1}{c|}{}                                & \multicolumn{1}{c|}{}                                & \multicolumn{1}{c|}{}                                & \multicolumn{1}{c|}{8} & \multicolumn{1}{c|}{(8,1)}          & \multicolumn{1}{c|}{(8,2)}          & \multicolumn{1}{c|}{(8,2)}          & \multicolumn{1}{c|}{(8,2)}          \\ \hline
\multicolumn{1}{c|}{\multirow{3}{*}{Motif 4 ($\sigma = 2$)}}   & \multicolumn{1}{c|}{\multirow{3}{*}{(6,1)}}          & \multicolumn{1}{c|}{\multirow{3}{*}{(6,1)}}          & \multicolumn{1}{c|}{\multirow{3}{*}{(6,1)}}          & \multicolumn{1}{c|}{\multirow{3}{*}{(7,1)}}          & \multicolumn{1}{c|}{4} & \multicolumn{1}{c|}{(4,2)}          & \multicolumn{1}{c|}{(4,2)}          & \multicolumn{1}{c|}{(4,2)}          & \multicolumn{1}{c|}{(4,2)}          \\ \cline{6-10} 
\multicolumn{1}{c|}{}                                          & \multicolumn{1}{c|}{}                                & \multicolumn{1}{c|}{}                                & \multicolumn{1}{c|}{}                                & \multicolumn{1}{c|}{}                                & \multicolumn{1}{c|}{6} & \multicolumn{1}{c|}{(5,3)}          & \multicolumn{1}{c|}{(5,3)}          & \multicolumn{1}{c|}{(5,3)}          & \multicolumn{1}{c|}{(5,3)}          \\ \cline{6-10} 
\multicolumn{1}{c|}{}                                          & \multicolumn{1}{c|}{}                                & \multicolumn{1}{c|}{}                                & \multicolumn{1}{c|}{}                                & \multicolumn{1}{c|}{}                                & \multicolumn{1}{c|}{8} & \multicolumn{1}{c|}{(6,4)}          & \multicolumn{1}{c|}{(7,3)}          & \multicolumn{1}{c|}{(7,3)}          & \multicolumn{1}{c|}{(7,3)}          \\ \hline
\end{tabular}
\caption{
    Comparison between \emph{funBIalign} and \emph{SCRIMP-MP}. Cases in which all occurrences are correctly identified without the addition of any extra portions are in bold. 
}
\label{tab:MP_bis}
\end{table}

\section{Case studies}
\label{sec:casestudies}

\subsection{Case study 1: Food price inflation}
\label{sec:casestudies_inflation}

The $19$ curves considered in this case study represent the food price inflation measurements for the following geographical regions:
\begin{itemize}
    \item Eastern Africa
    \item Middle Africa
    \item Northern Africa
    \item Southern Africa
    \item Western Africa
    \item Northern America
    \item Central America
    \item Caribbean
    \item South America
    \item Central Asia
    \item Eastern Asia
    \item Southern Asia
    \item South-eastern Asia
    \item Western Asia
    \item Eastern Europe
    \item Northern Europe
    \item Southern Europe
    \item Western Europe
    \item Oceania
\end{itemize}  

\begin{figure}[H]
\includegraphics[width=\linewidth]{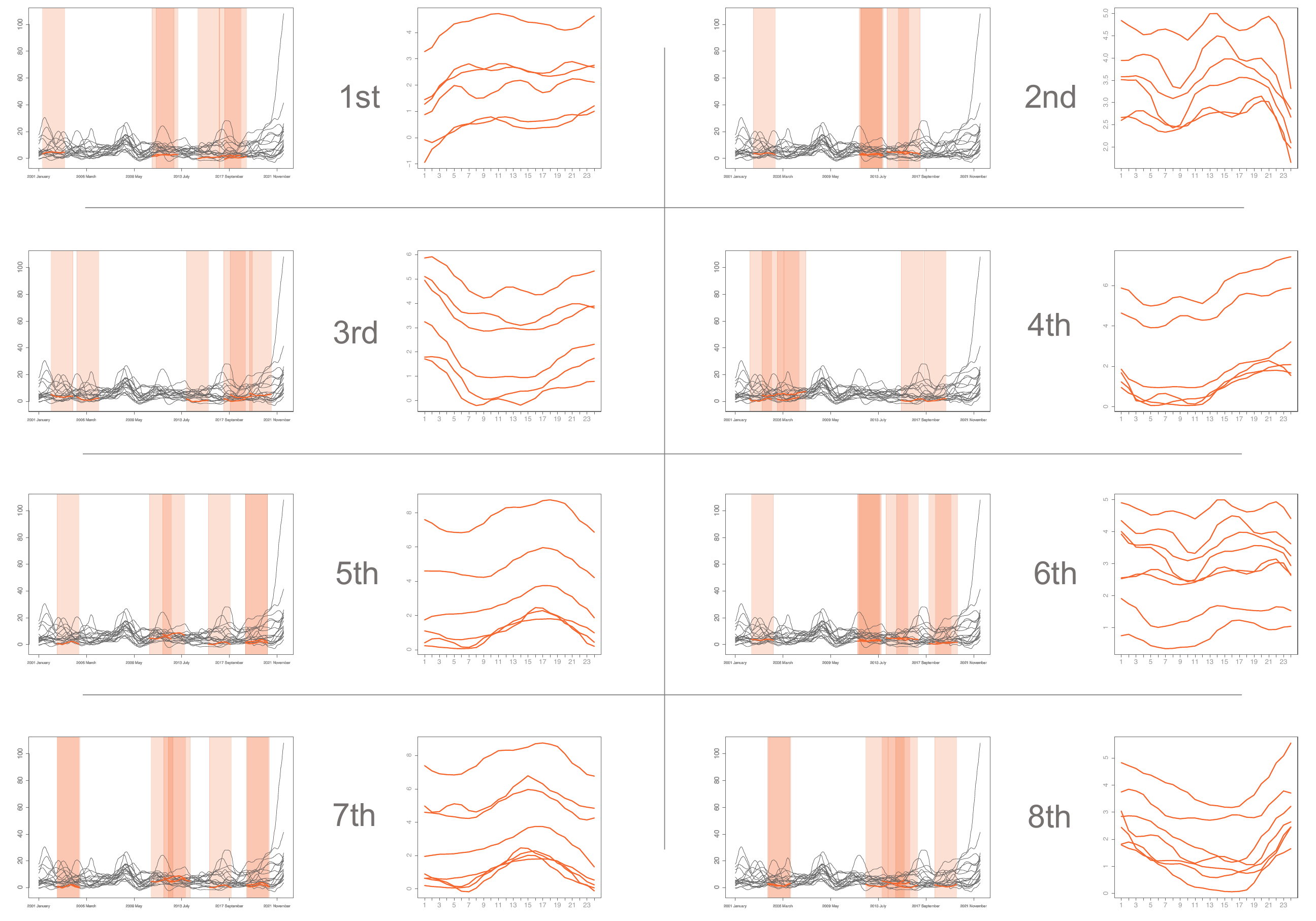}
\caption{
    Best $8$ motifs according to the adjusted fMSR criterion in the curves for the $19$ geographical regions. It is possible to notice how some motif presents occurrences with shared or overlapping domains.
}
\label{fig:scores_best8_regions}
\end{figure}

\begin{figure}[H]
\includegraphics[width=\linewidth]{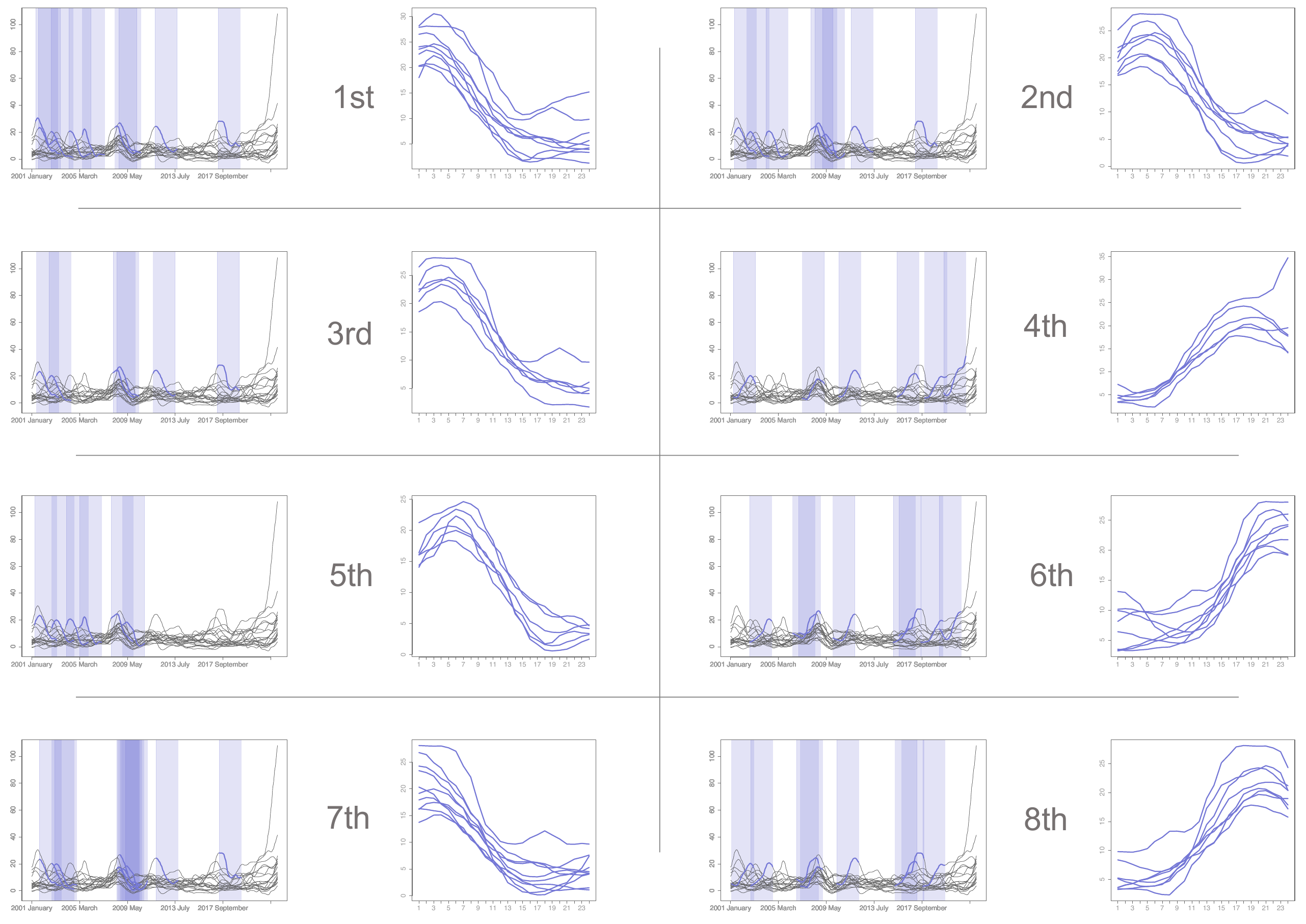}
\caption{
    Best $8$ motifs according to the variance criterion for the curves for the $19$ geographical regions. It is possible to notice how some motif presents occurrences with shared or overlapping domains.
}
\label{fig:var_best8_regions}
\end{figure}

\clearpage
\subsection{Case study 2: Temperature changes}
\label{sec:casestudies_temperature}

The $19$ curves used in this case study represent the temperature change with respect to a baseline climatology for the following regions:
\begin{itemize}
    \item Eastern Africa
    \item Middle Africa
    \item Northern Africa
    \item Southern Africa
    \item Western Africa
    \item Northern America
    \item Central America
    \item Caribbean
    \item South America
    \item Central Asia
    \item Eastern Asia
    \item Southern Asia
    \item South-eastern Asia
    \item Western Asia
    \item Eastern Europe
    \item Northern Europe
    \item Southern Europe
    \item Western Europe
    \item Oceania
\end{itemize}